\documentclass[twocolumn,final]{revtex4}
\usepackage{amsmath}
\usepackage{mathtools}
\usepackage{amssymb}
\usepackage{color}
\usepackage{graphicx}
\usepackage{hyperref}
\usepackage{url}
\newcommand{\al}{\alpha}
\newcommand{\be}{\beta}

\newcommand{\fr}{\frac}
\newcommand{\Ga}{\Gamma}
\newcommand{\ga}{\gamma}

\newcommand{\lf}{\left}

\newcommand{\pa}{\partial}

\newcommand{\rg}{\right}

\newcommand{\Si}{\Sigma}

\newcommand{\si}{\sigma}

\newcommand{\te}{\theta}

\newcommand{\LN}{\text{LN}}
\newcommand{\IGa}{\text{IGa}}
\newcommand{\GIGa}{\text{GIGa}}
\newcommand{\GGa}{\text{GGa}}
\newcommand{\Gadist}{\text{Ga}}

\begin{document}
\title{A Model for Stock Returns and Volatility}
\author{Tao Ma}
\affiliation{Department of Physics, University of Cincinnati, Cincinnati, OH 45221-0011}
\author{R.A. Serota}
\email{serota@ucmail.uc.edu}
\affiliation{Department of Physics, University of Cincinnati, Cincinnati, OH 45221-0011}
\begin{abstract}
We prove that Student's \emph{t}-distribution provides one of the better fits to returns of S\&P component stocks and the generalized inverse gamma distribution best fits VIX and VXO volatility data. We further argue that a more accurate measure of the volatility may be possible based on the fact that stock returns can be understood as the product distribution of the volatility and normal distributions. We find Brown noise in VIX and VXO time series and explain the mean and the variance of the relaxation times on approach to the steady-state distribution. 
\end{abstract}
\maketitle

\section{Introduction}

The generalized inverse gamma (GIGa) function (Appendix \ref{GIGa_Scale}) belongs to a family of distributions (Appendix \ref{GIGa_LN}), which includes inverse gamma (IGa), lognormal (LN), gamma (Ga) and generalized gamma (GGa). The remarkable property of GIGa is its power-law tail; for a general three-parameter case, the power-law exponent is given by the negative $1+\al\ga$, so that $\GIGa(x; \al,\be,\ga)\propto x^{-1-\al\ga}$, $x\rightarrow \infty$. GIGa emerges as a steady state distribution in a number of problems, from a network model of economy, \cite{ma2013distribution} to ontogenetic mass growth, \cite{west2012} to response times in human cognition. \cite{ma12RT} This common feature can be traced to a birth-death phenomenological model subject to stochastic perturbations (Appendix \ref{Birth_Death}). Here we argue that the GIGa distribution best describes stock volatility distribution and the product distribution (Appendix \ref{Product_Distribution}) of GIGa and normal (N) distributions, GIGa*N, best describes stock returns distribution.

Numerically, we used the maximum likelihood method to determine the best parameters for each of the distributions in the above family of distributions and found that GIGa provides the best fit for VIX and VXO volatility data. We also found that among product distributions of the above family with normal distribution, GIGa's product with N gives the best fit to the stock returns distribution. Furthermore, among the better GIGa*N fits are those with $\ga \approx 2$.

In general, product distribution GIGa*N has $|x|^{-1-\al\ga}$ tails [left and right]. For $\ga = 2$, the product distribution $\GIGa(\al,\be,2)*\text{N}$ for stock returns is Student's \emph{t}-distibution, which has $|x|^{-1-2\al}$ tails. \cite{praetz1972, platen2008, gerig2009}. Accordingly, our starting point is the geometric Brownian motion model of stock price, \cite{hull1997,gatheral2006} where the steady-state distribution of stock returns is given by the product distribution of volatility and normal distributions. Further, the instantaneous variance of volatility (or square stochastic volatility - the terms used interchangeably) is described by the Nelson diffusion limit (NDL) of GARCH$(1,1)$ model of stock volatility \cite{nelson1990, duan1995}, whose stochastic term is uncorrelated from that in the equation for stock price; in the steady state, it is distributed as IGa, that is GIGa with  $\ga = 1$.

This paper is organized as follows. In Sec. II, we discuss stochastic stock and volatility models. In Sec. III, we fit VIX and VXO, including direct evaluation of their power law tail exponents by log-log plot. We also address Brown noise observed in the VIX/VXO time series. In Sec. IV, we discuss numerical results of fitting returns of S\&P component stocks \footnote{DJIA components are fitted in the same fashion leading to identical conclusions, which is described elsewhere.} based on log-likelihood and discuss white noise in stock return series. In Sec. V, we summarize our key findings.

\section{Stochastic stock and volatility models}

The widely accepted equation for stock price is given by
\begin{equation}\label{dSS}
\fr{dS}{S} = \mu dt + \si dW_1 . 
\end{equation}
where $\mu$ is a constant and $\si$ volatility. 
The equation for the instantaneous volatility variance (square volatility) can be written in the following general form:
\begin{equation}\label{dV}
dV =\tilde{ f}(V) dt + \tilde{ g}(V) dW_2 . 
\end{equation}
Here $dW_1$ and $dW_2$ are Wiener processes correlated by $\langle dW_1 dW_2 \rangle = \rho dt$. Substituting $V=\si^2$ and using Ito calculus, we obtain the volatility equation 
\begin{equation}\label{dsigma}
d\si = f(\si) dt + g(\si) dW_2 . 
\end{equation}
The Fokker-Planck equation for the distribution function of $\si$, $P(\si,t)$, is given by
\begin{equation}
\fr{\pa}{\pa t} P(\si,t) 
= \fr{1}{2}\fr{\pa^2}{\pa \si^2} [g^2(\si) P(\si,t)]
-\fr{\pa}{\pa \si}[f(\si) P(\si,t)] . 
\end{equation}
It has a stationary (steady-state) solution given by
\begin{equation}\label{P_integral}
P(\si) = \fr{2}{g^2} \exp({\int \fr{2f}{g^2} d\si}) .
\end{equation}

In what follows, we shall assume that $dW_1$ and $dW_2$ are uncorrelated, that is $\rho=0$. A number of possible forms of $f(\si)$ and $g(\si)$ are discussed in Appendix \ref{SDE_Volatility}; see also \cite{wiggins1987}. Here we concentrate on one particular form 
\begin{equation}\label{ST:eq:GIGa_SDE}
d\si = J(\te \si^{1-\ga}-\si)dt + \Si \si dW_2 ,
\end{equation}
The stationary (see Appendix \ref{Relax_Time} for discussion of relaxation times) solution of this equation is given by
\begin{equation}\label{ST:eq:GIGa_from_SDE}
\begin{split}
P(\si) &=\GIGa(\si ; \al,\be,\ga) \\
&=\text{GIGa}\lf(\si ;  \lf(1+\fr{2J}{\Si^2}\rg) \ga^{-1} , \lf({\te}\fr{2J}{{\Si^2}}\ga^{-1}\rg)^{1/\ga},\ga\rg) ,
\end{split}
\end{equation}
where the parameter $\te$ can be expressed using the mean $\overline{\si}$ as 
\begin{equation}\label{ST:eq:GIGa_SDE_theta}
\te 
= \fr{\ga\Si^2}{2J} 
\lf[ \fr{\overline{\si} \Ga((1+\fr{2J}{\Si^2})\ga^{-1})}{\Ga(\fr{2J}{\Si^2} \ga^{-1})} \rg]^\ga . 
\end{equation}
In particular, when $\ga=1$, $\te=\overline{\si}$.

A case of particular importance is $\gamma=2$, in which case the equation for the volatility variance is that of $\gamma=1$ and reads as follows:
\begin{equation}\label{ST:eq:IGa_SDE_V}
dV = \tilde{ J}(\overline{V}-V)dt + \tilde{\Si} V dW_2 .
\end{equation}
Its stationary solution is given by the IGa distribution,
\begin{equation}
P(V) = 
\text{IGa}\lf(V; 1+\fr{2\tilde{J}}{\tilde{\Si}^2}, \overline{V} \fr{2\tilde{J}}{\tilde{\Si}^2}\rg) . 
\end{equation}
Using $V=\si^2$ and Ito calculus, we obtain
\begin{equation}
d\si = \lf(\fr{\tilde{J}}{2}\overline{V} \si^{-1} - 
\lf(\fr{\tilde{J}}{2} + \fr{\tilde{\Si}^2}{8}\rg) \si\rg) dt + 
\fr{\tilde{\Si}}{2} \si dW_2 .
\end{equation}
On comparison with Eq. (\ref{ST:eq:GIGa_SDE}), we find the following parameter correspondence:
\begin{equation}
\begin{split}
&J = \fr{\tilde{J}}{2} + \fr{\tilde{\Si}^2}{8} \\
&\te = \overline{V}\fr{\tilde{J}}{2}\lf(\fr{\tilde{J}}{2} + \fr{\tilde{\Si}^2}{8}\rg)^{-1} \\
&\Si = \fr{\tilde{\Si}}{2} \\
&\ga=2 .
\end{split}
\end{equation}
Substitution into Eq. (\ref{ST:eq:GIGa_from_SDE}), gives the distribution of $\si$ as, 
\begin{equation}\label{ST:eq:GIGa_SDE_V}
\text{GIGa}\lf(\si; 1+\fr{2\tilde{J}}{\tilde{\Si}^2}, \sqrt{\overline{V} \fr{2\tilde{J}}{\tilde{\Si}^2}}, 2\rg) ,
\end{equation}
It should be  emphasized that a simple change of the variate to its square root produces the following transformation: $\GIGa(\al,\be,\ga) \rightarrow \GIGa(\al,\sqrt{\be},2\ga)$ and in particular $\IGa(\al,\be,1) \rightarrow \GIGa(\al,\sqrt{\be},2)$, which is consistent with (\ref{ST:eq:IGa_SDE_V}) and (\ref{ST:eq:GIGa_SDE_V}). From  (\ref{ST:eq:GIGa_SDE_theta}), the mean of $\si$ is given by
\begin{equation}\label{ST:eq:IGa_V_mean_sigma}
\sqrt{\overline{V}} \sqrt{\fr{2\tilde{J}}{\tilde{\Si}^2}}
\fr{\Ga\lf( \fr{2\tilde{J}}{\tilde{\Si}^2} + \fr{1}{2} \rg)}
{\Ga\lf( \fr{2\tilde{J}}{\tilde{\Si}^2} + 1 \rg)} ,
\end{equation}
which is a monotonically increasing function which approaches $\sqrt{\overline{V}}$ as $2\tilde{J}/\tilde{\Si}^2\rightarrow \infty$.

Turning to Eq. (\ref{dSS}), we observe that the stationary distribution of stock returns is a product distribution $P(\si)*\text{N}$. In Appendix \ref{Product_Distribution}, we consider both formalism of the product distribution and various cases of $P(\si)$. Here we concentrate specifically on 
\begin{equation}
\text{GIGa}(\al,\be,2) * \text{N}(0,1)
= \fr{\Gamma\lf(\fr{1}{2}+\al\rg)}{\sqrt{2\pi} \be\Gamma(\al)} \lf(\fr{2\be^2}{z^2+2\be^2}\rg)^{\fr{1}{2}+\al} ,
\end{equation}
which is the generalized Student's \emph{t}-distribution T$(0,{\be}/{\sqrt\al}, 2\al)$ \cite{jackman2009}. 

It should be mentioned that by Ito calculus and Eq. (\ref{dSS})
\begin{equation}\label{dlogS}
d\log S 
= (\mu - \fr{1}{2}\si^2) dt + \si dW_1 . 
\end{equation}
In numerical calculations of stock returns, it is actually the $\Delta\log S$ that is being evaluated. However, it is clear that the premise of the stationary distribution being the product distribution of the volatility distribution and the normal distribution remains in force.

\section{Market volatility}

\begin{figure}[htp]
\centering
\includegraphics[width=0.23\textwidth]{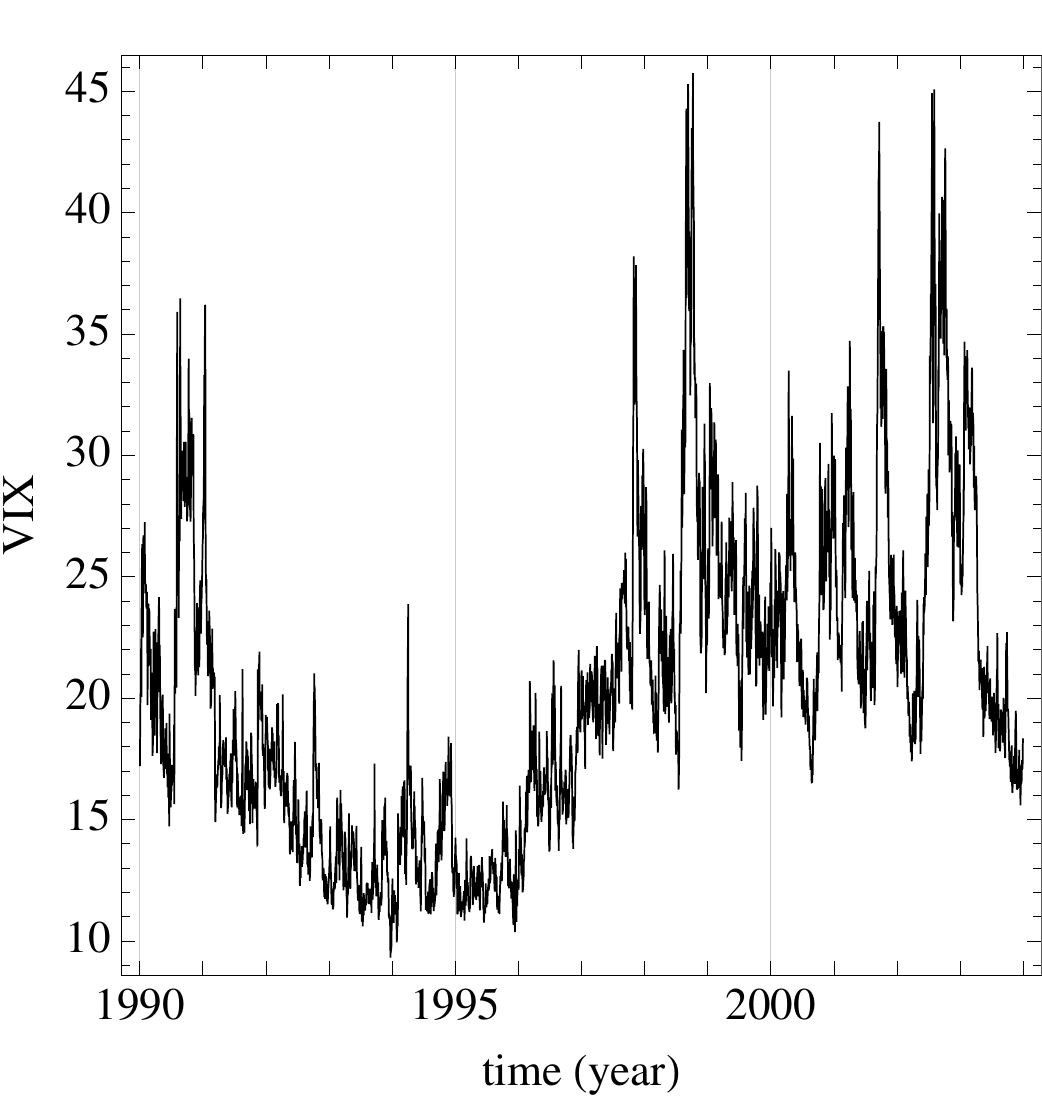}
\includegraphics[width=0.23\textwidth]{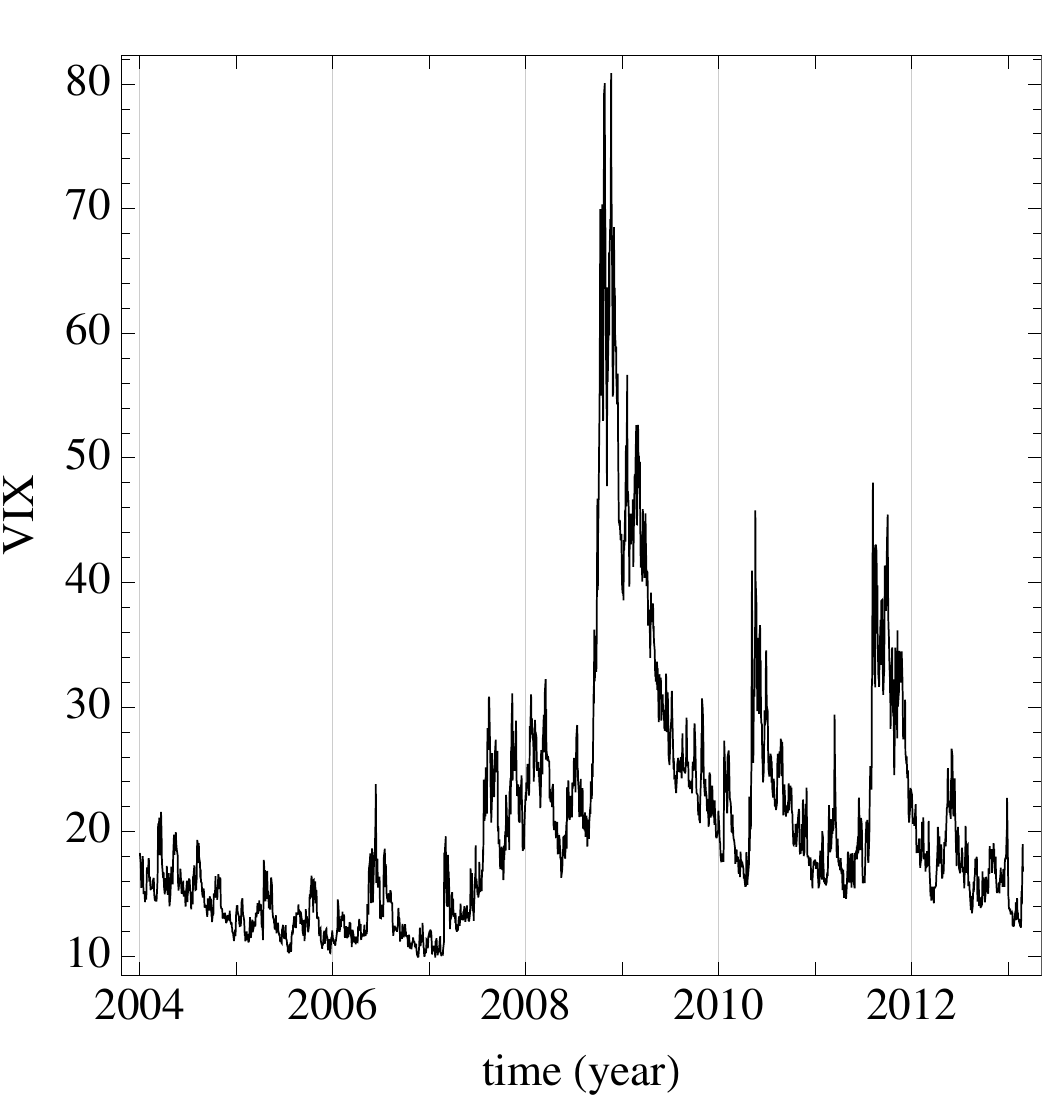}\\
\includegraphics[width=0.23\textwidth]{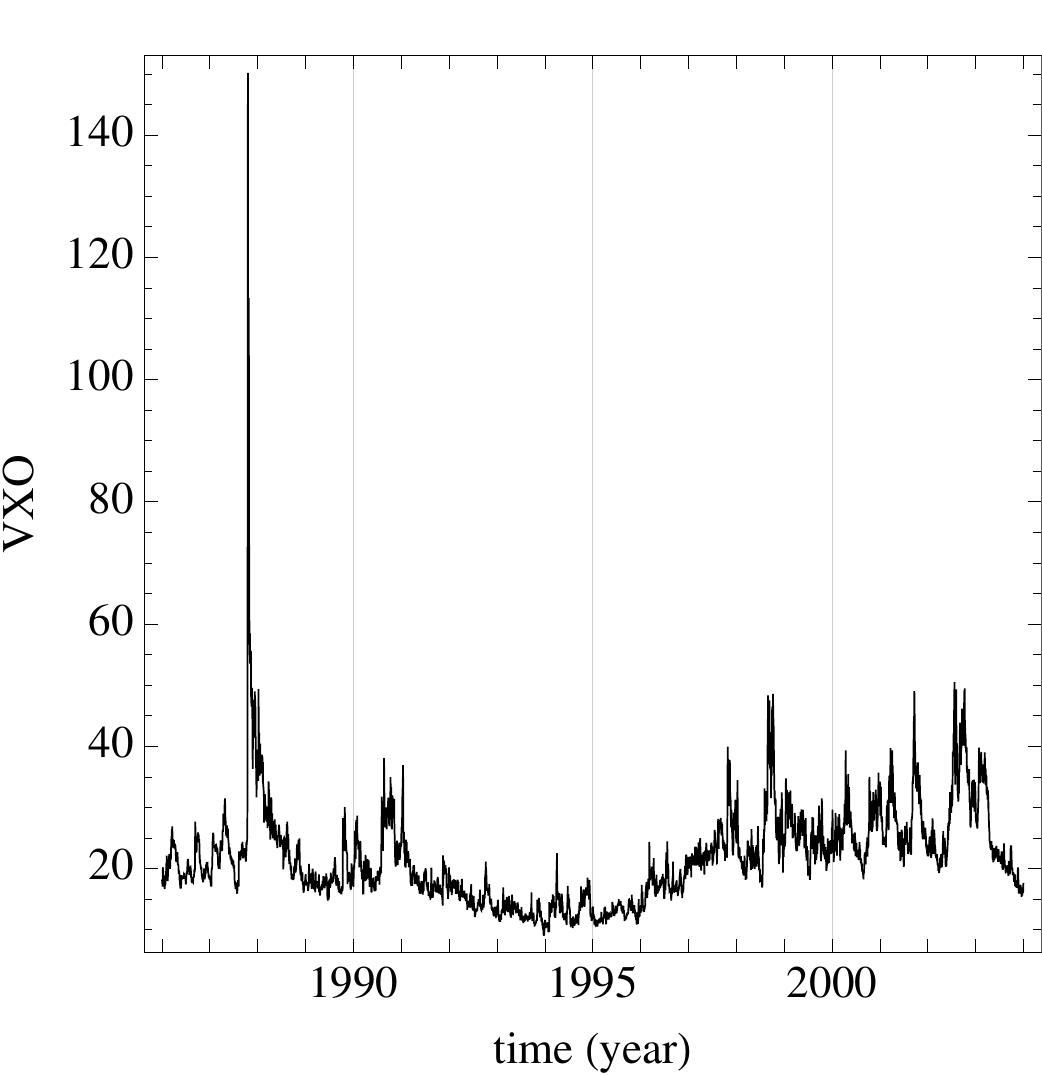}
\includegraphics[width=0.24\textwidth]{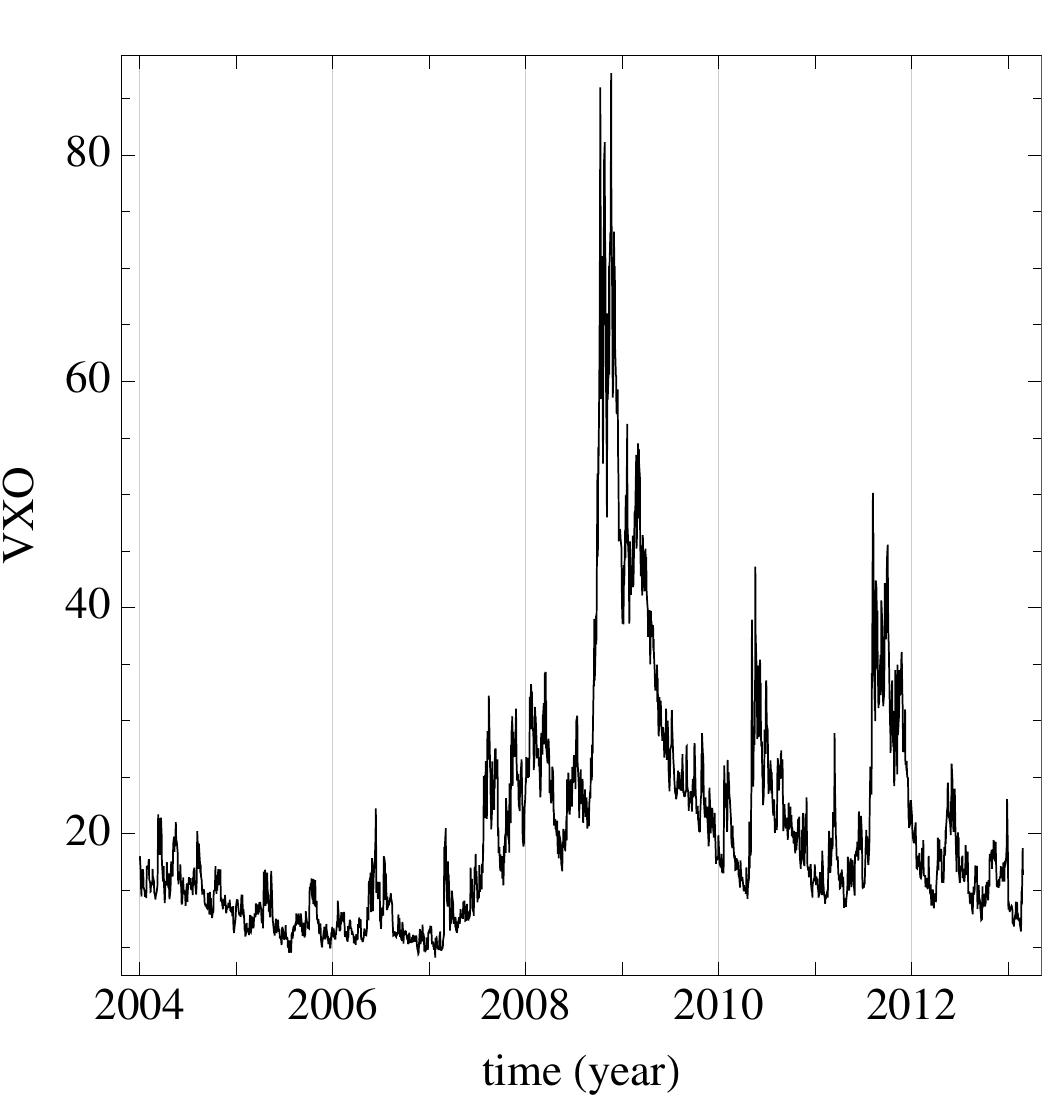}
\caption{VIX and VXO plots. }
\label{ST:fig:VIX:VXO:listplot}
\end{figure}

We analyze the Chicago Board Options Exchange (CBOE) volatility index \cite{CBOE_website, VIX_CBOE, VXO_CBOE}. On September 22, 2003, CBOE decided to change the manner in which it calculated the volatility index from VXO to VIX. However, both methods were applied to both the old and new data. Following CBOE convention \cite{CBOE_website}, the VIX/VXO data from 1990 to 2004 are called vixarchive/vxoarchive and from 2004 to present vixcurrent/vxocurrent. In Fig. \ref{ST:fig:VIX:VXO:listplot}, we show the time series of the indices.

\begin{figure}[htp]
\centering
\includegraphics[width=0.46\textwidth]{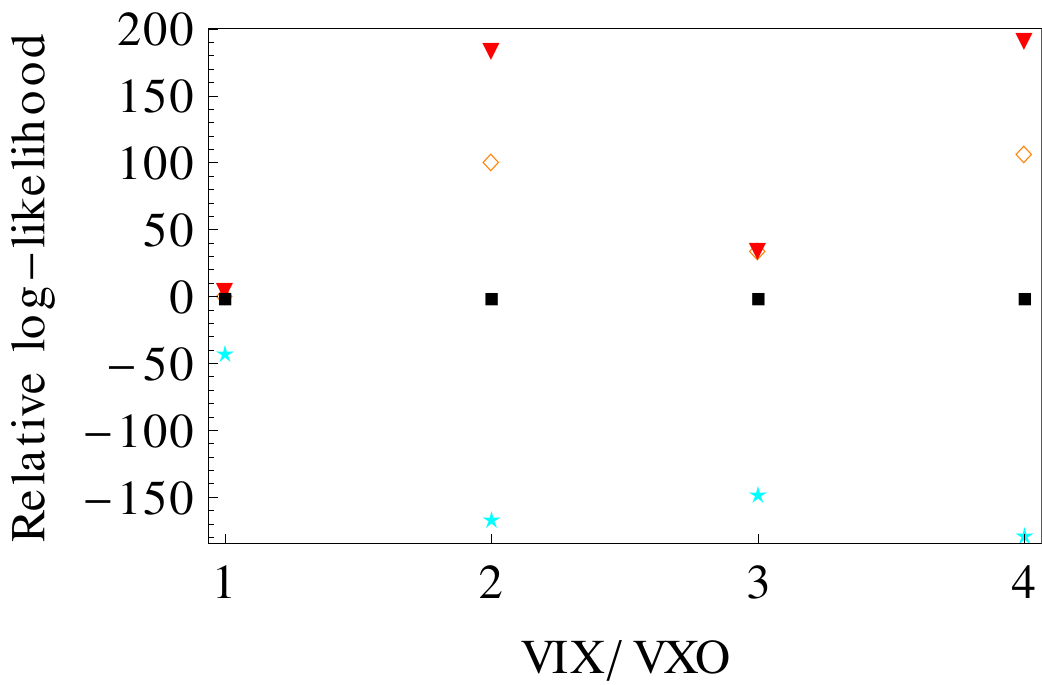}
\caption{Log-likelihood of fitting distribtuion relative to LN. Red down-pointing triangles: GIGa; orange diamonds: IGa; cyan stars: Ga; and black squares: LN. 1 at the x-axis is for vixarchive, 2 for vixcurrent, 3 for vxoarchive, and 4 for vxocurrent.}
\label{ST:fig:VIX:VXO:loglikelihood}
\end{figure}

We  apply the maximum likelihood estimation method (Appendix \ref{Max_Likelihood}) to find the best fitting parameters of  IGa, GIGa, Ga, and LN summarized in Table \ref{ST:tbl:VIX_dist_parameter}. Comparison of loglikelihood in Fig. \ref{ST:fig:VIX:VXO:loglikelihood} shows that the goodness of fit decreases in the following order: GIGa, IGa, LN, and Ga.

\begin{figure}[htp]
\centering
\includegraphics[width=0.345\textwidth]{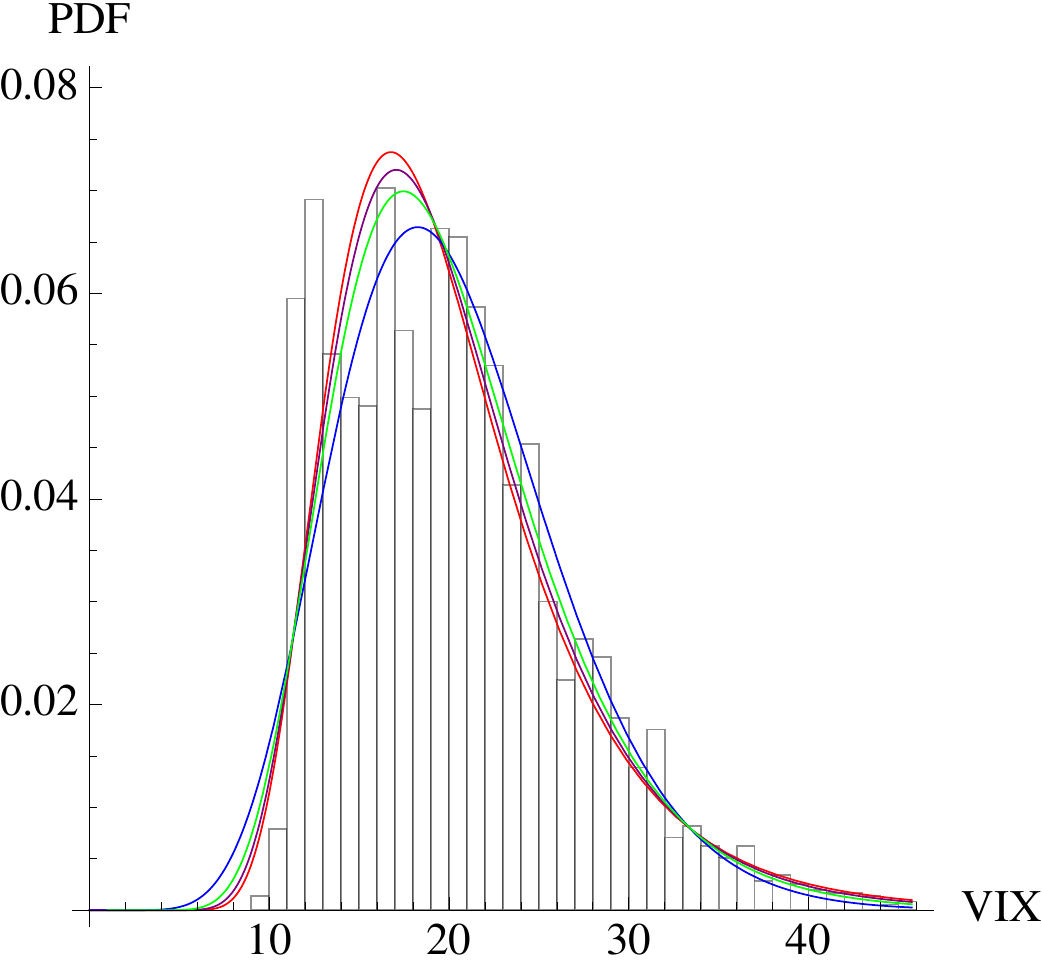}
\includegraphics[width=0.345\textwidth]{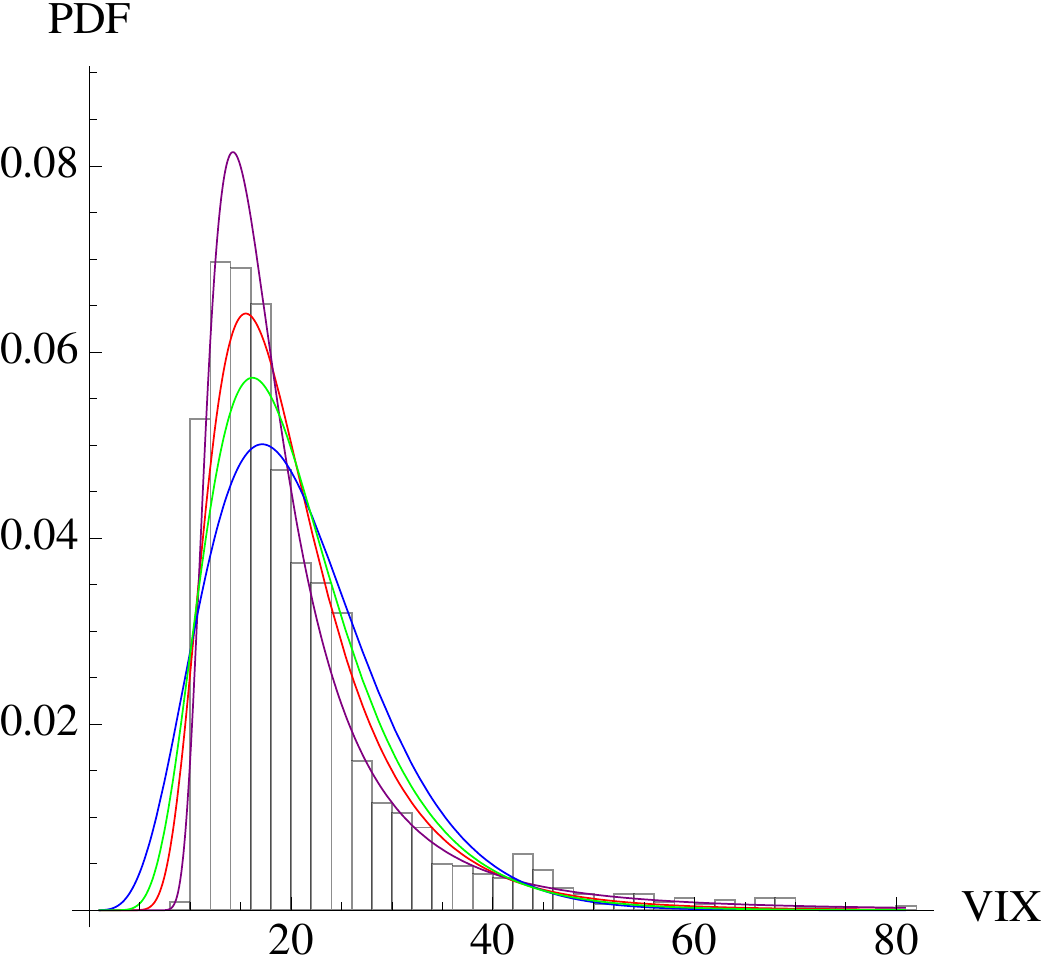}\\
\includegraphics[width=0.345\textwidth]{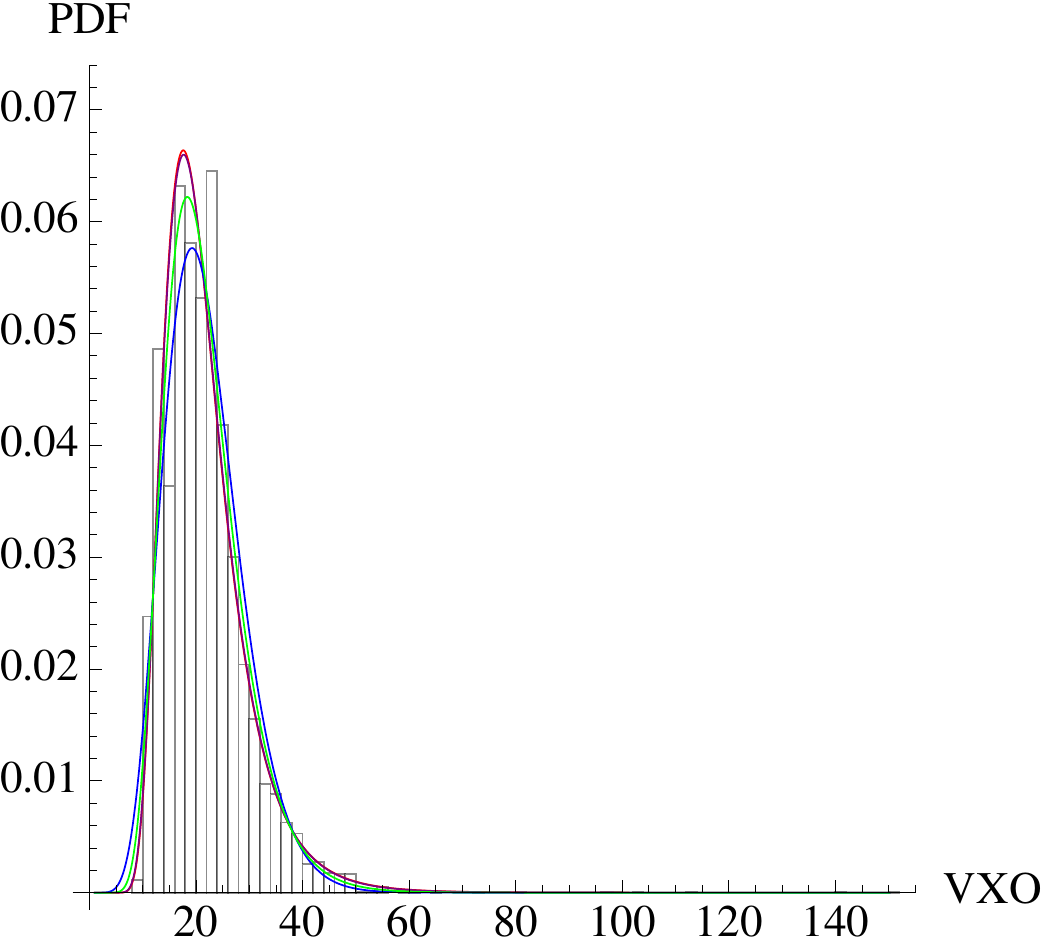}
\includegraphics[width=0.345\textwidth]{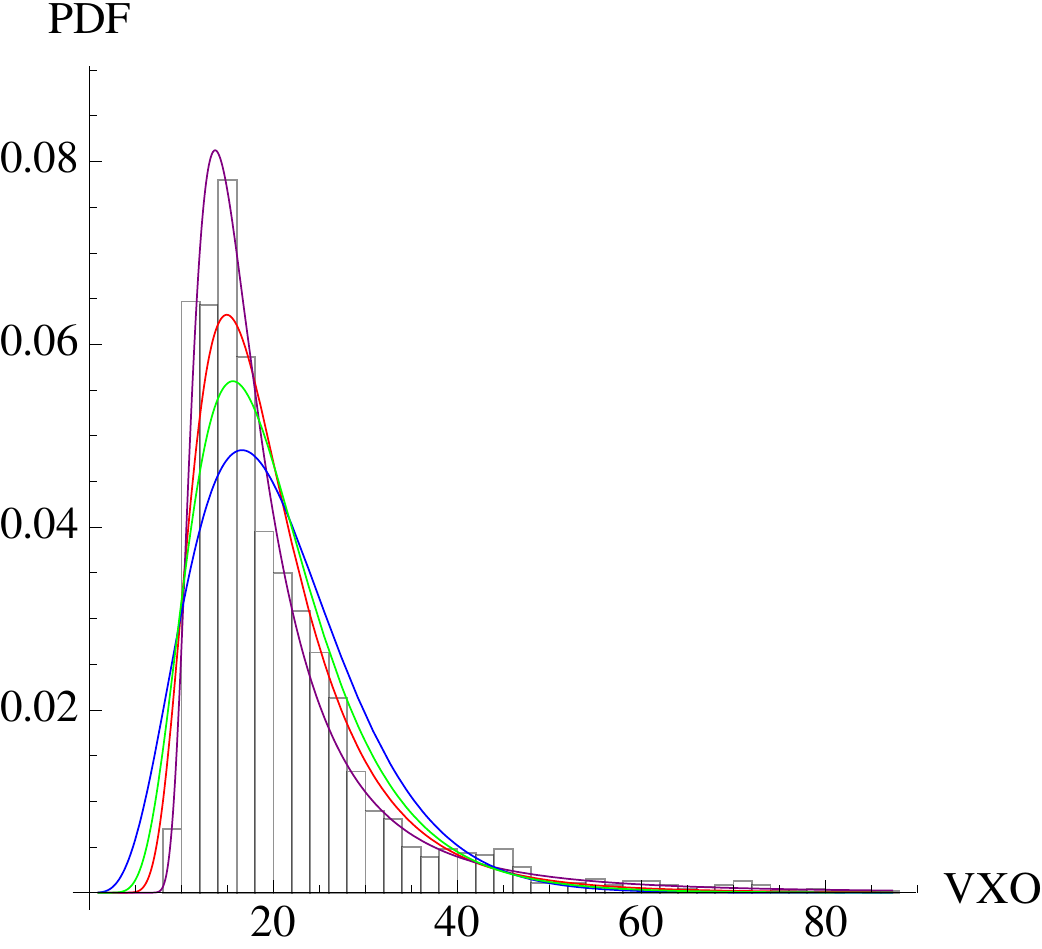}
\caption{Histogram plots of VIX and VXO. Top down: vixarchive, vixcurrent, vxoarchive, and vxocurrent. Red: IGa; purple: GIGa; blue: Ga; and green: LN. }
\label{ST:fig:VIX:VXO:histogram}
\end{figure}

\begin{figure}[htp]
\centering
\includegraphics[width=0.23\textwidth]{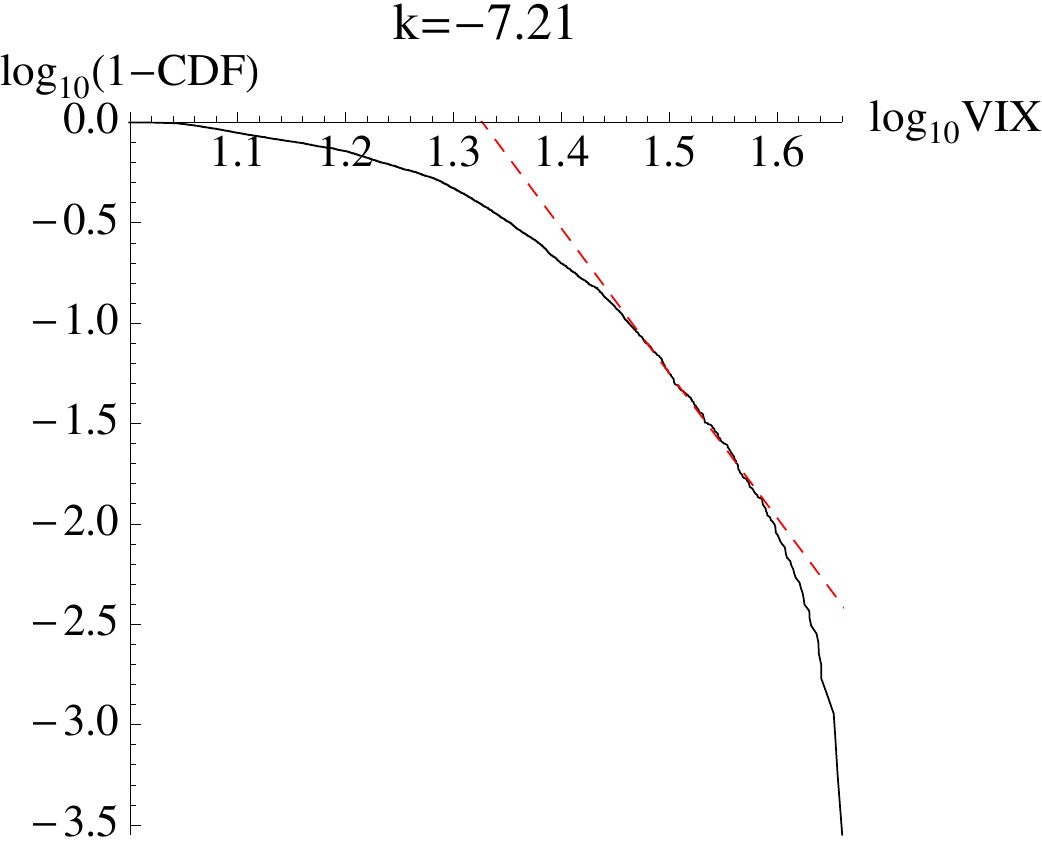}
\includegraphics[width=0.23\textwidth]{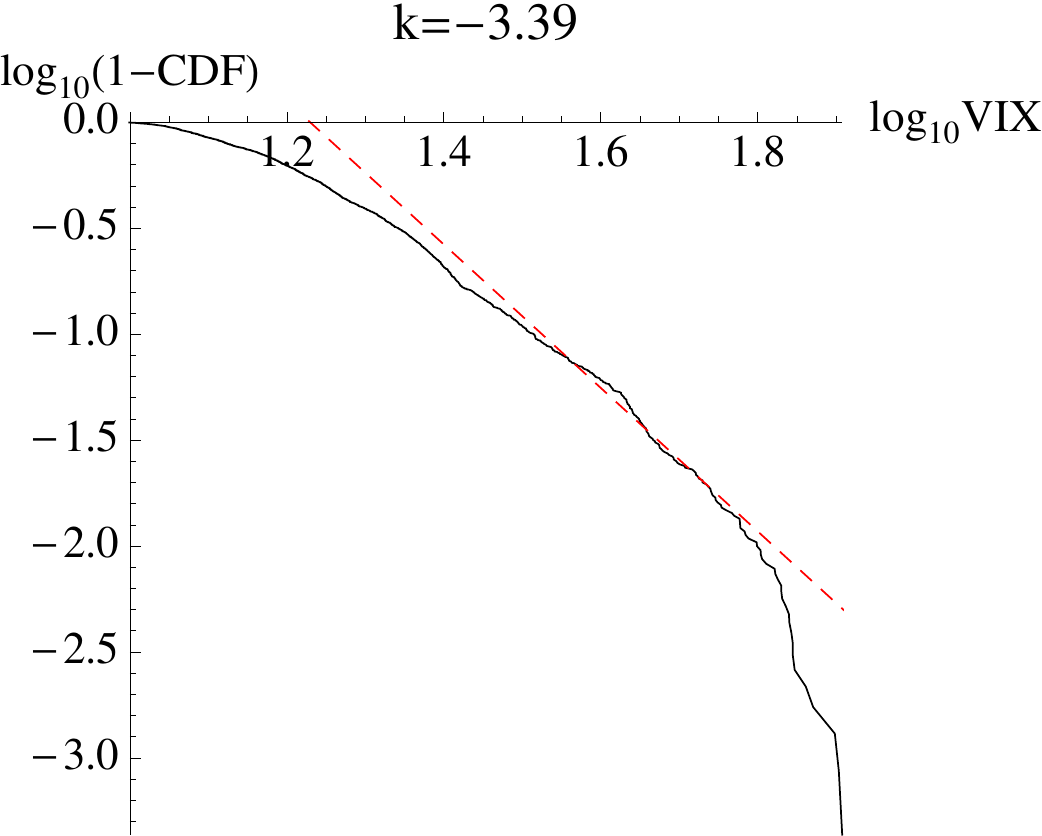}\\
\includegraphics[width=0.23\textwidth]{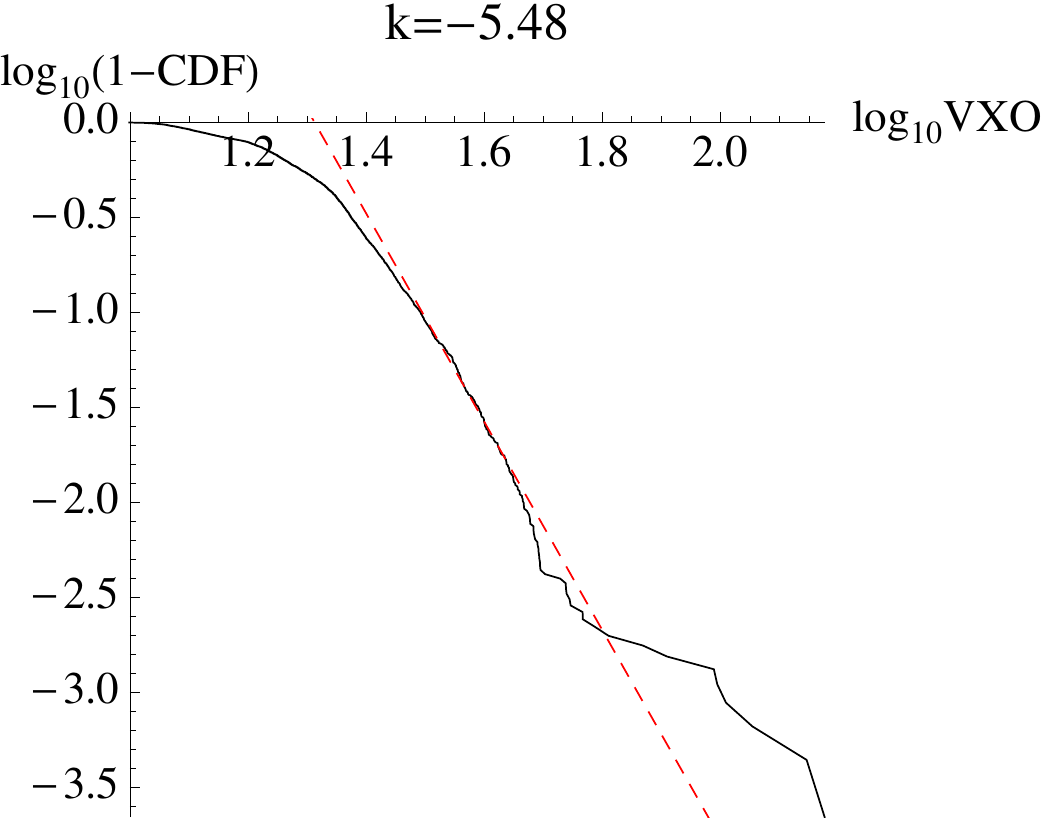}
\includegraphics[width=0.23\textwidth]{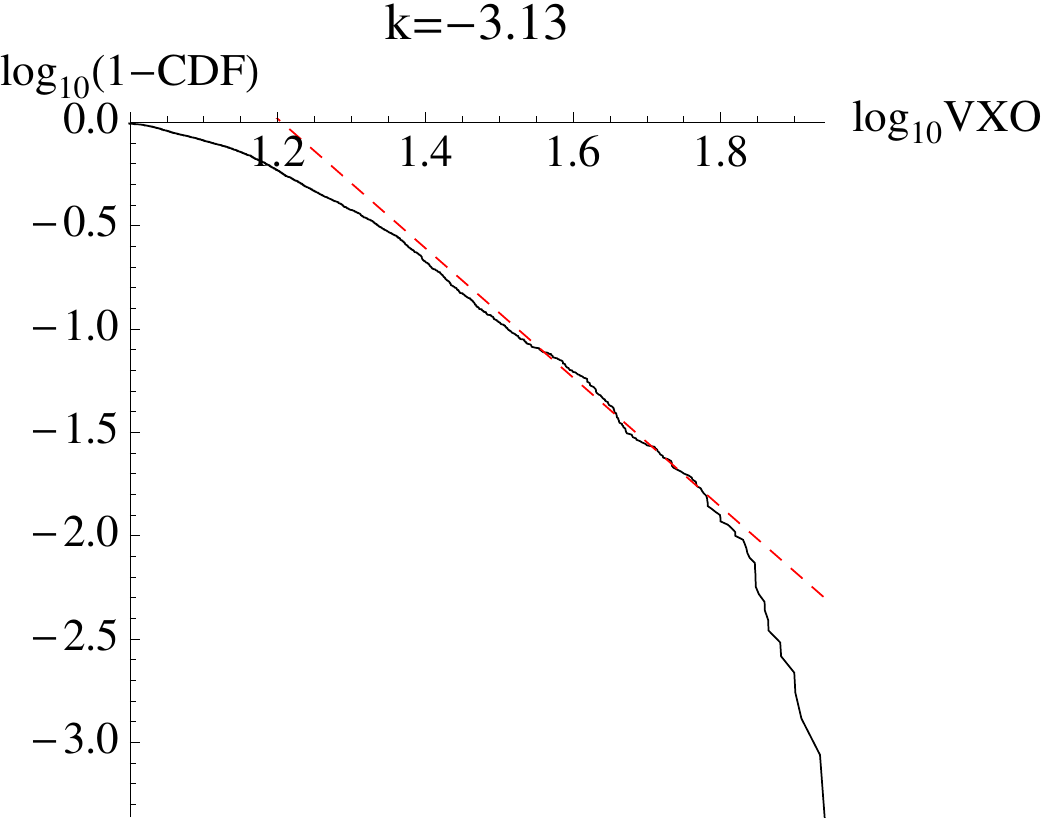}
\caption{Log-log plots of VIX and VXO. Black curves: log-log plots; red dashed lines: linear fit of the log-log plots from $1-CDF=0.01$ to $0.1$. }
\label{ST:fig:VIX:VXO:loglogplot}
\end{figure}

In Fig. \ref{ST:fig:VIX:VXO:histogram}, we plot  histograms of vixarchive, vixcurrent, vxoarchive, and vxocurrent respectively, fitted with the best GIGa, IGa, LN, and Ga. We also measure the exponent of the power law tail of VIX and VXO directly (Apendix \ref{Tail_Power}), as shown in Fig. \ref{ST:fig:VIX:VXO:loglogplot} and Table \ref{ST:tbl:VIX_exponent}.

\begin{figure}[htp]
\centering
\includegraphics[width=0.23\textwidth]{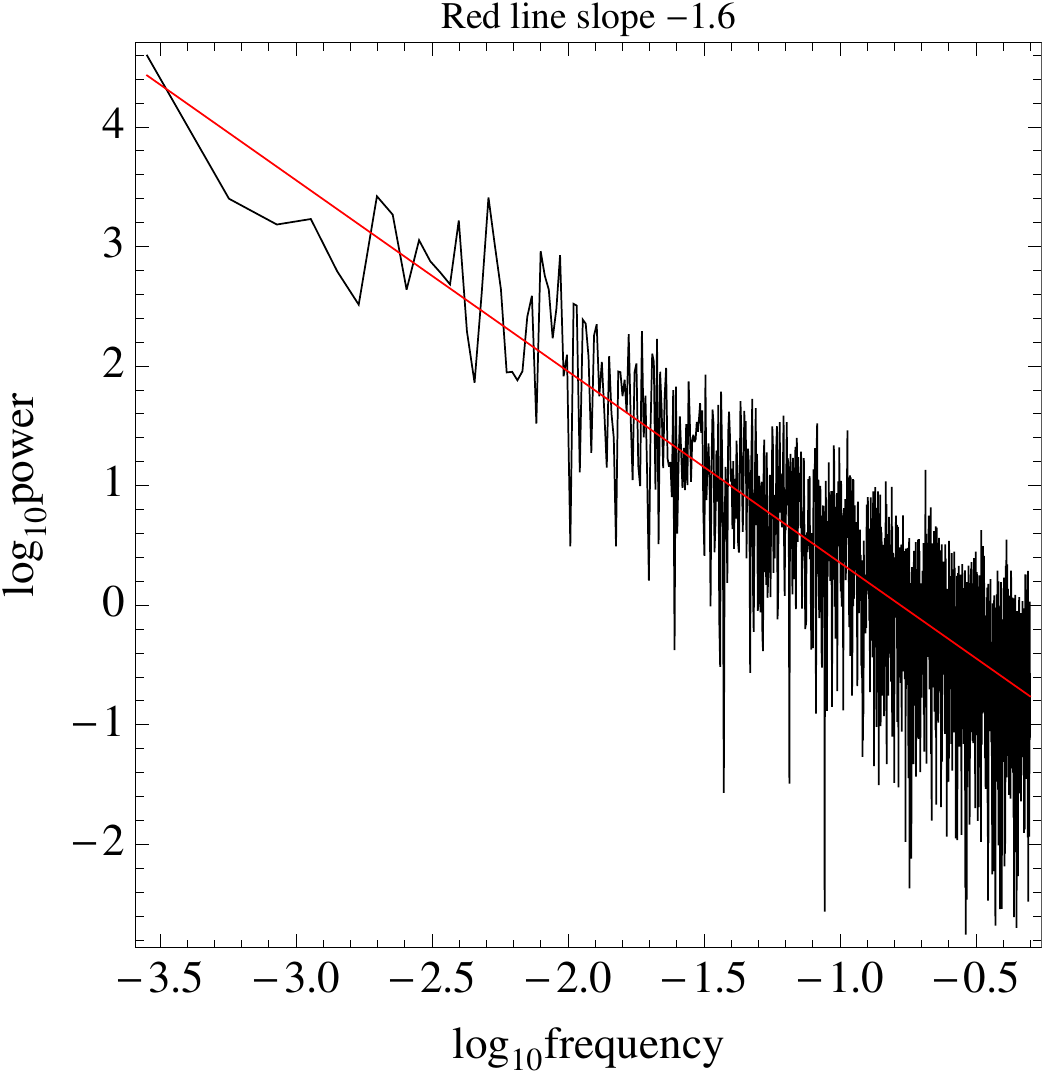}
\includegraphics[width=0.23\textwidth]{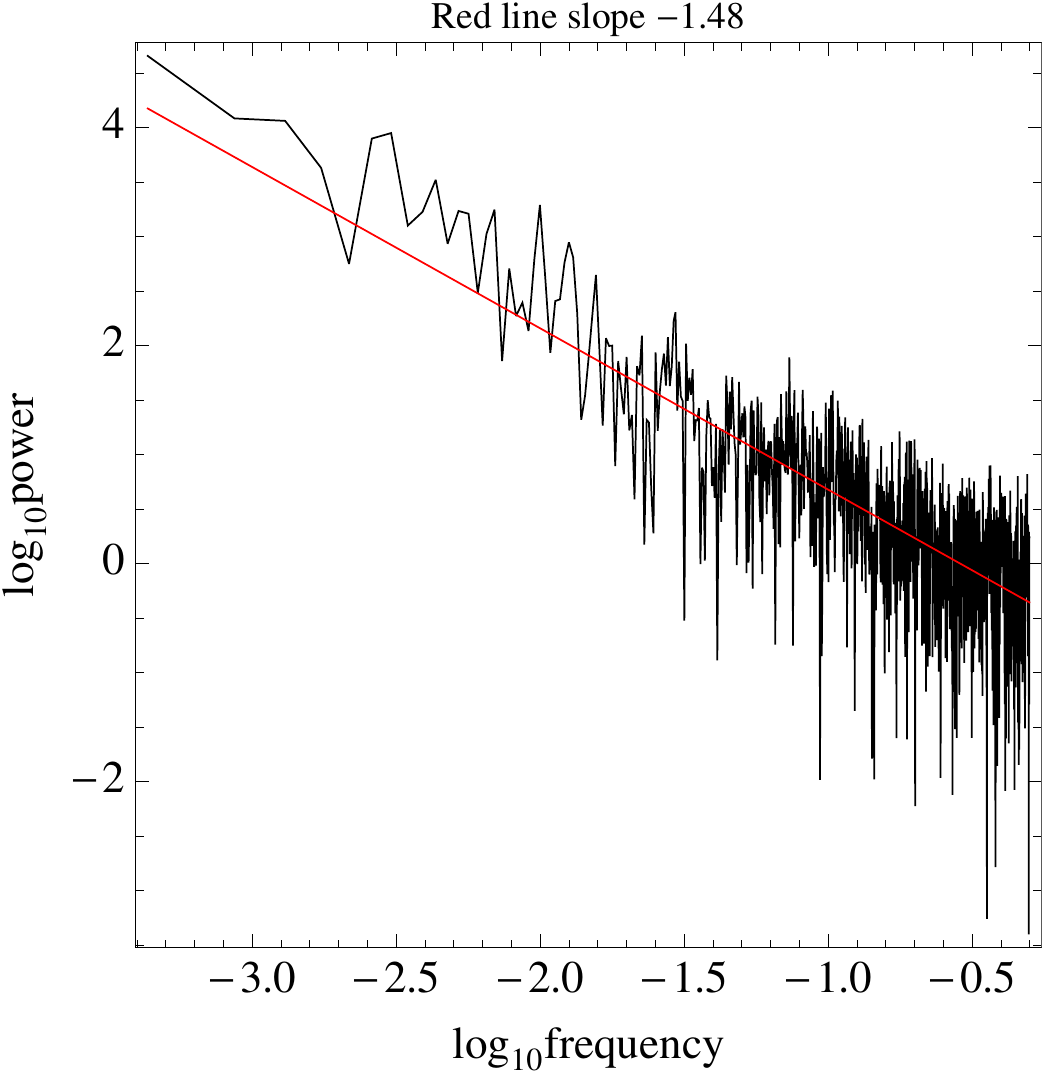}\\
\includegraphics[width=0.23\textwidth]{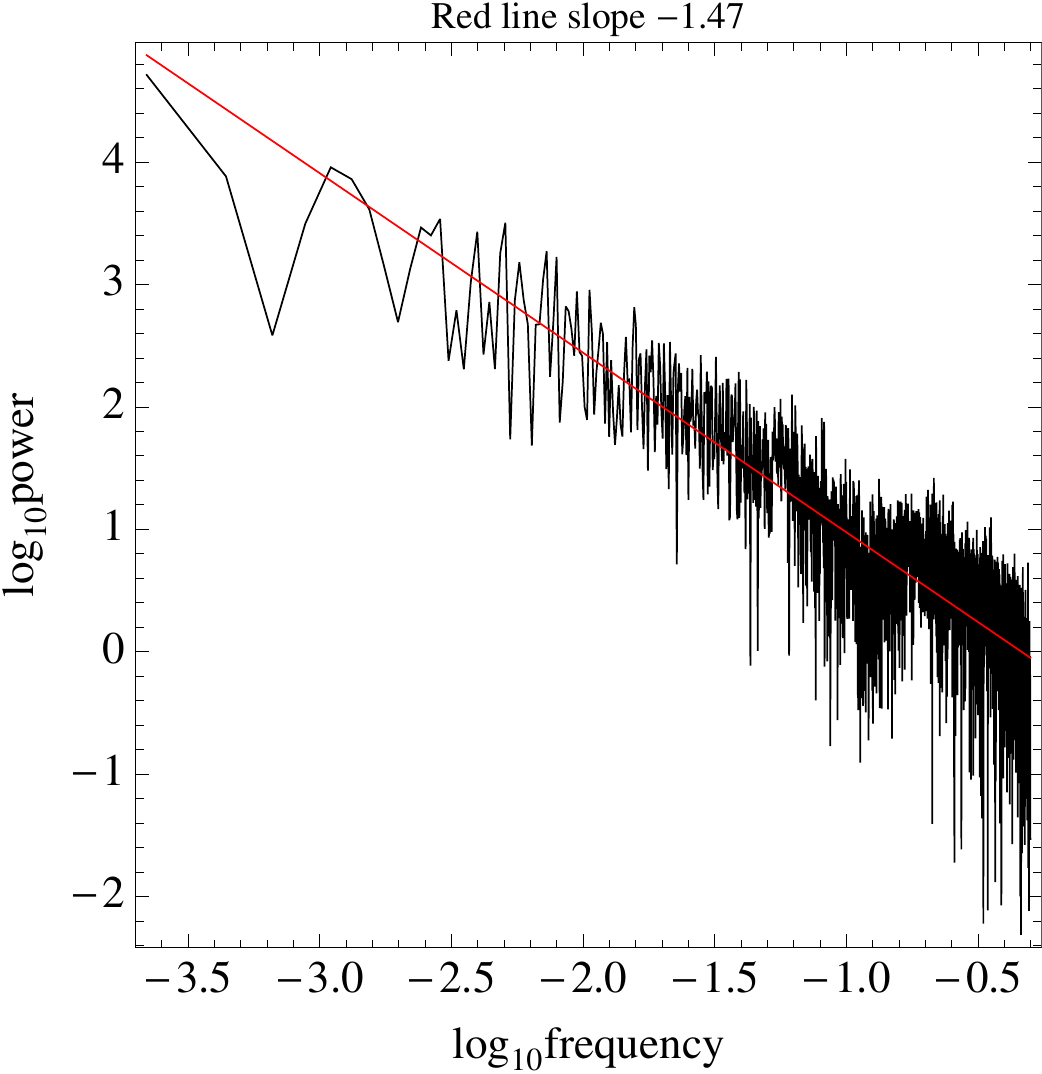}
\includegraphics[width=0.23\textwidth]{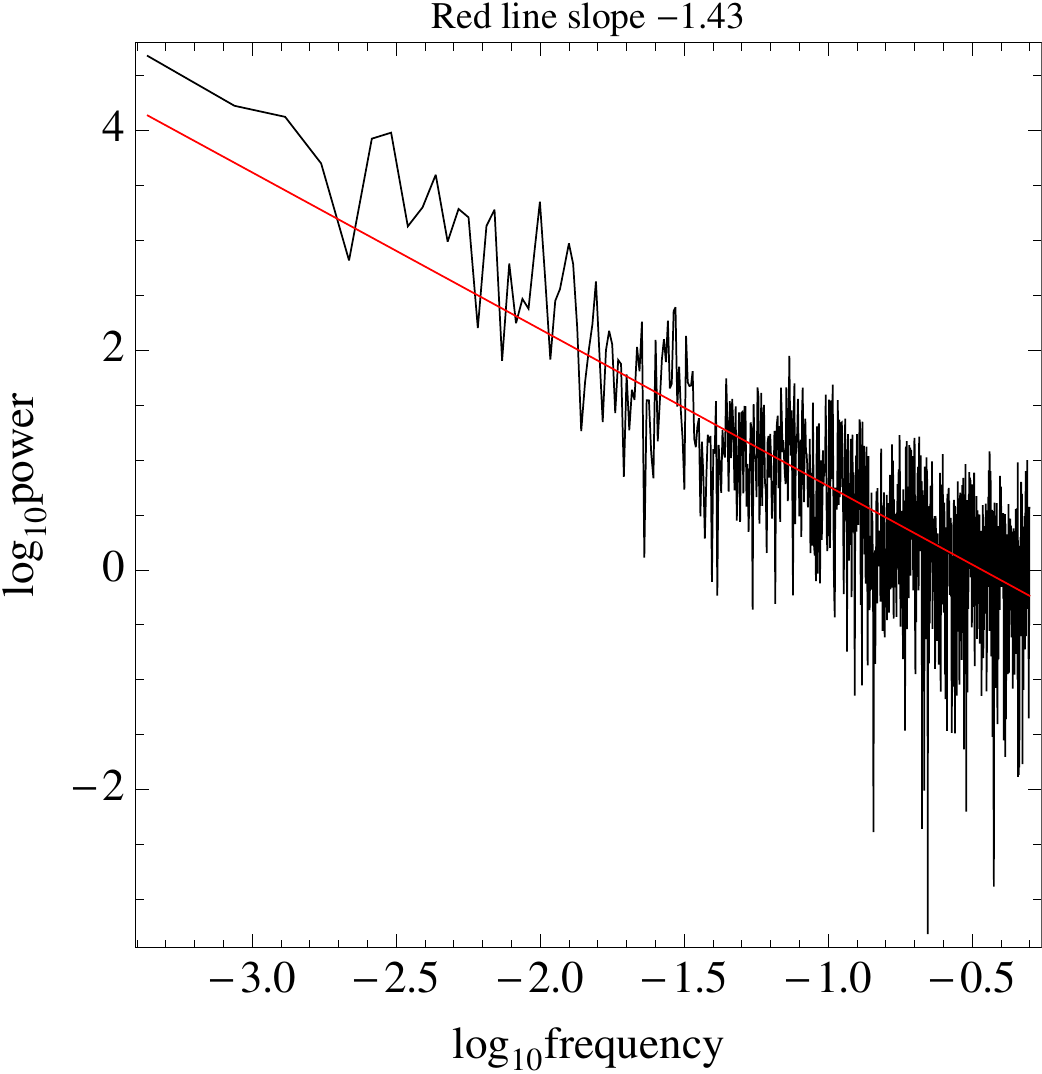}
\caption{Fourier transform of VIX and VXO time series. Jagged black lines: discrete Fourier transform; red lines: linear fit of the black lines.}
\label{ST:fig:VIX:VXO:DFT}
\end{figure}

\begin{figure}[htp]
\centering
\includegraphics[width=0.345\textwidth]{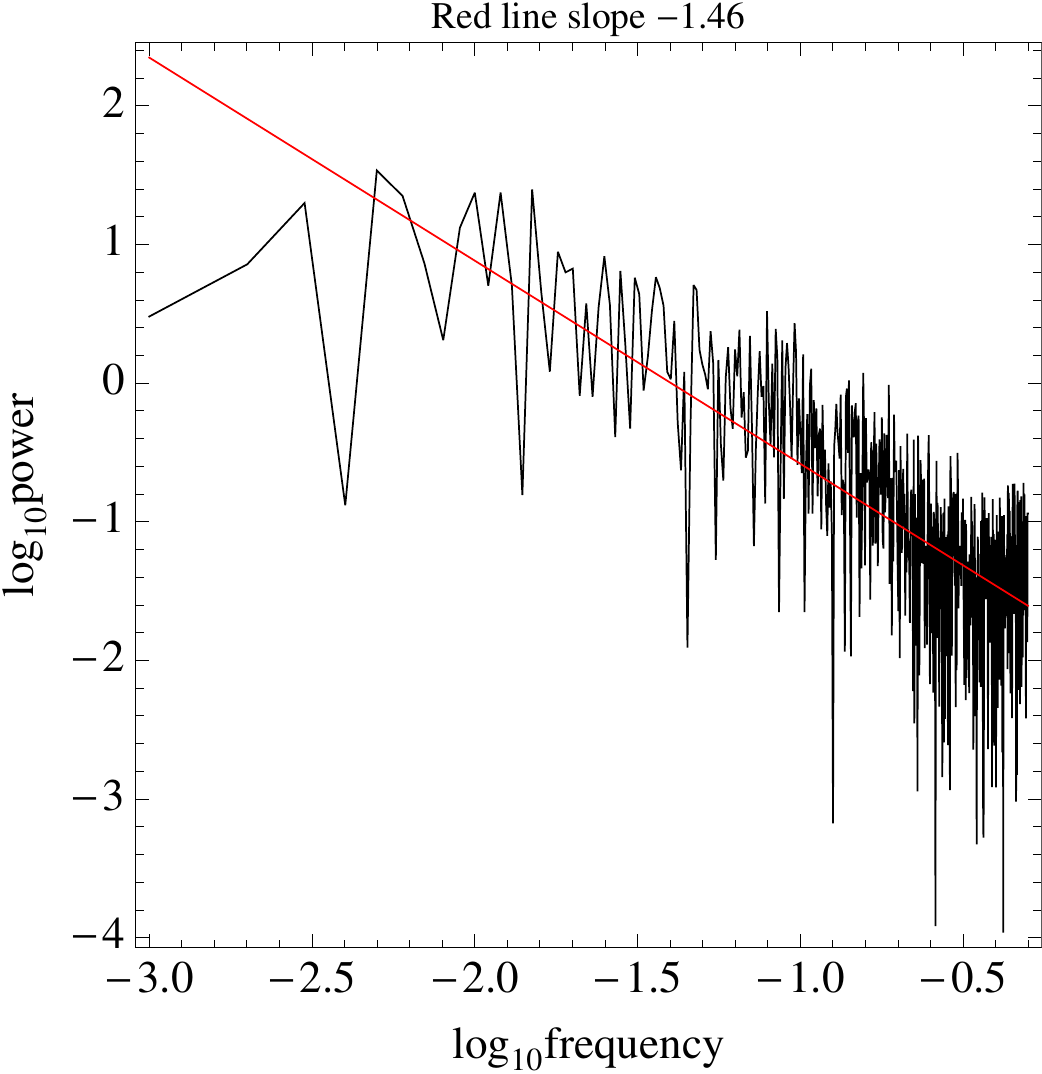}
\caption{Discrete Fourier transform of a time series of IGa process
with $J=0.1$, $\si=\sqrt{0.1}$ at time 1, 2, ..., 1000. Jagged black
line: discrete Fourier transform; red line: linear fit of the black
lines.}
\label{ST:fig:IGa:simulation:DFT}
\end{figure}

Finally, in Fig. \ref{ST:fig:VIX:VXO:DFT} we clearly observe Brown noise in the volatility time series. This is entirely consistent with the Brown noise observed in the time series of the GIGa process of Eq. (\ref{ST:eq:GIGa_SDE}); in Fig. \ref{ST:fig:IGa:simulation:DFT} we show Brown noise for an IGa process, $\ga=1$.

\begin{table}[h]
\centering
\begin{tabular}{|l|l|l|l|}
\hline
Data & IGa: $\al,\be$ & GIGa: $\al,\be,\ga$ & LN: $\mu,\si$  \\ \hline
vixarchive & $10.7, 196$ & $33.7, 1.03\times 10^4,0.557$ & $2.96,
0.311$ \\ \hline
vixcurrent & $7.22, 127$ & $0.721, 14.1, 3.96$  & $2.94, 0.398$ \\ \hline
vxoarchive & $9.63,187$ & $11.7, 295, 0.905$    & $3.02, 0.330$ \\ \hline
vxocurrent & $6.59, 113$ & $0.678, 13.4, 3.94$  & $2.92, 0.419$ \\ \hline
\end{tabular}
\caption{Parameters of fitting distributions.}\label{ST:tbl:VIX_dist_parameter}
\end{table}

\begin{table}[h]
\centering
\begin{tabular}{|l|l|l|l|}
\hline
Data & IGa: $-\al$ & GIGa: $-\al\ga$ & Slope of log-log plot \\ \hline
vixarchive & -10.7 & -18.8 & -7.21 \\ \hline
vixcurrent & -7.22 & -2.86 & -3.39 \\ \hline
vxoarchive & -9.63 & -10.6 & -5.48 \\ \hline
vxocurrent & -6.59 & -2.67 & -3.13 \\ \hline
\end{tabular}
\caption{Power-law tail exponents of fitting distributions and log-log
plot.}\label{ST:tbl:VIX_exponent}
\end{table}

\section{Stock returns}\label{ST:sec:stock_return_I}

We analyze component stock prices (at close) of major indices. S\&P 100 and S\&P 500 lists can be found at the Standard \& Poor's website.\cite{SP_website} The historical daily prices of component stocks of S\&P 100 and S\&P 500 are downloaded from Yahoo! Finance.\cite{YahooFinance_website} The final date of S\&P 100 and S\&P 500 stocks used here is March 25, 2013. The daily S\&P 500 and DJIA data is downloaded from the Research Division of the Federal Reserve Bank of St. Louis.\cite{FedStLouis_website1,FedStLouis_website2} For S\&P 500, we ignored five stocks that have less than 200 stock datapoints each. 

\begin{figure}[htp]
\centering
\includegraphics[width=0.23\textwidth]{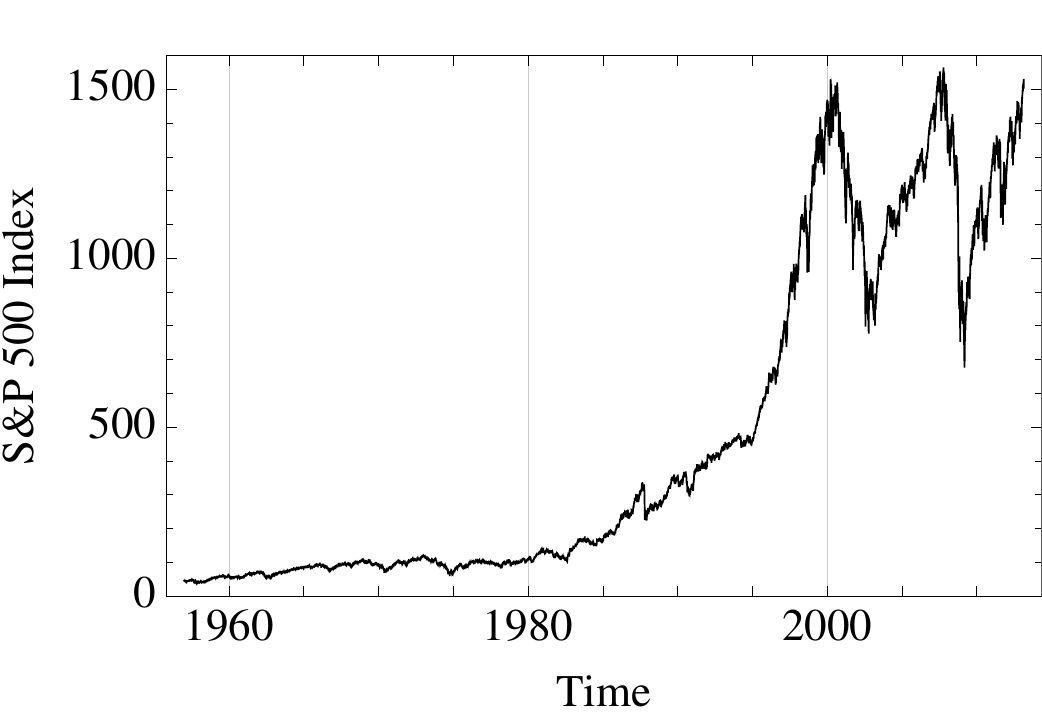}
\includegraphics[width=0.23\textwidth]{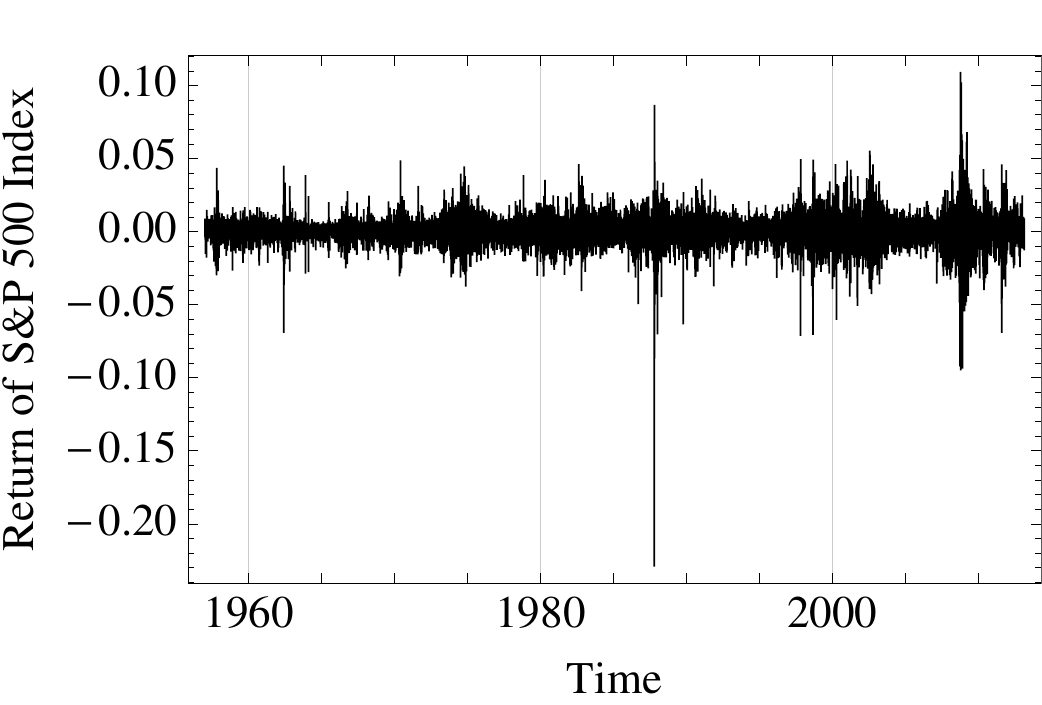}
\includegraphics[width=0.46\textwidth]{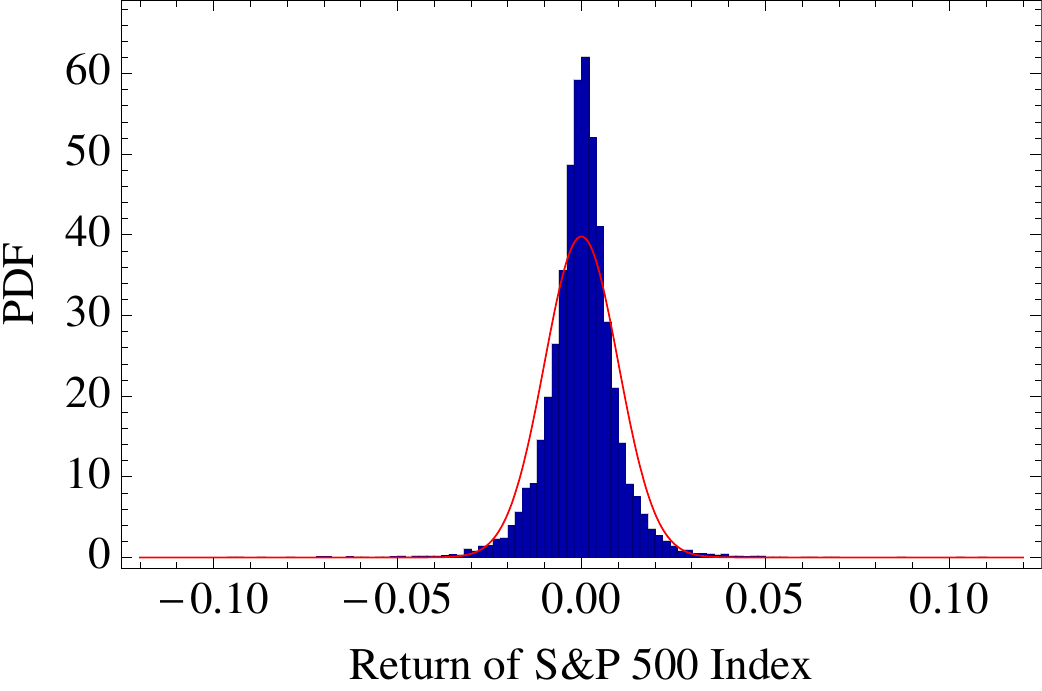}
\caption{Top left: historical curve of daily S\&P 500 index. Top
right: historical curve of daily return of S\&P 500 index. Bottom:
histogram of return of S\&P 500 index. The red line is a fit of normal
distribution. In the plots, return is computed from $\log S_{tomorrow}
-\log S_{today}$ without other operations. }
\label{ST:fig:SP500:stock_return_general}
\end{figure}

\begin{figure}[htp]
\centering
\includegraphics[width=0.23\textwidth]{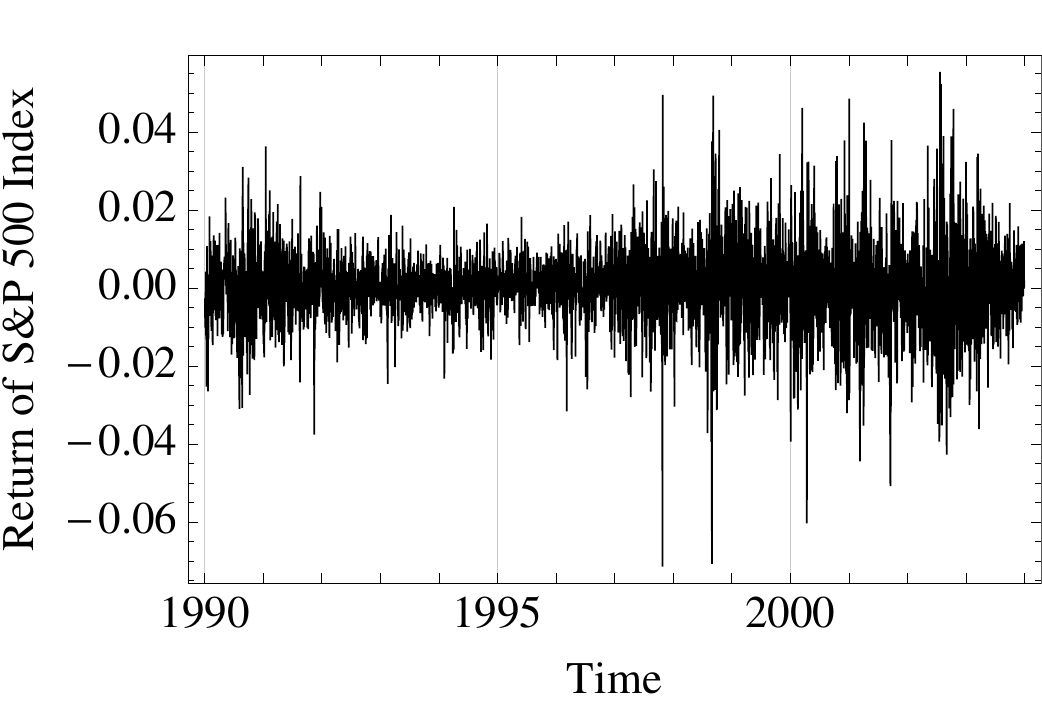}
\includegraphics[width=0.23\textwidth]{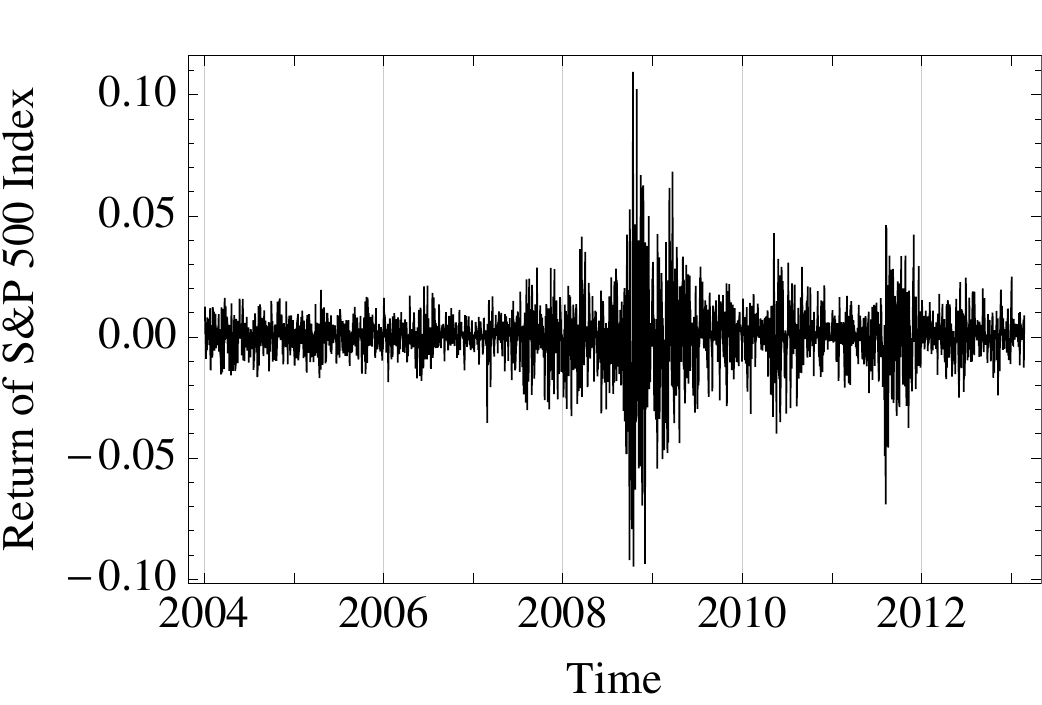}\\
\includegraphics[width=0.23\textwidth]{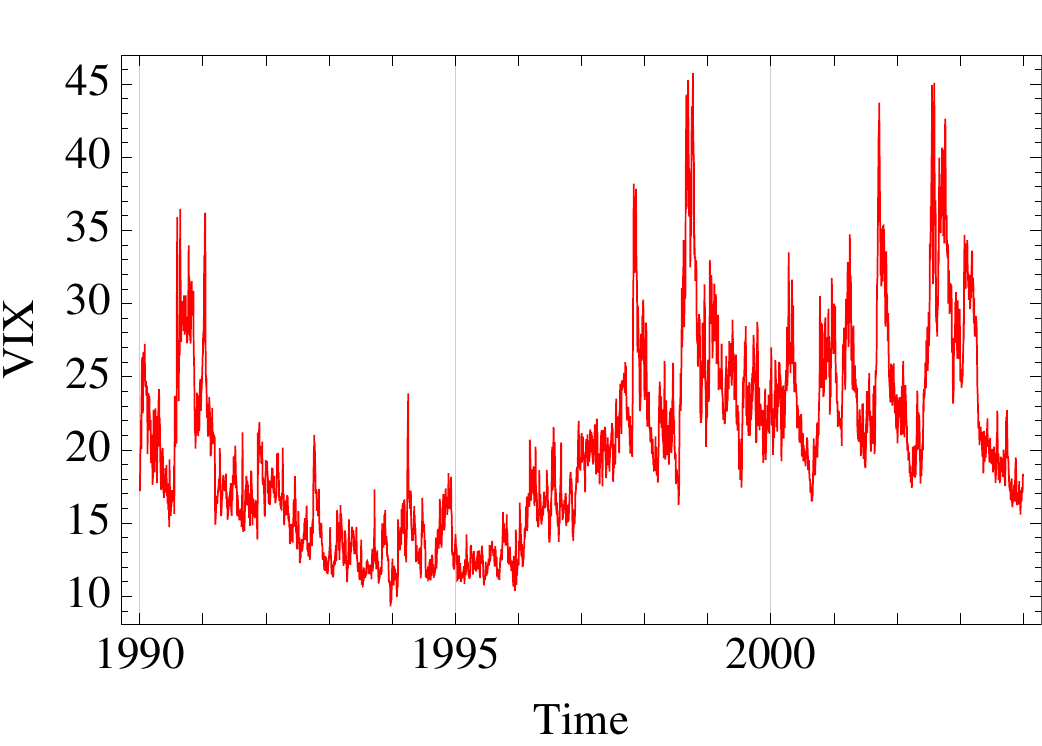}
\includegraphics[width=0.23\textwidth]{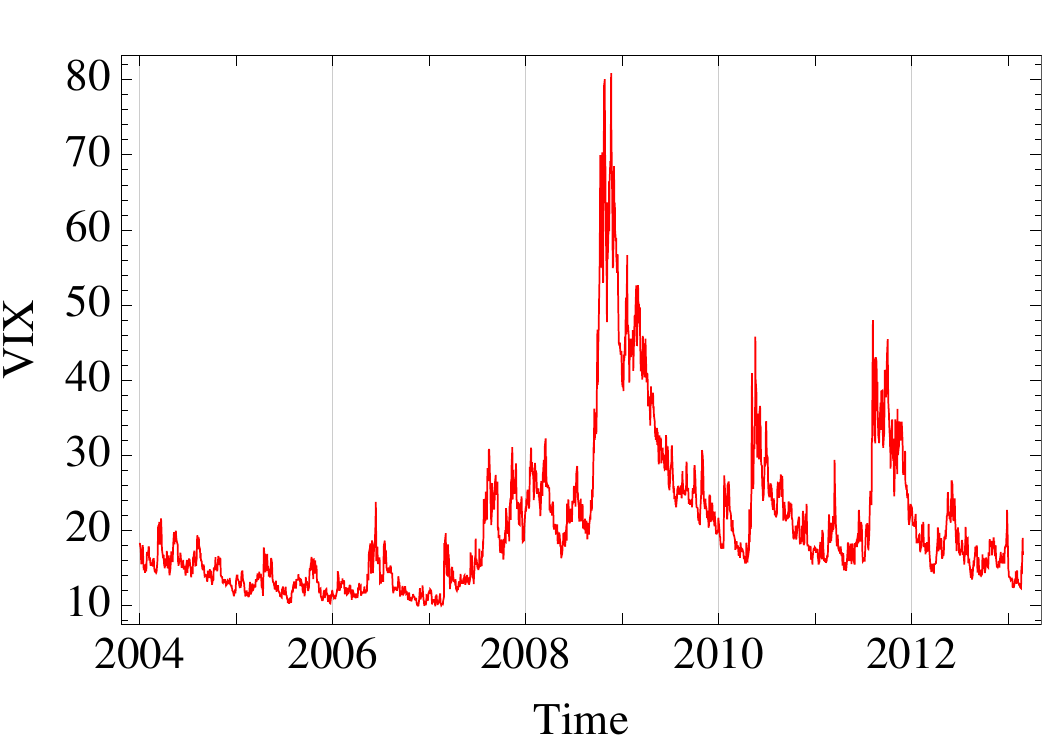}\\
\includegraphics[width=0.23\textwidth]{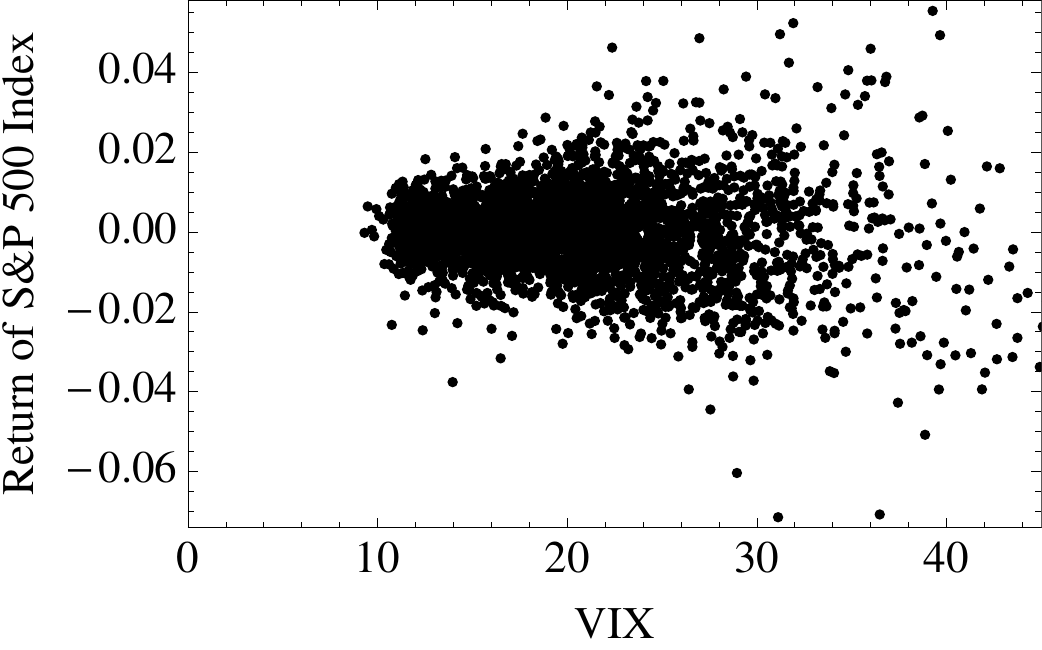}
\includegraphics[width=0.23\textwidth]{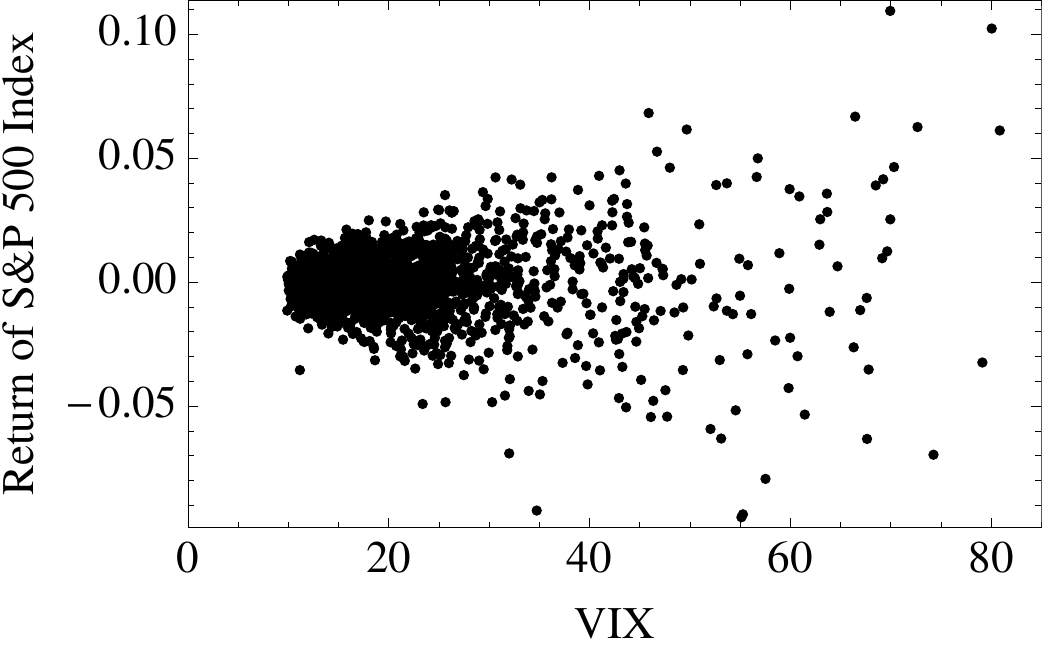}
\caption{Left three plots: return and VIX from 01/02/1990 to
12/31/2003. Right three plots: return and VIX from 01/02/2004 to
02/26/2013.}
\label{ST:fig:SP500:return:VIX}
\end{figure}

\begin{figure}[htp]
\centering
\includegraphics[width=0.23\textwidth]{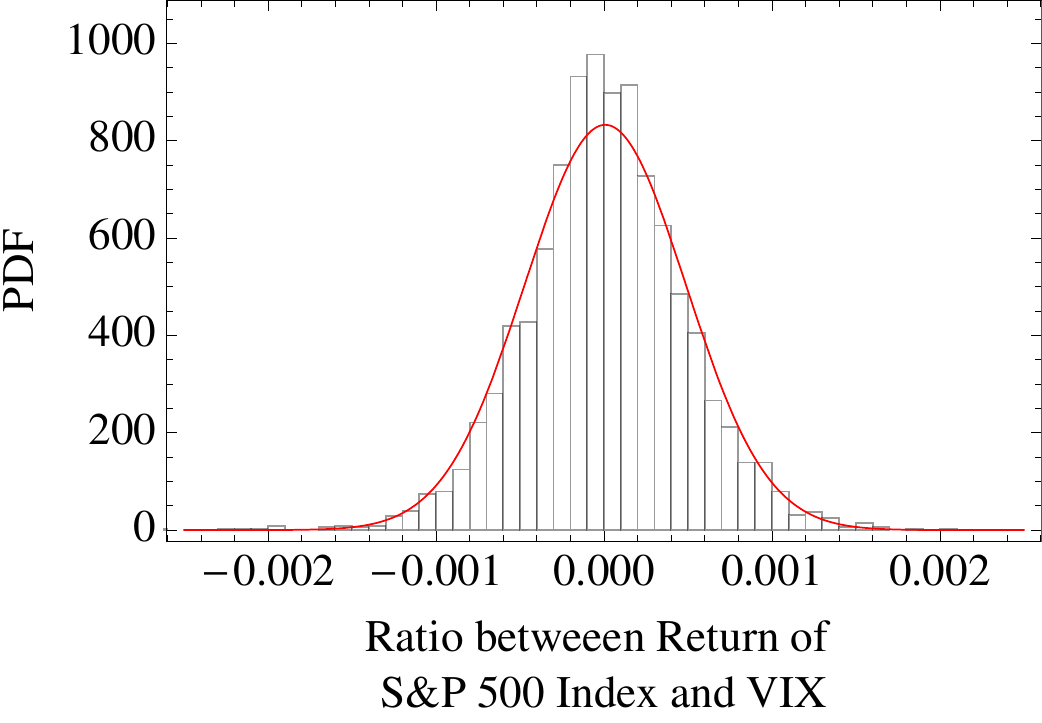}
\includegraphics[width=0.23\textwidth]{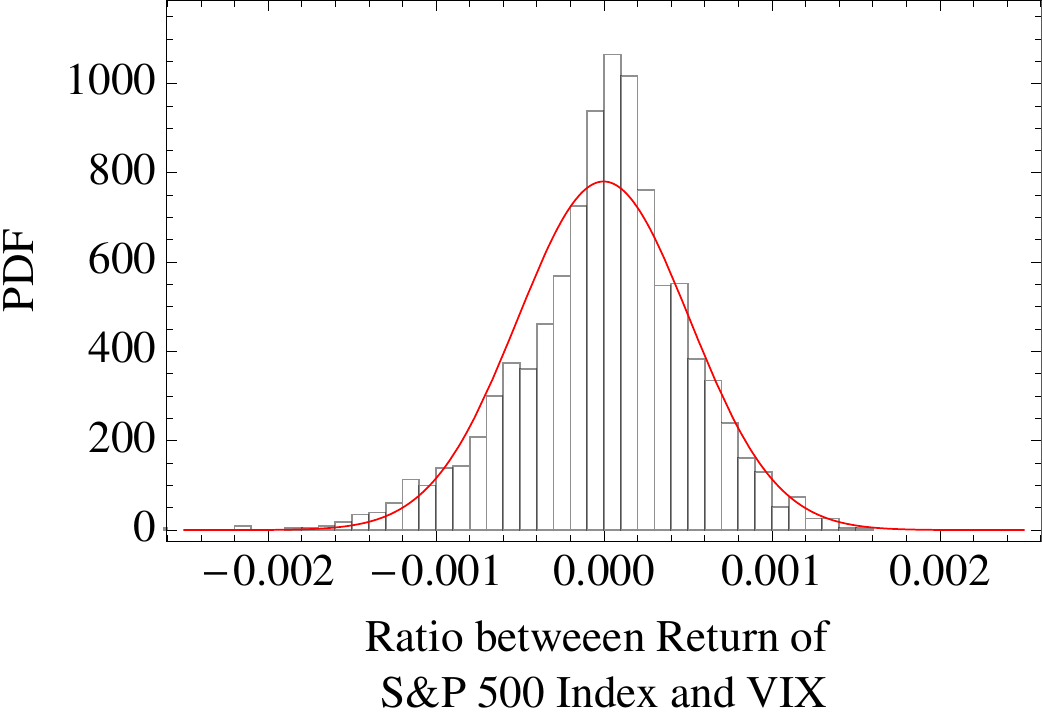}
\caption{Ratio of return and VIX. Left: from 01/02/1990 to 12/31/2003.
Right: from 01/02/2004 to 02/26/2013.}
\label{ST:fig:SP500:return:VIX:ratio}
\end{figure}

We start with a simple test of the stochastic volatility model,  Eqs. (\ref{dSS}) and (\ref{ST:eq:GIGa_SDE}). In Fig. \ref{ST:fig:SP500:stock_return_general}, we show stock returns of S\&P 500 index and their distribution. Average daily return, $\log(S_{tomorrow})-\log(S_{today}) \approx (S_{tomorrow}-S_{today})/S_{today} \approx 0.00025$ corresponds to 6.4\% annual return. For constant volatility, one would expect a normal distribution for stock returns. However, as is obvious from the figure, normal distribution is not a good fit. On the other hand, the stochastic volatility model indicates that it is the ratio of stock return to volatility that should be normal. Visual inspection of Fig. \ref{ST:fig:SP500:return:VIX} and the fit in Fig. \ref{ST:fig:SP500:return:VIX:ratio} give initial validation to the model.

\begin{figure}[htp]
\centering
\includegraphics[width=0.46\textwidth]{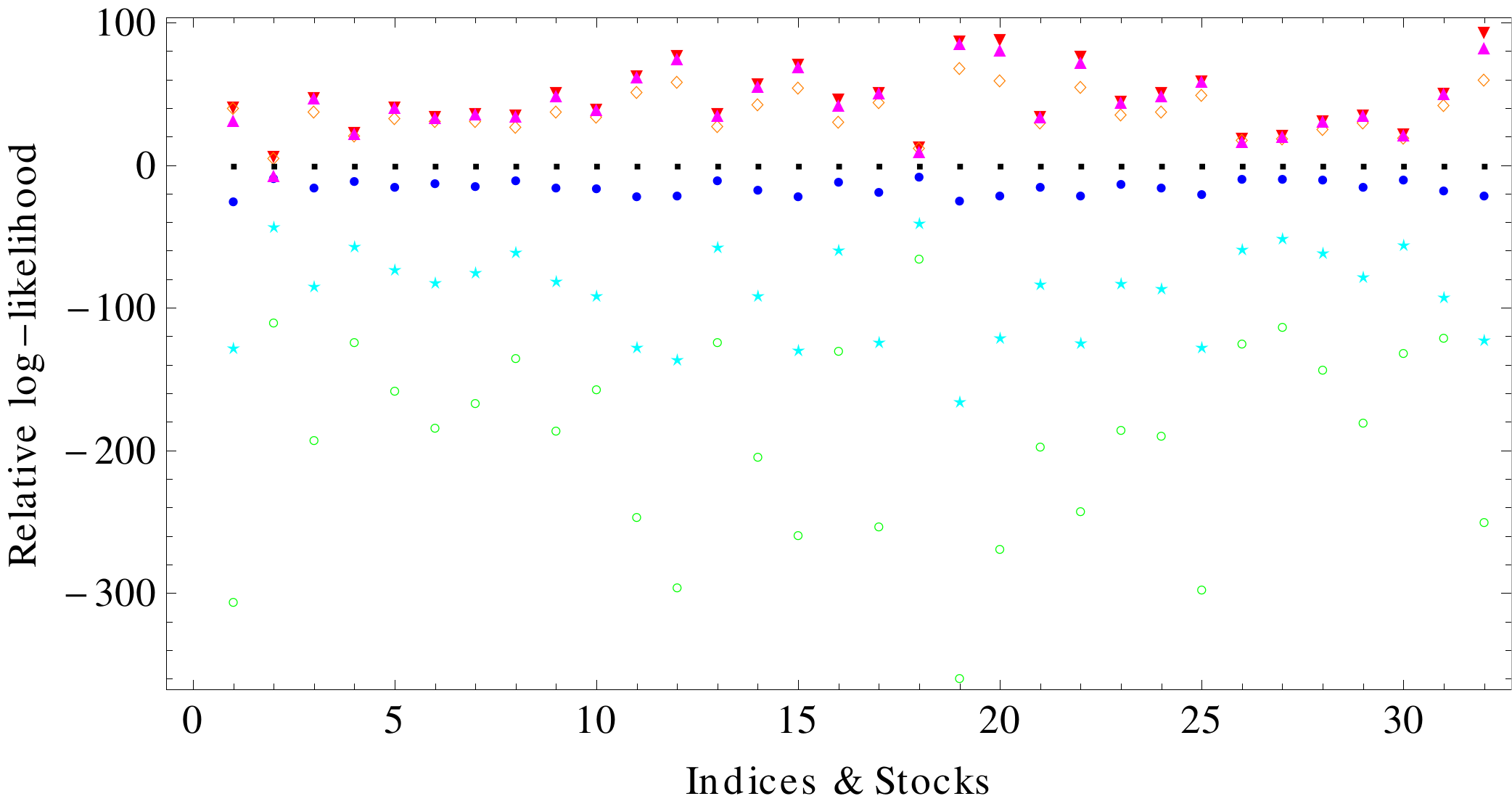}
\caption{Log-likelihood of DJIA, S\&P 500 index, and DJIA component stocks. From 1 to 32, the $x$-coordinates correspond to DJIA, S\&P 500 index, AA, AXP, BA, BAC, CAT, CSCO, CVX, DD, DIS, GE, HD, HPQ, IBM, INTC, JNJ, JPM, KO, MCD, MMM, MRK, MSFT, PFE, PG, T, TRV, UNH, UTX, VZ, WMT, and XOM. All the values of log-likelihood are relative to the LN distribution. Red down-pointing triangles: GIGa*N, magenta up-pointing triangles: GIGa$(\al,\be,2)$*N, orange diamonds: IGa*N, black squares: LN*N, blue dots: GGa*N computed from the simplex method with iteration numbers as 1000, cyan stars: Ga*N, and green circles: GGa$(\al,\be,2)$*N. The parameter $\al\ga$ of GIGa (Red) is 3.79, 5.99,
3.35, 3.61, 3.7, 2.5, 3.76, 2.92, 
3.58, 3.78, 3.4, 3.02, 3.16, 3.28, 
3.13, 3.01, 3.72, 3.63, 3.13, 2.91, 
3.84, 3.2, 2.95, 3.38, 3.37, 3.95, 
3.23, 2.97, 3.93, 3.76, 3.17, 3.21. The parameter $\ga$ of GIGa (Red) is 
1.17, 0.69,
2.32, 1.64, 2.11, 1.6, 1.89, 2.6, 
2.89, 1.89, 2.23, 2.9, 2.53, 2.66, 
2.68, 4.3, 1.93, 1.17, 2.63, 4.27, 
1.76, 3.59, 2.49, 3.02, 2.17, 1.36, 
1.6, 2.24, 1.86, 1.67, 2.08, 5.36.
}
\label{ST:fig:DJIA:stock:loglikelihood}
\end{figure}

\begin{figure}[htp]
\centering
\includegraphics[width=0.46\textwidth]{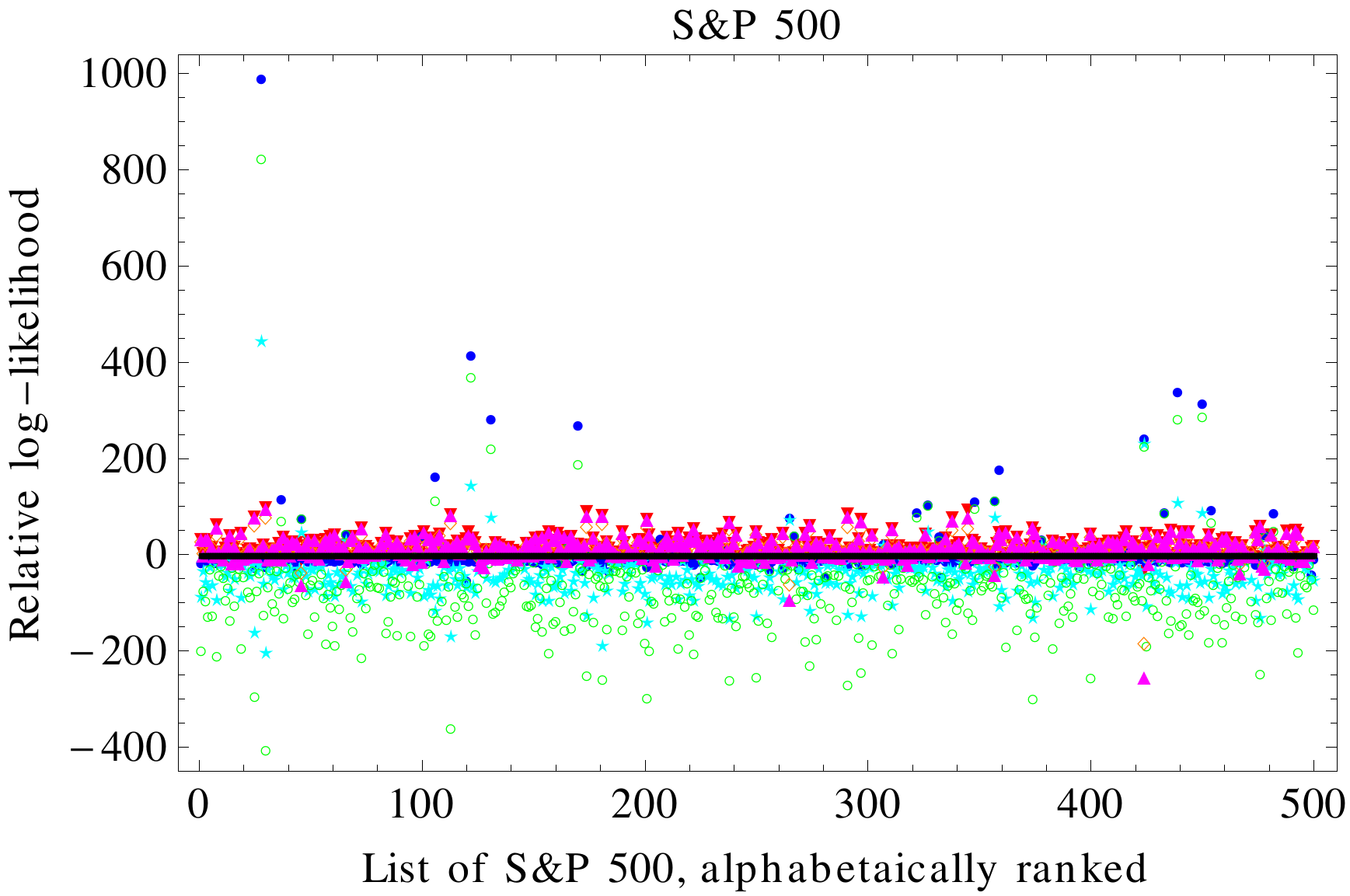}
\includegraphics[width=0.46\textwidth]{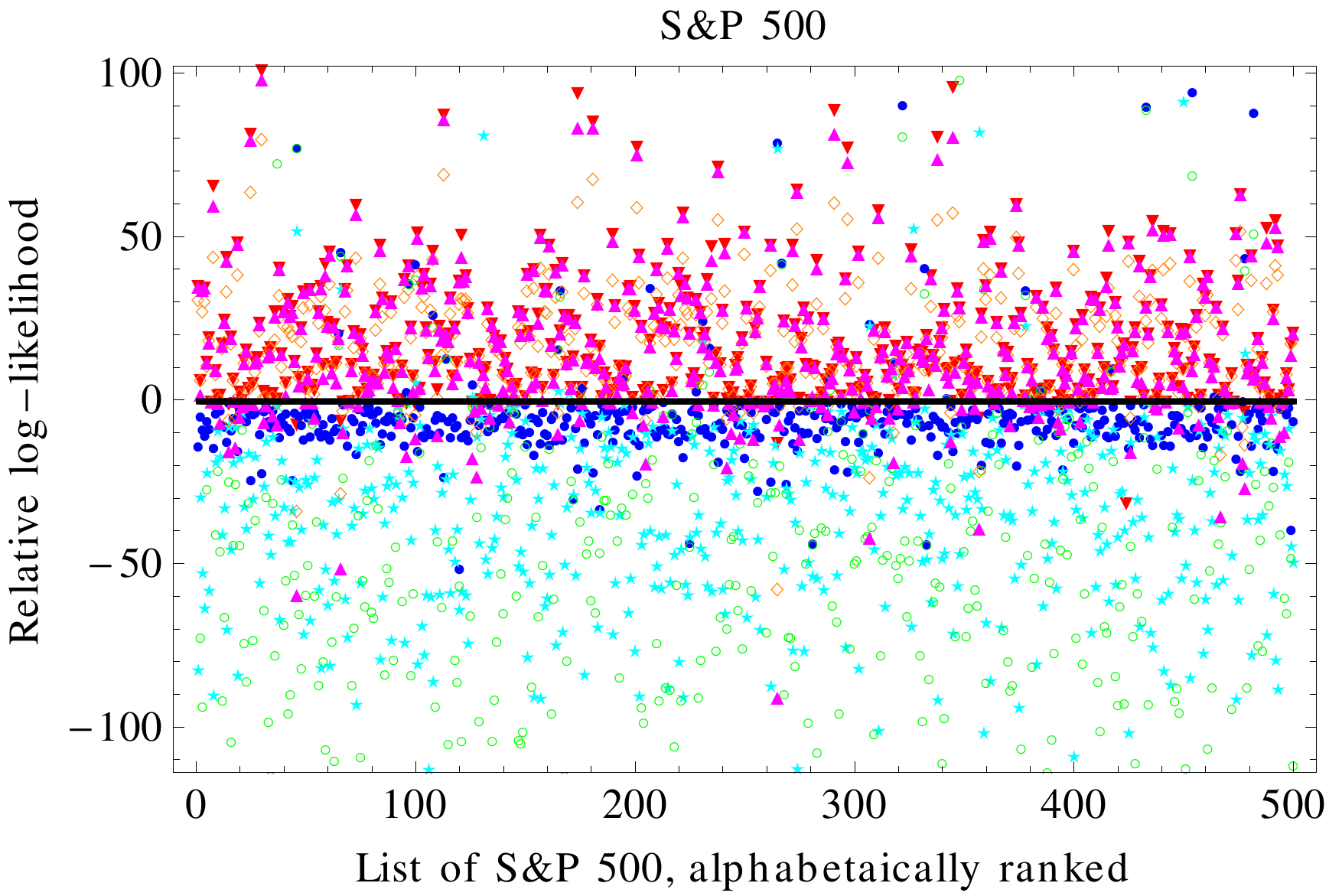}
\caption{Log-likelikelood of S\&P 500. All the values of log-likelihood are relative to the LN distribution. Red down-pointing triangles: GIGa*N, magenta up-pointing triangles: GIGa$(\al,\be,2)$*N, orange diamonds: IGa*N, black squares: LN*N, blue dots: GGa*N computed from the simplex method with iteration numbers as 1000, cyan stars: Ga*N, and green circles: GGa$(\al,\be,2)$*N. Top plot for all the range of log-likelihood and bottom plot for a shorter range. }
\label{ST:fig:SP500:stock:loglikelihood}
\end{figure}

\begin{figure}[htp]
\centering
\includegraphics[width=0.46\textwidth]{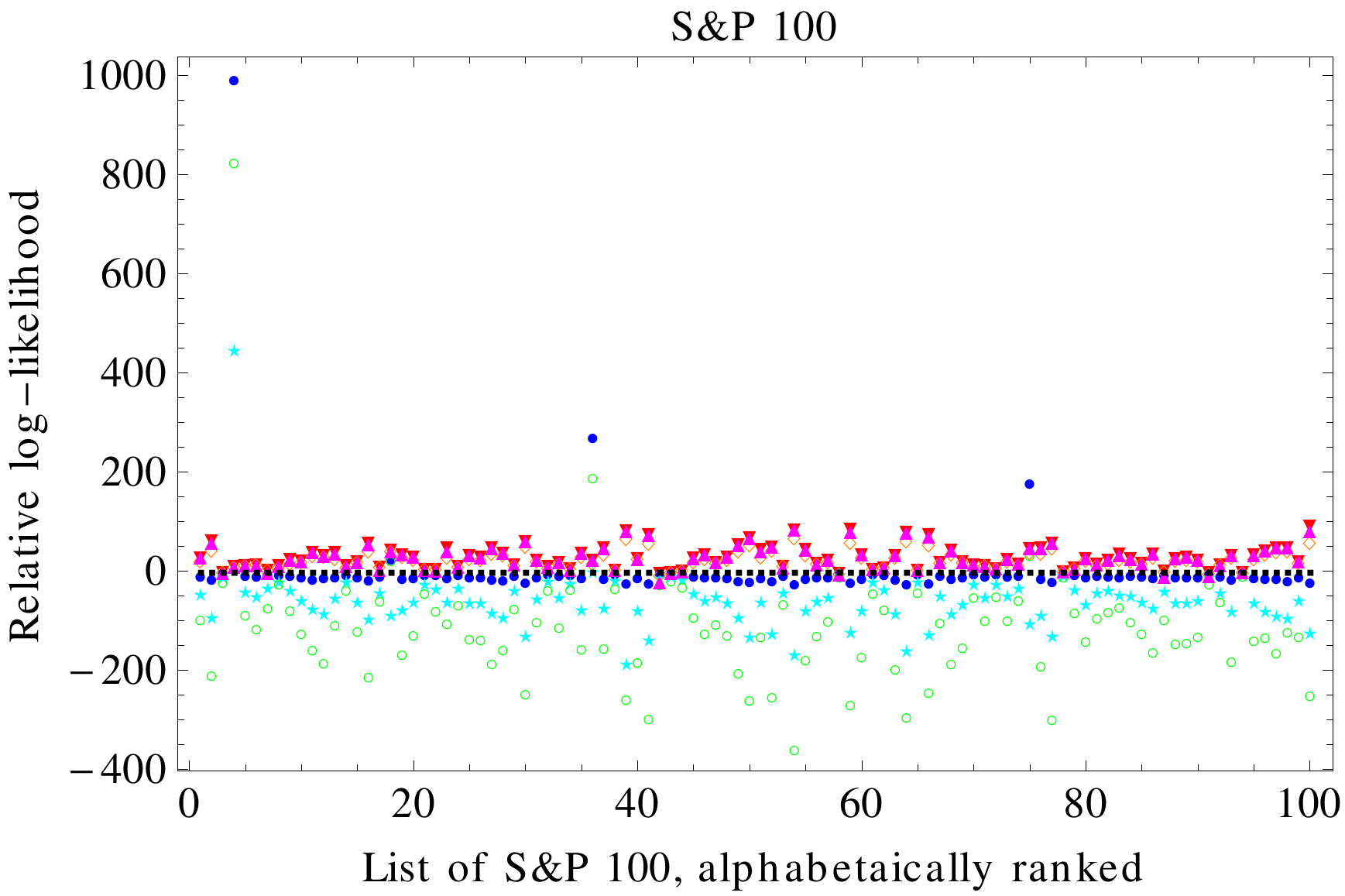}
\includegraphics[width=0.46\textwidth]{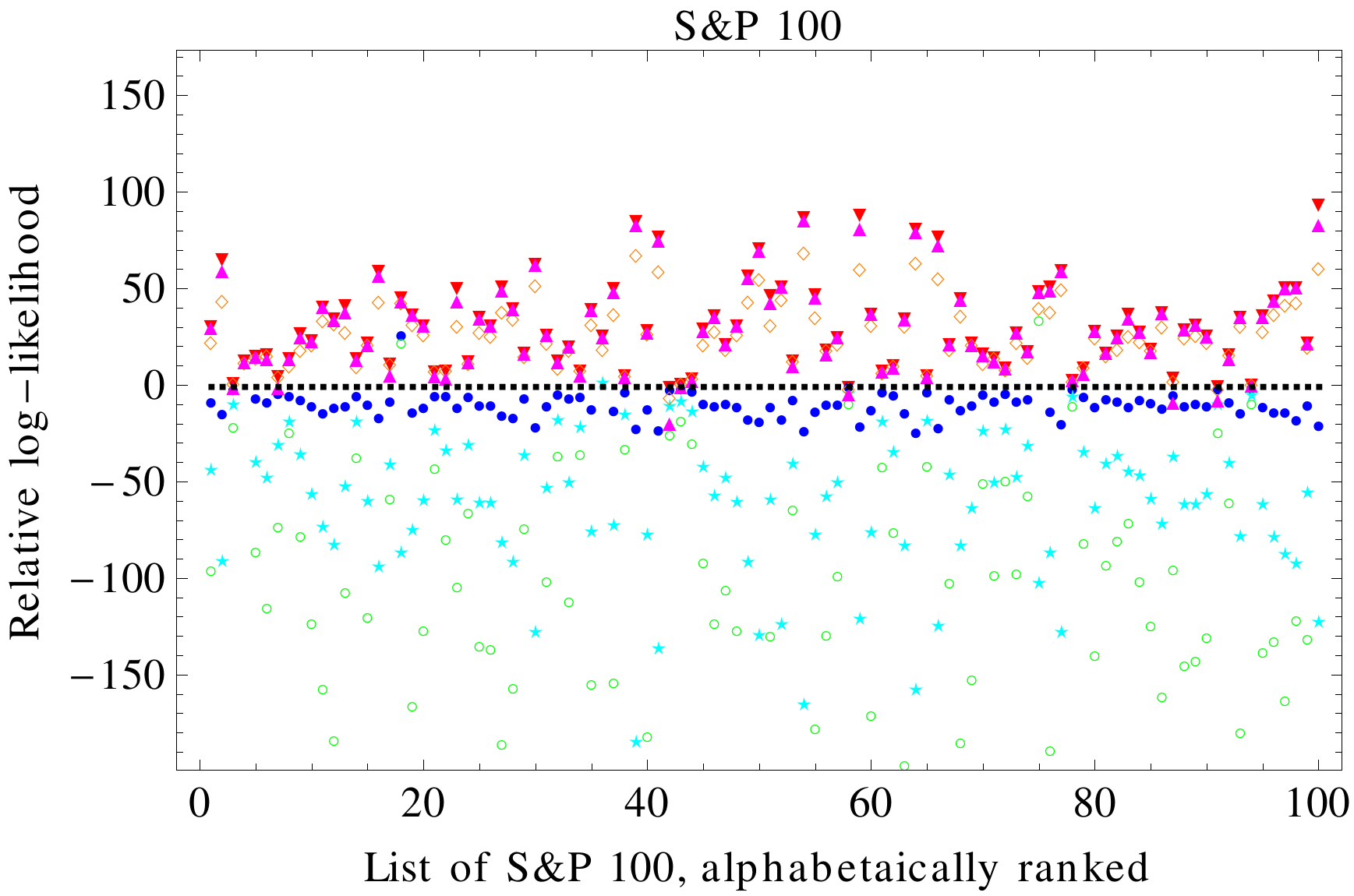}
\caption{Log-likelikelood of S\&P 100. All the values of log-likelihood are relative to the LN distribution. Red down-pointing triangles: GIGa*N, magenta up-pointing triangles: GIGa$(\al,\be,2)$*N, orange diamonds: IGa*N, black squares: LN*N, blue dots: GGa*N computed from the simplex method with iteration numbers as 1000, cyan stars: Ga*N, and green circles: GGa$(\al,\be,2)$*N. Top plot for all the range of log-likelihood and bottom plot for a shorter range. }
\label{ST:fig:SP100:stock:loglikelihood}
\end{figure}

\begin{figure}[htp]
\centering
\includegraphics[width=0.23\textwidth]{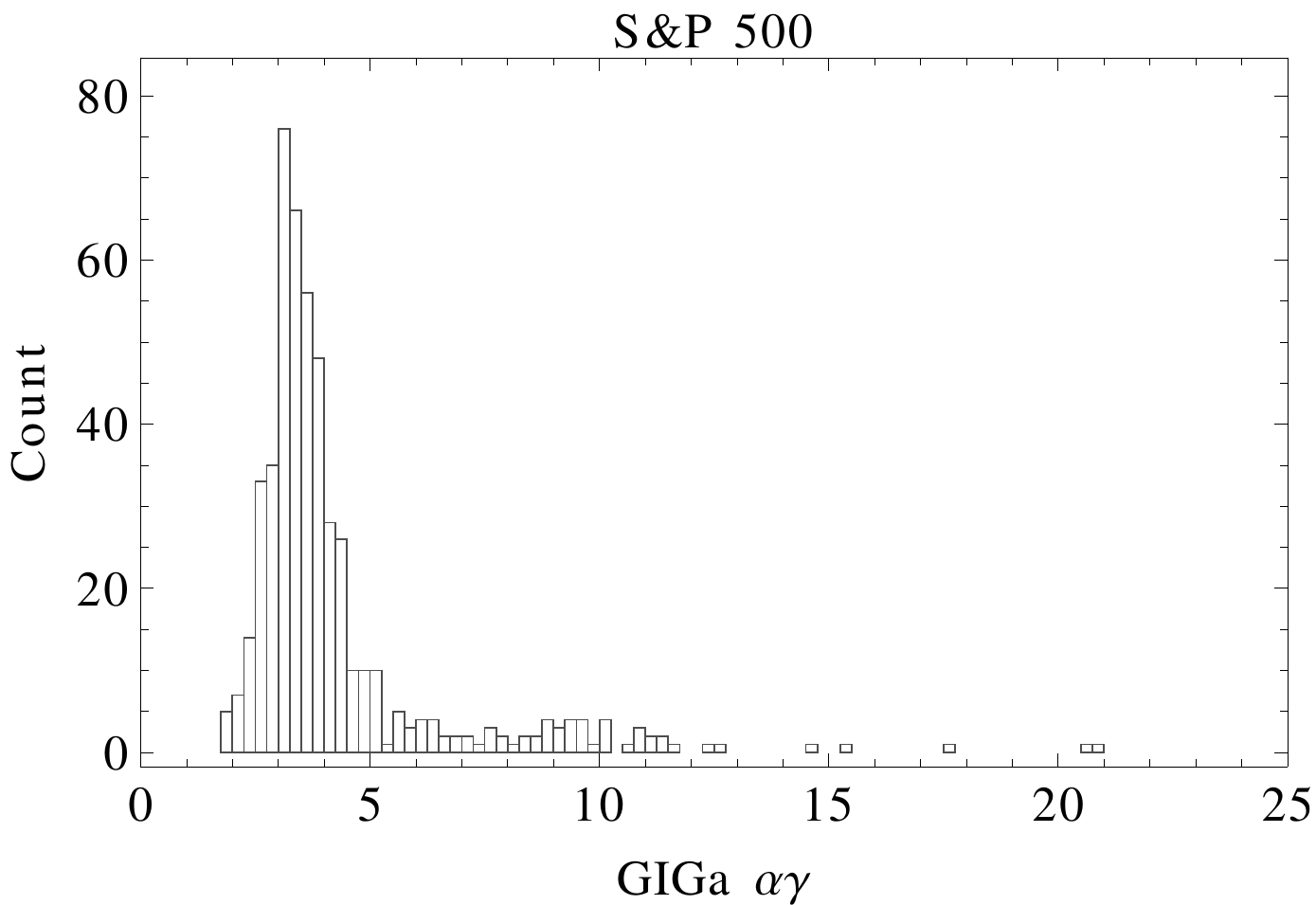}
\includegraphics[width=0.23\textwidth]{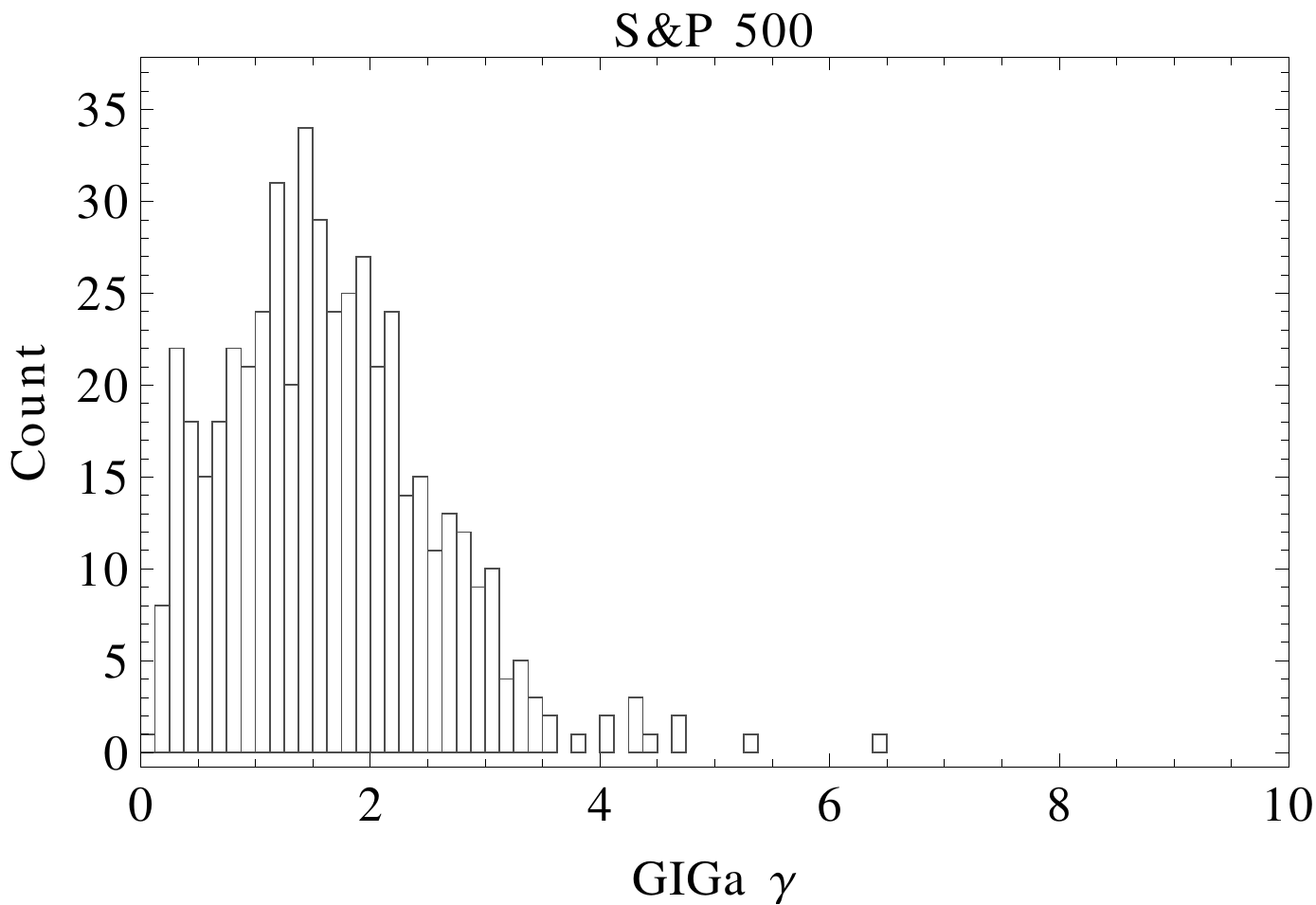}
\includegraphics[width=0.23\textwidth]{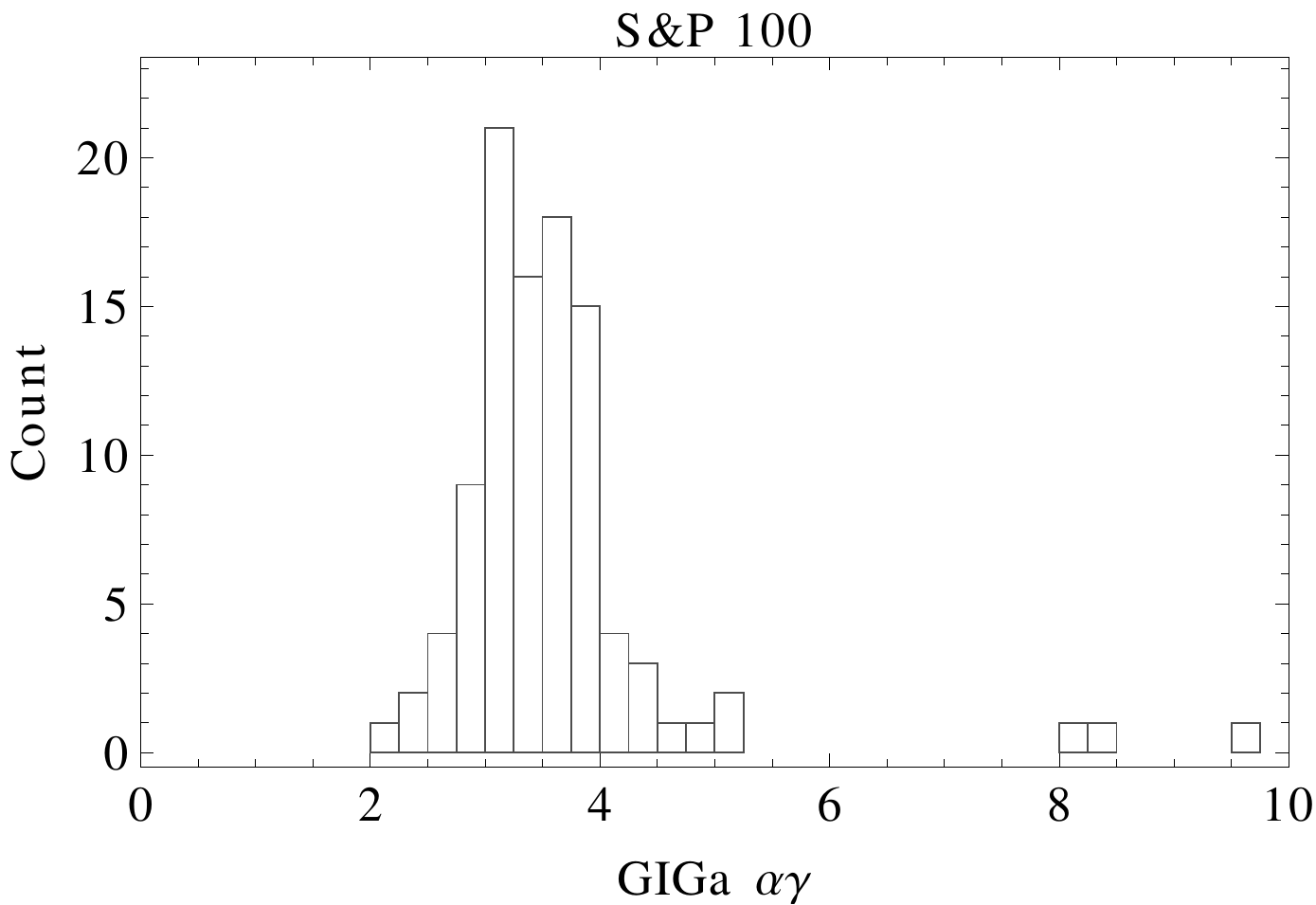}
\includegraphics[width=0.23\textwidth]{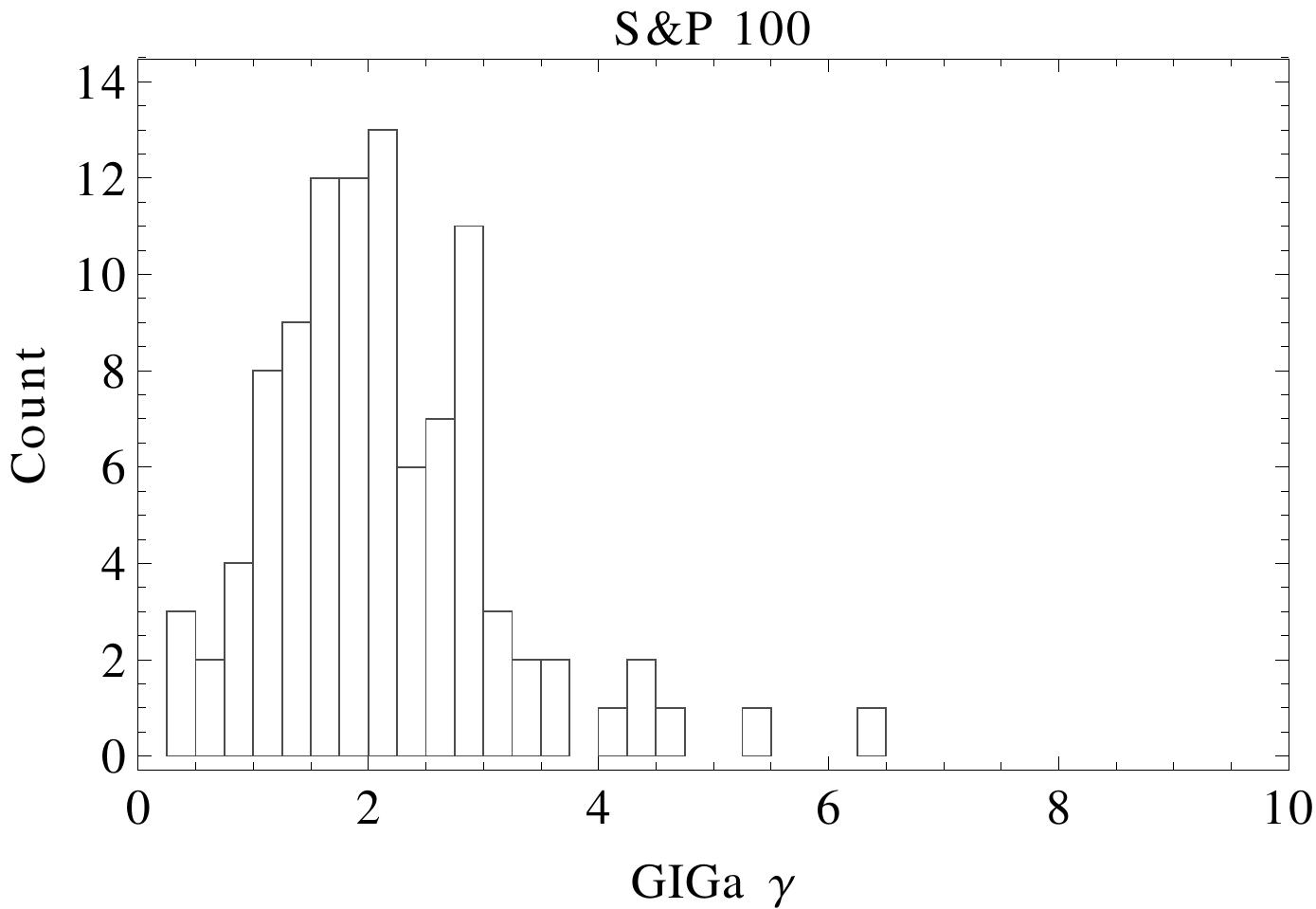}
\caption{Parameters of GIGa used in fitting S\&P 500 (top) and S\&P 500 (bottom). $\al\ga$ is the thereotical exponent of the power-law tails in stock return. $\ga$ is the control parameter. If $\ga=2$, the volatility variance is described by IGa, which can be generated by a mean-field theory of (\ref{ST:eq:IGa_SDE_V}). \cite{Bouchaud2000,ma2013distribution} The mean of $\al\ga$ is 4.3 and the median is 3.6. The mean of $\ga$ is 1.9 and the median is 1.5 for S\&P 500 and 2.1 and 2.0 for S\&P 100.}
\label{ST:fig:SP500:stock:GIGa:parameter}
\end{figure}

\begin{figure}[htp]
\centering
\includegraphics[width=0.23\textwidth]{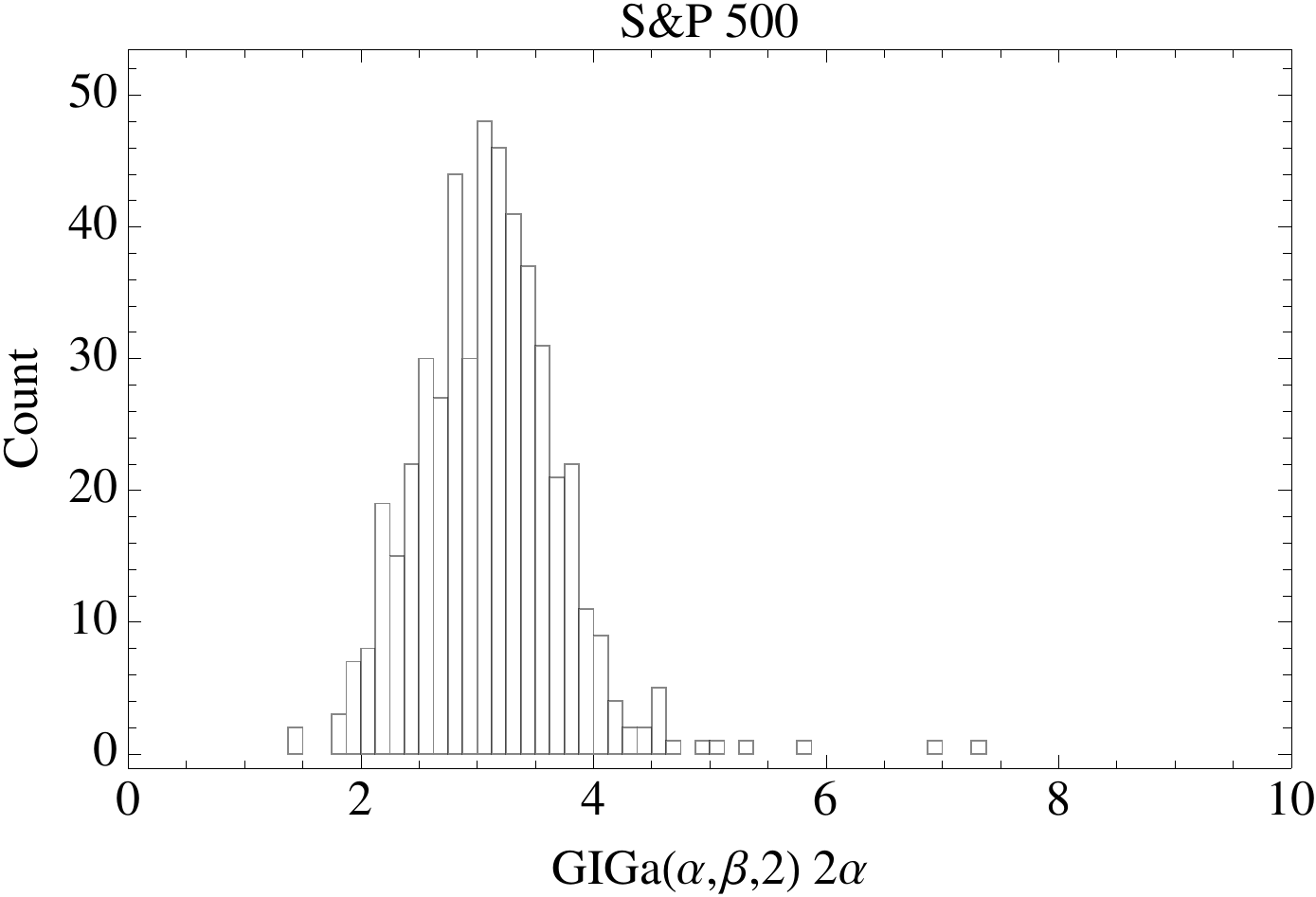}
\includegraphics[width=0.23\textwidth]{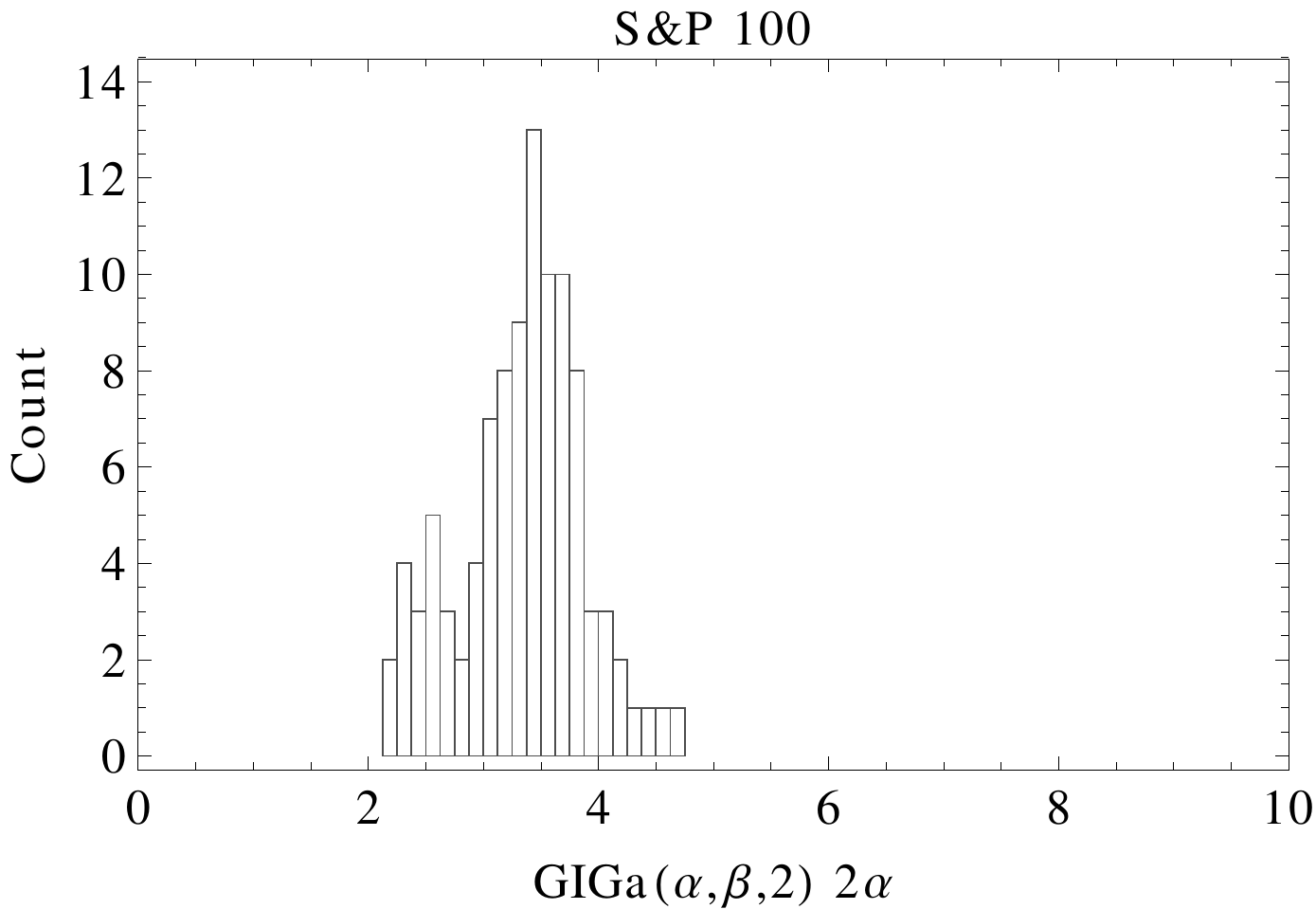}
\caption{Parameter $\al$ of GIGa$(\al,\be,2)$ used in fitting S\&P 500 (left) and S\&P 100 (right). $2\al$ is the theoretical exponent of the power-law tails in stock return. The mean of $2\al$ is 3.2 and the median is 3.1 for S\&P 500 and 3.3 and 3.4 for S\&P 100.}
\label{ST:fig:SP500:stock:GIGa:ga2:parameter}
\end{figure}

We proceed to rigorously analyze stock returns data using maximum likelihood estimation. In our analysis, the stock returns are detrended and scaled into unit STDEV. The log-likelihood of product distribution is evaluated by numerical quadrature and maximized by the simplex algorithm. The numerical quadrature and the simplex algorithm are checked to be correct for the special case $\GIGa(\al,\be,2)*N$, which is generalized Student's \emph{t}-distribution, whose maximum likelihood estimation can be computed directly. \cite{MaThesis2013} Figs. \ref{ST:fig:DJIA:stock:loglikelihood}-\ref{ST:fig:SP100:stock:loglikelihood} convincingly show that the product distribution $\GIGa(\al,\be,\ga)*\text{N(0,1)}$ fits the stock return best. Distributions of best fit parameters are shown in Figs. \ref{ST:fig:SP500:stock:GIGa:parameter} and \ref{ST:fig:SP500:stock:GIGa:ga2:parameter}. 

\begin{figure}[htp]
\centering
\includegraphics[width=0.46\textwidth]{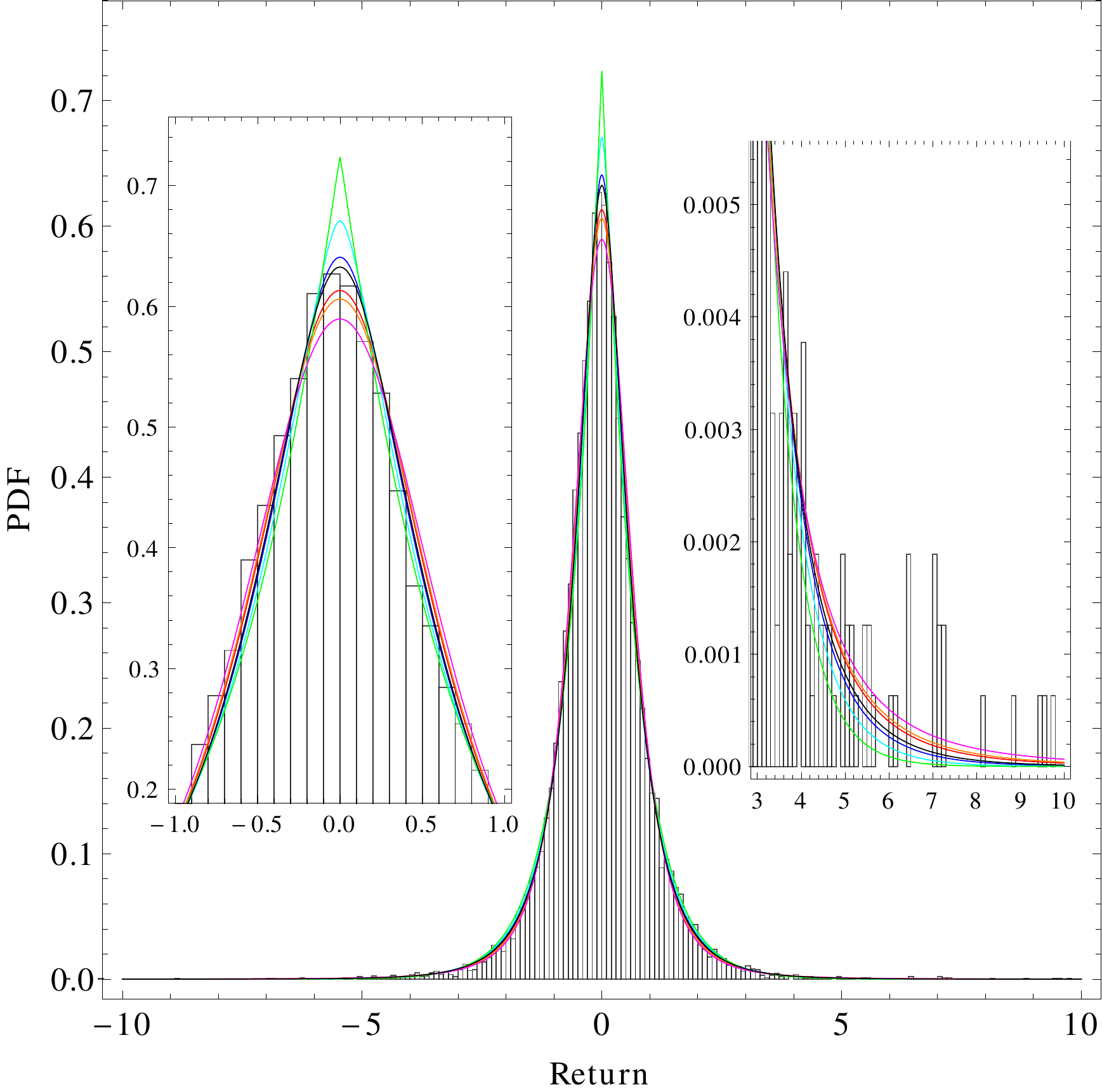}
\includegraphics[width=0.46\textwidth]{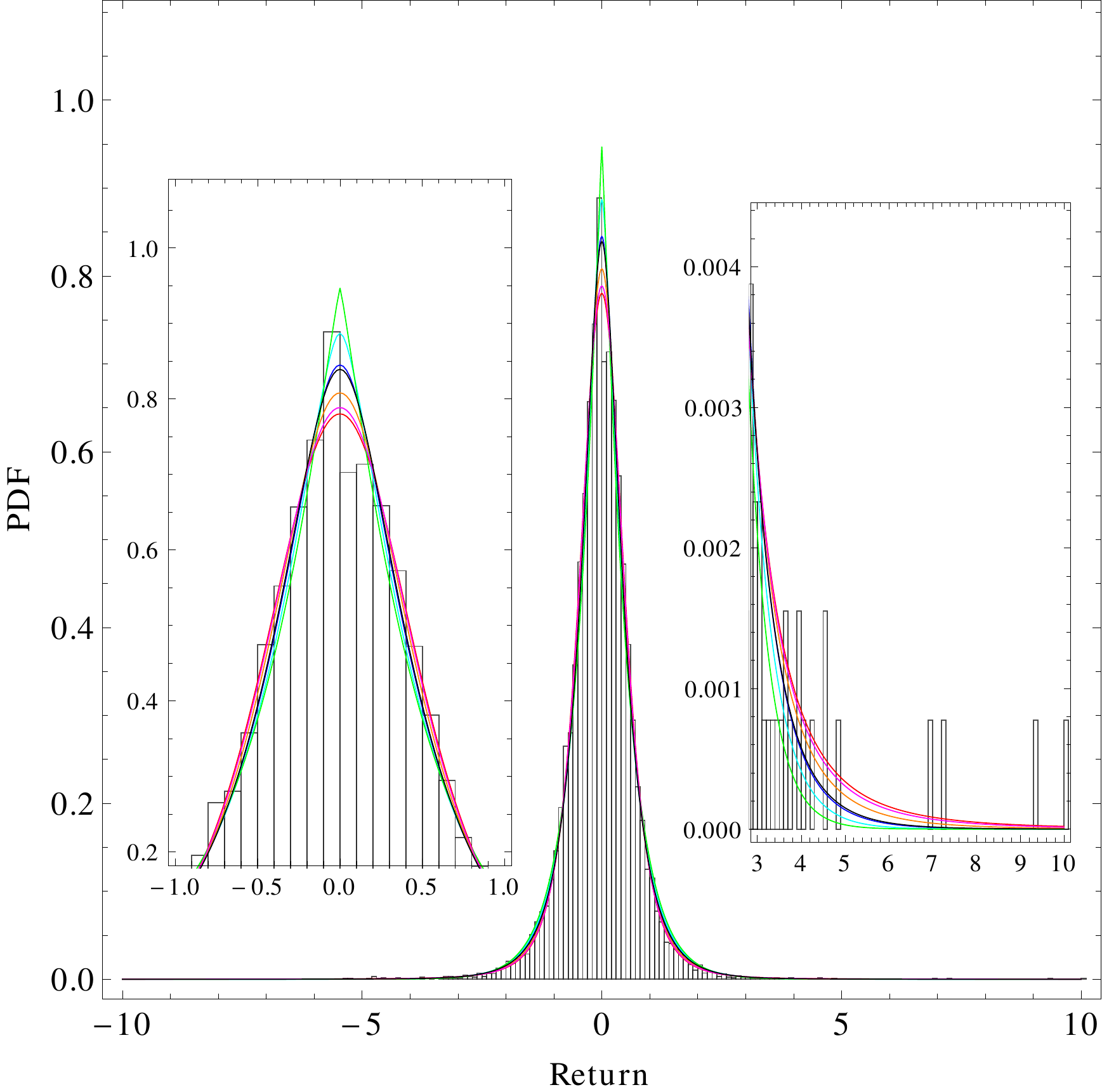}
\caption{Fitting of stock return rate of S\&P 500 (top) and IBM (bottom). Red: GIGa*N, magenta: GIGa$(\al,\be,2)$*N (Student's \emph{t}-distribution), orange: IGa*N, black: LN*N, blue: GGa*N computed from the simplex method with iteration numbers as 1000, cyan: Ga*N, and green: GGa$(\al,\be,2)$*N. }
\label{ST:fig:SP500:stock_return_fit}
\end{figure}

In Fig. \ref{ST:fig:SP500:stock_return_fit}, we show  histograms of S\&P 500 index and IBM respectively and their fitting by product distributions. Clearly, the product distribution of GIGa and normal distribution is better able to capture the tail events.

\begin{figure}[htp]
\centering
\includegraphics[width=0.23\textwidth]{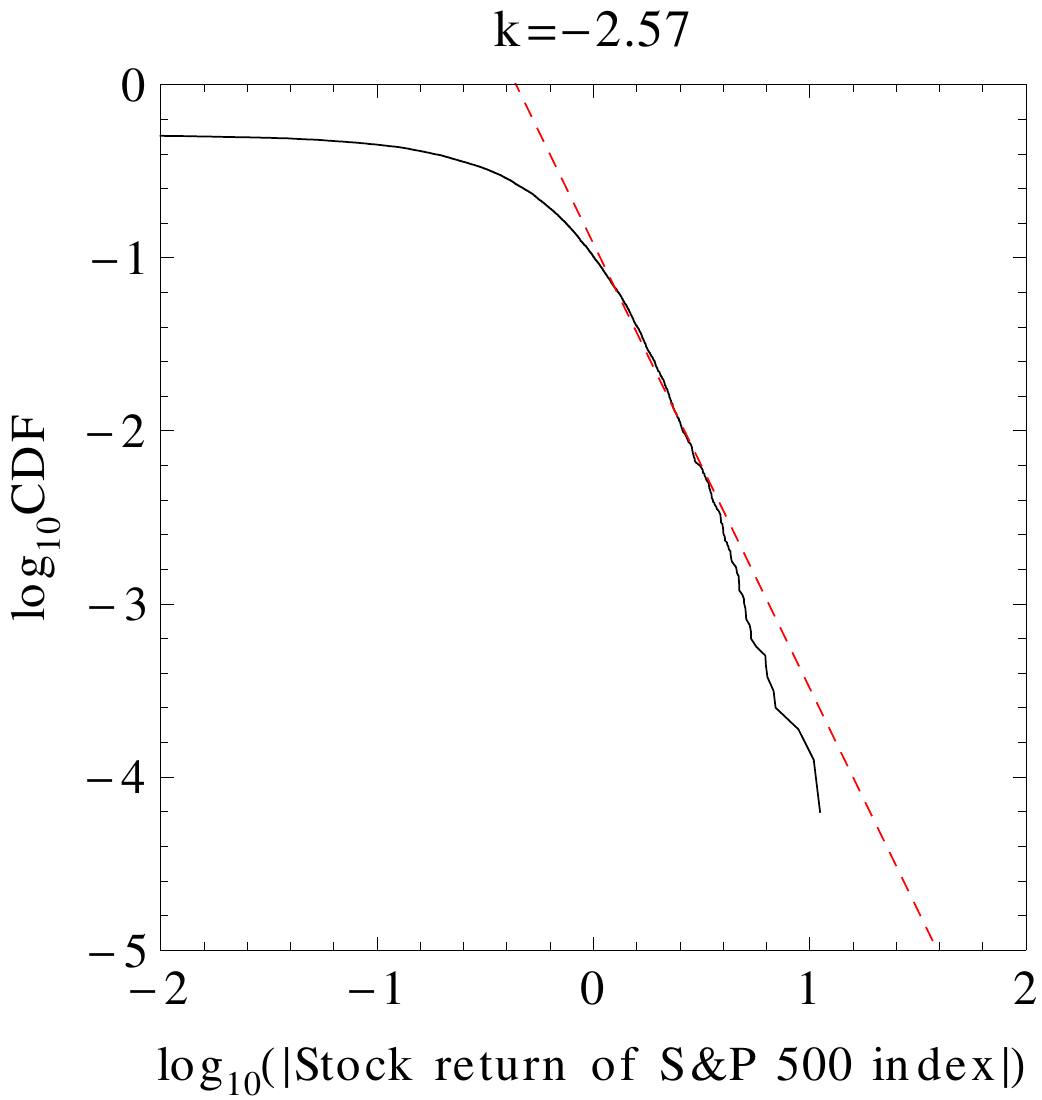}
\includegraphics[width=0.23\textwidth]{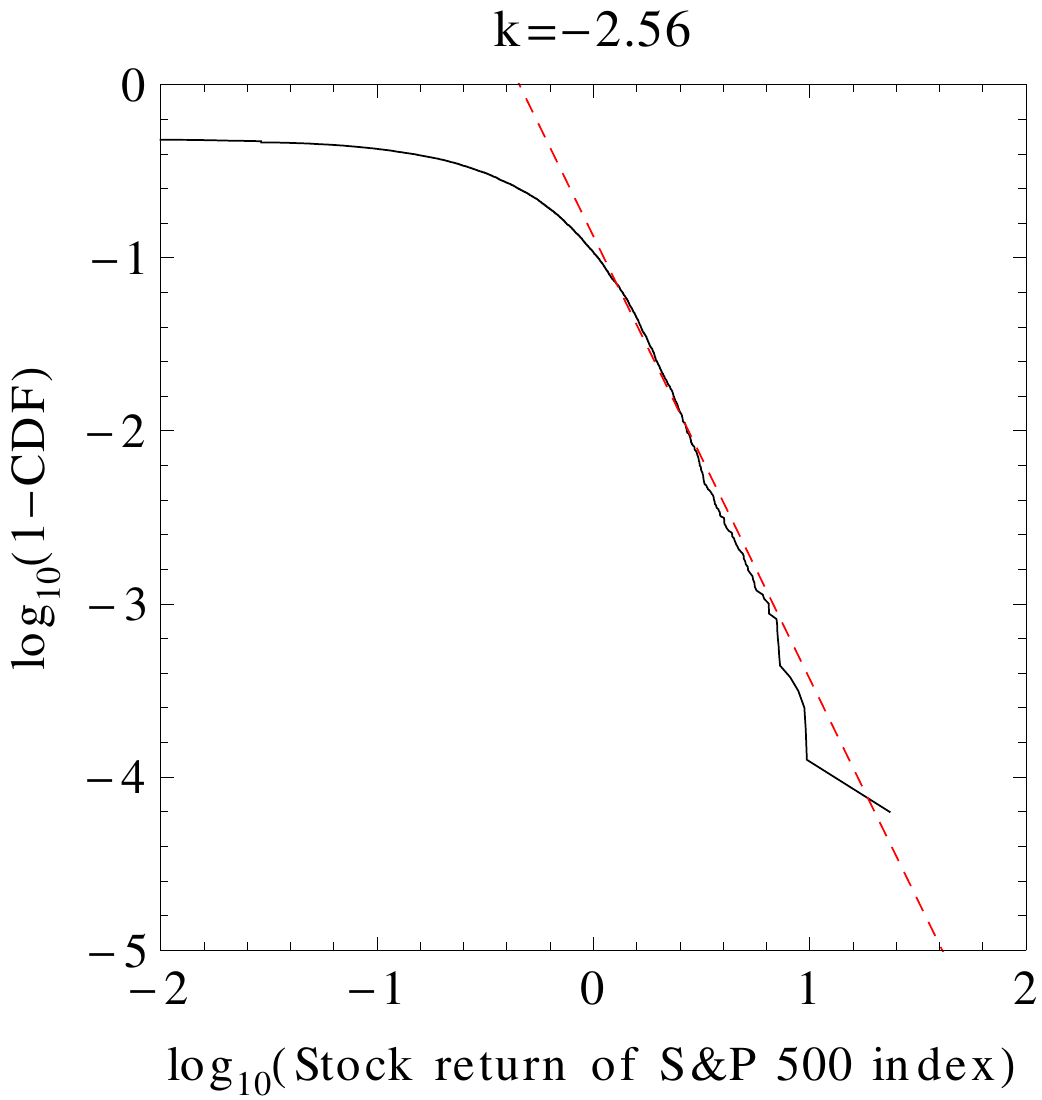}
\caption{Log-log plot of stock return rate of S\&P 500. Left: the left tail of negative stock return. The $x$-axis is logarithm of absolute values of stock return and the $y$-axis is CDF. Right: the right tail of (positive) stock return.}
\label{ST:fig:SP500:stock_return_loglogplot}
\end{figure}

\begin{figure}[htp]
\centering
\includegraphics[width=0.23\textwidth]{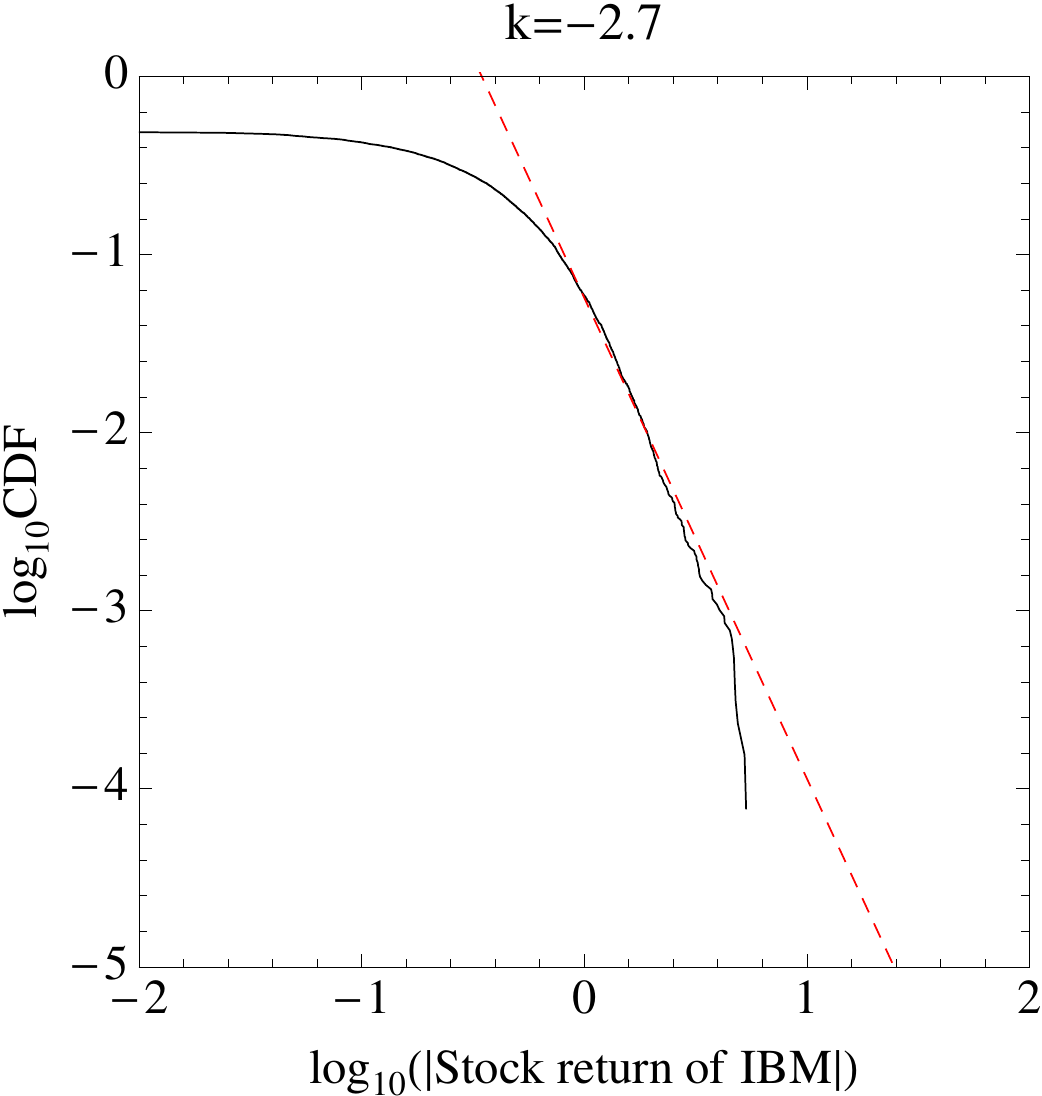}
\includegraphics[width=0.23\textwidth]{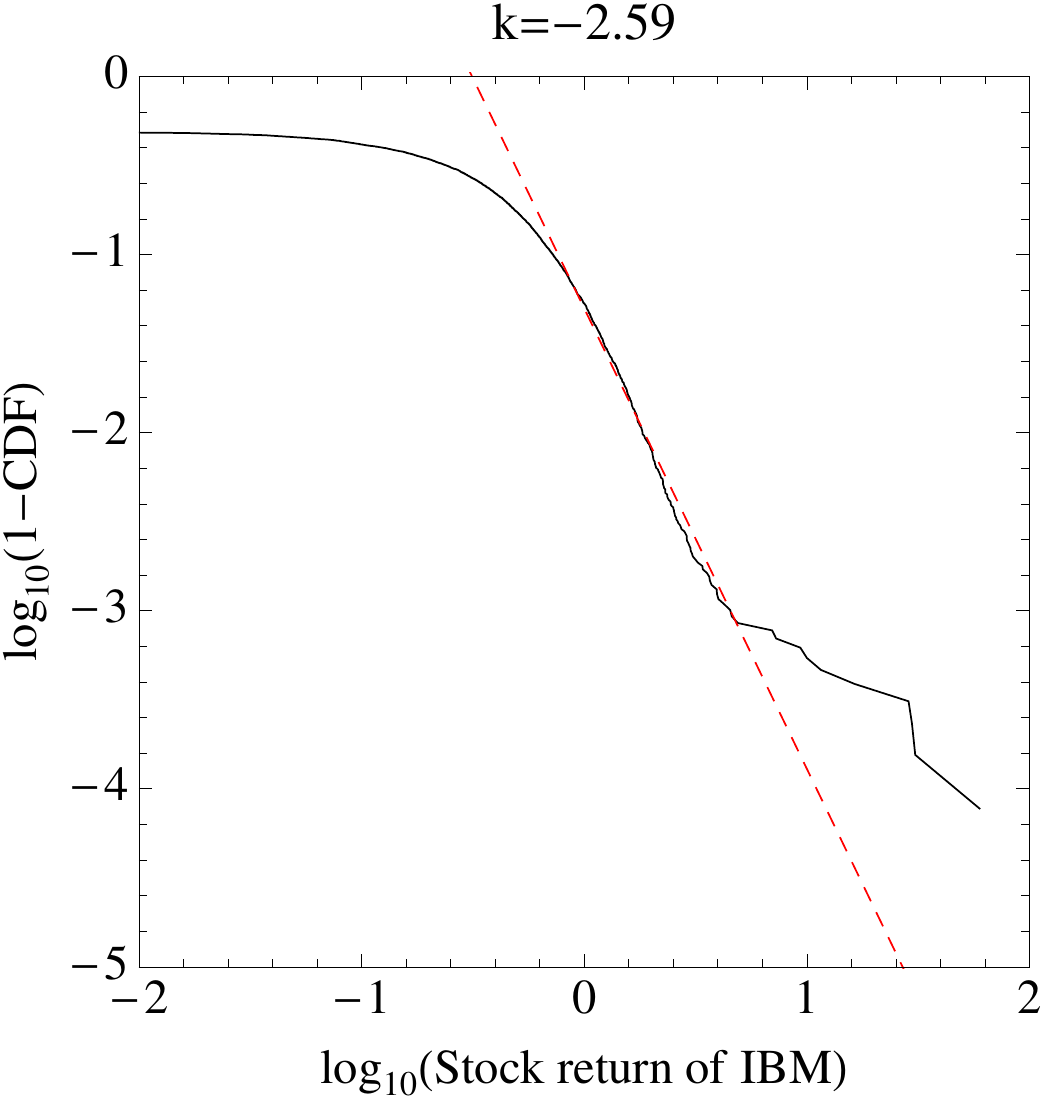}
\caption{Log-log plot of stock return rate of IBM. Left: the left tail of negative stock return. The $x$-axis is logarithm of absolute values of stock return and the $y$-axis is CDF. Right: the right tail of (positive) stock return.}
\label{ST:fig:IBM:stock_return_loglogplot}
\end{figure}

\begin{figure}[htp]
\centering
\includegraphics[width=0.23\textwidth]{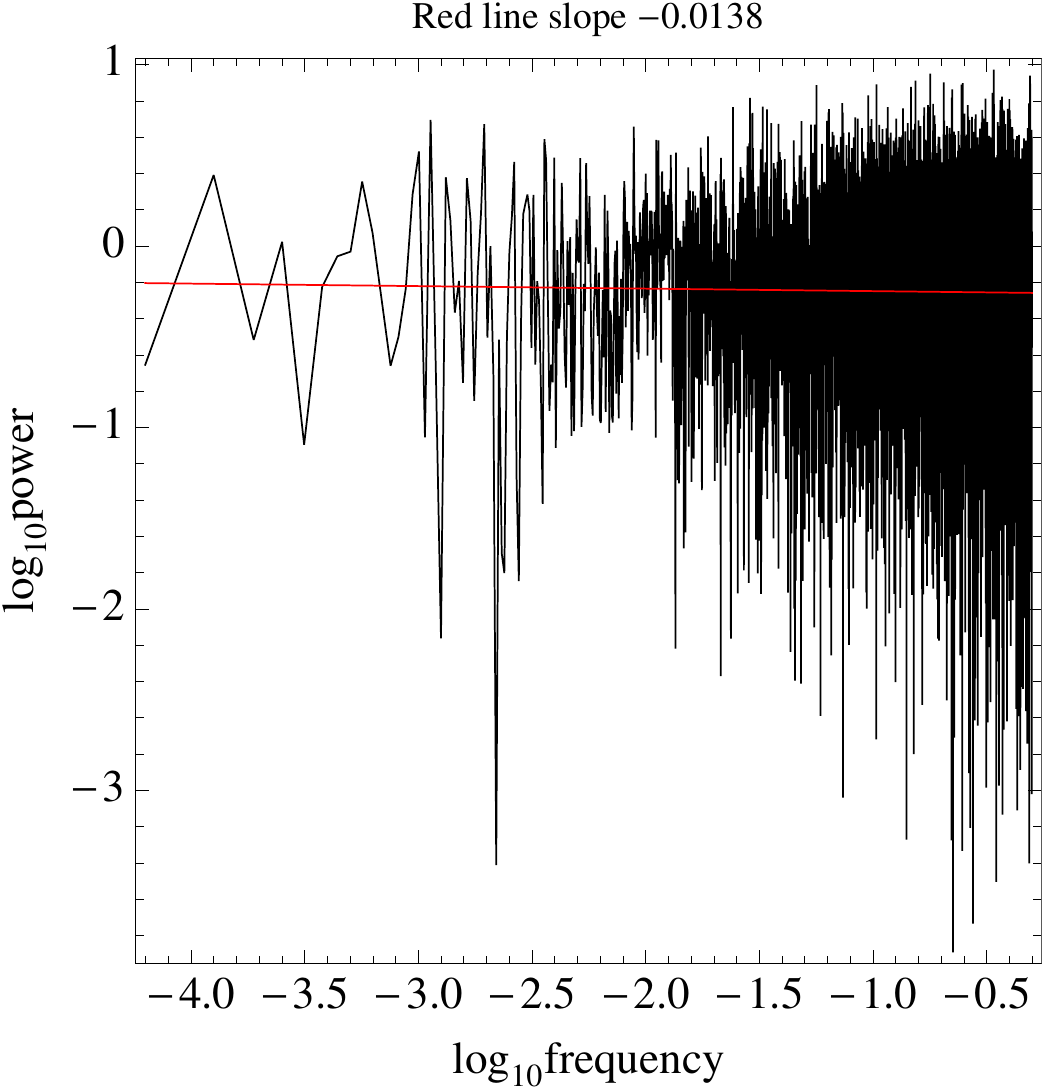}
\includegraphics[width=0.23\textwidth]{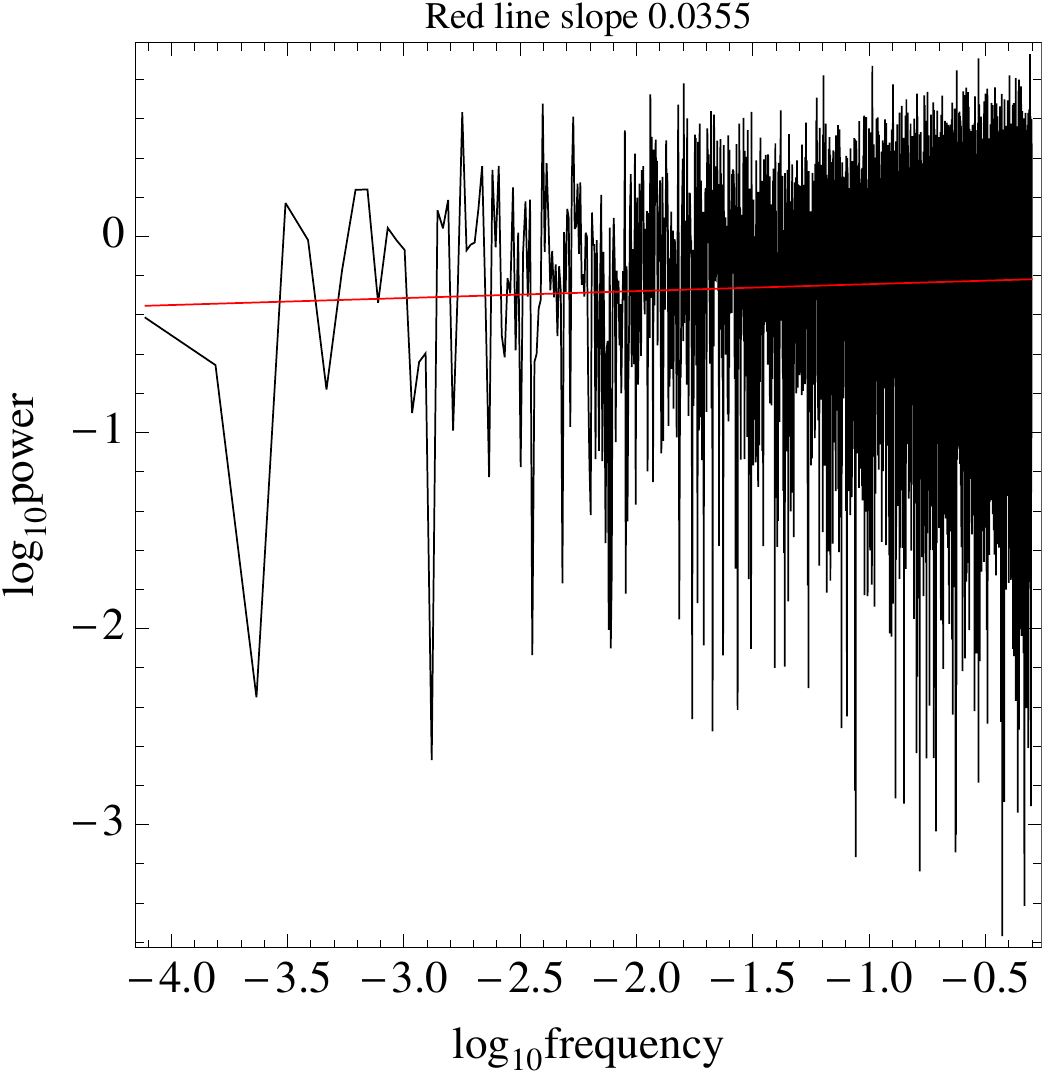}
\caption{Fourier transform of stock return of S\&P 500 (left) and IBM (right).}
\label{ST:fig:SP500:stock_return_DFT}
\end{figure}

In Figs. \ref{ST:fig:SP500:stock_return_loglogplot} and \ref{ST:fig:IBM:stock_return_loglogplot}, we do direct fitting of tails of stock return of S\&P 500 index and IBM respectively (Appendix  \ref{Tail_Power}). Obviously the stock return is fat-tailed. However, the tail exponents obtained here deviate from those obtained by GIGa*N fitting in Fig. \ref{ST:fig:DJIA:stock:loglikelihood}. In Fig. \ref{ST:fig:SP500:stock_return_DFT}, we show the Fourier transform of stock return series of S\&P 500 index and IBM respectively. It exhibits white noise as opposed to the Brown noise of VIX and VXO in Fig. \ref{ST:fig:VIX:VXO:DFT}. 

Notice that in the Heston model \cite{heston1993,gatheral2006} (Appendix \ref{SDE_Volatility}) stock returns are given by GGa$(\al,\be,2)*\text{N(0,1)}$ and cannot generate power-law tail. This contradicts both our and previous results \cite{Stanley1996,Cizeau1997,Liu1999,Gopikrishnan1999,Plerou1999} of $\approx$ 3 to 5 for the tail exponent.

\section{Summary}

We demonstrated that that $\text{GIGa}(\al, \be, \ga)$ provides the best fit to volatility distribution and the product distribution $\text{GIGa}(\al, \be, \ga)*\text{N(0,1)}$ to stock return distribution. Furthermore, we showed that $\ga=2$ is near the median/mode of the $\ga$-distribution of best fits. For $\ga=2$, the stock return distribution is the generalized Student's \emph{t}-distribution T$(0,{\be}/{\sqrt\al}, 2\al)$. Numerical evaluation of parameters of fitting distributions was done with the maximum likelihood estimation method.

Importance of $\ga=2$ puts our findings in excellent agreement with the stochastic volatility and stock return model defined by Eqs.  (\ref{dSS}) and (\ref{ST:eq:GIGa_SDE}). This model fully accounts for the power law tails observed in the volatility and stock return distributions. Additional evidence comes from the fact that Fourier transform of both empirical and simulated time series exhibit Brown noise for volatility and white noise for stock returns (volatility couples to Wiener noise in (\ref{dSS}) while stock return in (\ref{ST:eq:GIGa_SDE}) does not, which accounts for the difference). 

Lastly we argue that since the stock returns have been accumulated over much longer period of time and the definitions of VIX and VXO have changed over time, a better definition of a steady-state volatility would be a distribution whose product with the normal distribution gives distribution of stock returns. This, in turn, may lead to a better approach to calculating volatility than the currently adopted standard.

\appendix

\section{Properties of GIGa distribution}\label{GIGa_Scale}
\begin{figure}[htp]
\centering
\includegraphics[width=0.23\textwidth]{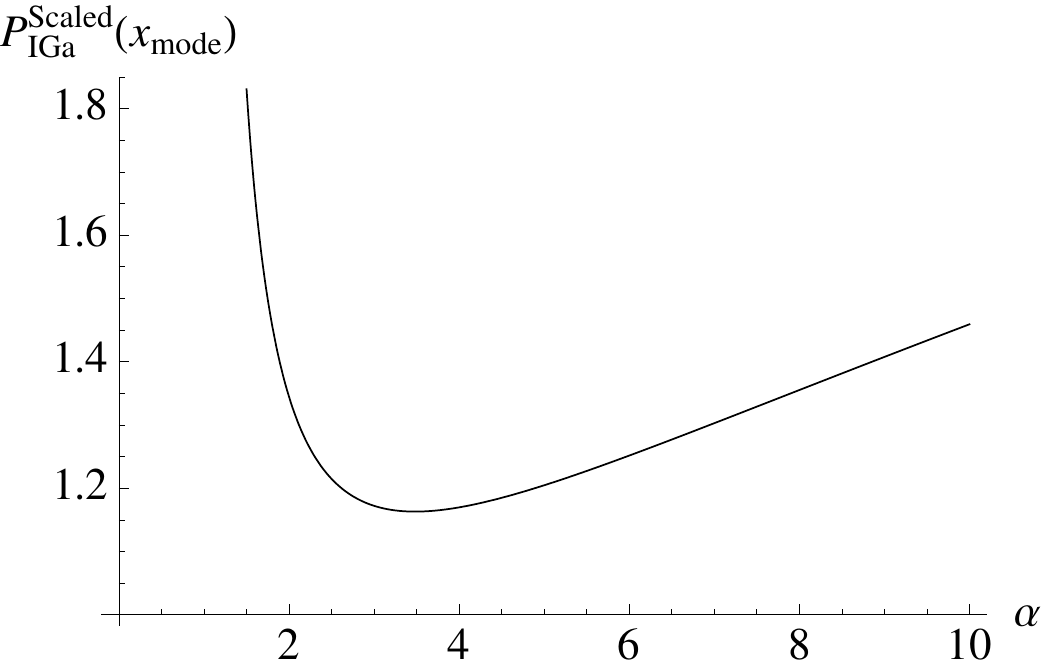}
\includegraphics[width=0.23\textwidth]{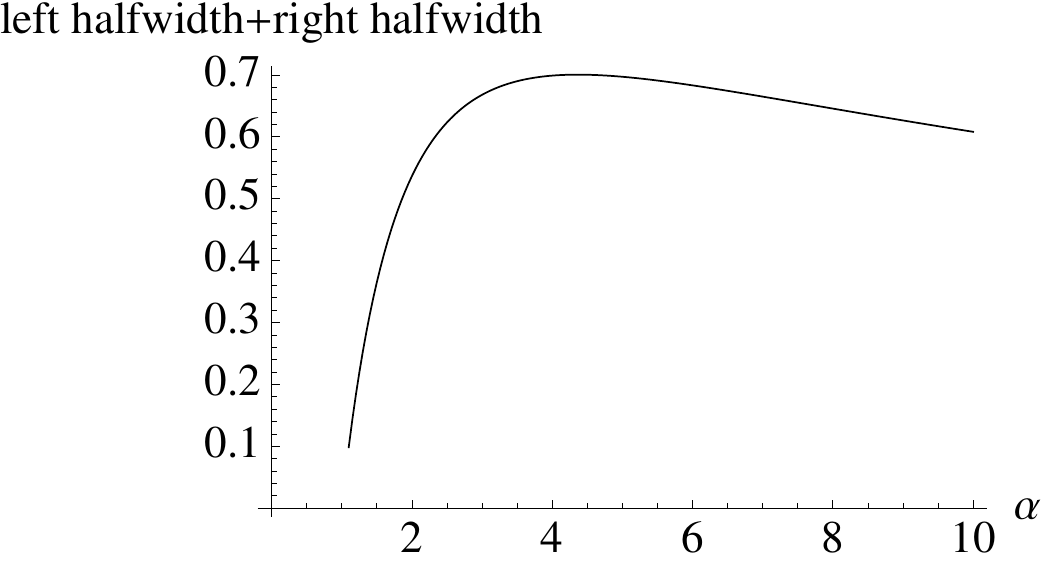}
\caption{Mode and half-width of scaled IGA as a function of $\al$}\label{ST:fig:IGa_PDF_mode}
\end{figure}

\begin{figure}[htp]
\centering
\includegraphics[width=0.345\textwidth]{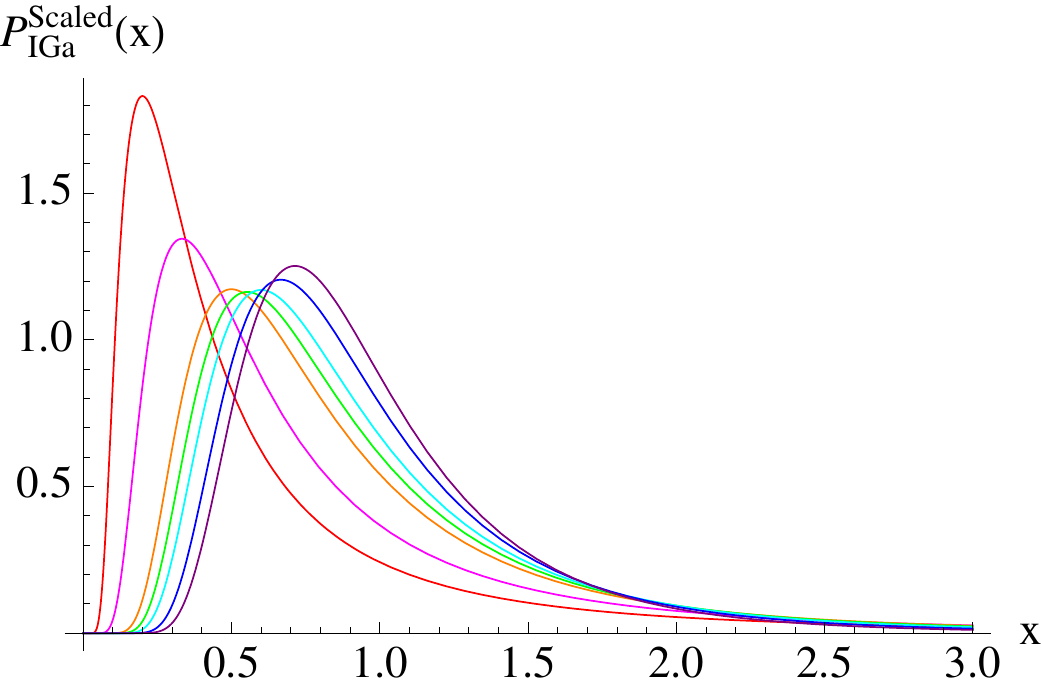}
\caption{PDF of IGa distributions. From left to right, $\al=1.5, 2, 3, 3.48, 4, 5$, and $6$, corresponding to red, magenta, orange, green, cyan, blue, and purple lines.}
\label{ST:fig:IGa_PDF_list}
\end{figure}

We begin with the $\ga=1$ limit of GIGa, namely IGa distribution PDF
\begin{equation}
P_{\IGa}(x) = \fr{1}{\be \Gamma(\al)} \exp\lf[ -\fr{\be}{x} \rg] \lf(\fr{\be}{x}\rg)^{1+\al} .
\end{equation}
Setting the mean to unity, the scaled distribution is 
\begin{equation}
P_{\IGa}^{\text{Scaled}}(x) = \fr{(\al-1)^\al \exp\lf( -\fr{\al-1}{x} \rg)}{\Gamma(\al) x^{1+\al}} . 
\end{equation}
The mode of the above distribution is $x_{\text{mode}} = (\al-1)/(\al+1)$. The modal PDF is
\begin{equation}
P_{\IGa}^{\text{Scaled}}(x_\text{mode}) = 
\fr{(1+\al)^{1+\al} \exp(-1-\al)}{\Gamma(\al)(\al-1)} , 
\end{equation}
which has a minimum at $\al \approx 3.48$ as shown in Fig. \ref{ST:fig:IGa_PDF_mode}. The change in PDF behavior on transition through this value is clearly observed Fig. \ref{ST:fig:IGa_PDF_list}. Also plotted in Fig. \ref{ST:fig:IGa_PDF_mode} is the half-width of the distribution. Clearly, it highly correlates with the modal PDF above.

Both minimum and maximum above clearly separate the regime of small $\al$: $\al\rightarrow 1$, where the approximate form of the scaled PDF is 

\begin{equation}
P_{\IGa}^{\text{Scaled}}(x) \approx
\fr{(\al-1) \exp\lf[ -\fr{\al-1}{x} \rg]}{x^2} ,
\end{equation}
whose mode is $(\al-1)/2$ and the magnitude of the maximum is $ 4 \exp[-(\al-1)^2/2]/(\al-1) \approx 4/(\al-1) $, from the regime of large $\al$, $\al\rightarrow \infty$, where 
\begin{equation}
P_{\IGa}^{\text{Scaled}}(x) \rightarrow \delta(x-1) .
\end{equation}

\begin{figure}[htp]
\centering
\includegraphics[width=0.345\textwidth]{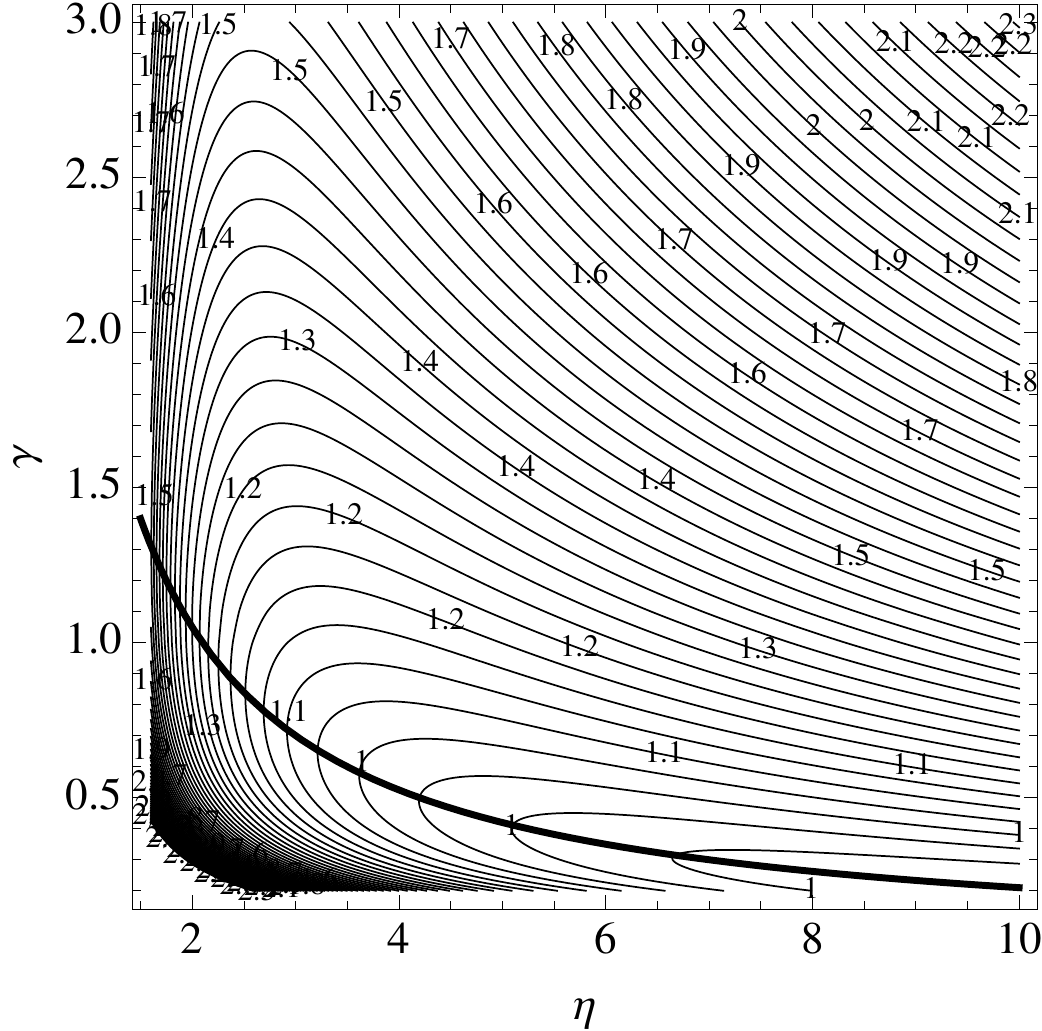}
\includegraphics[width=0.345\textwidth]{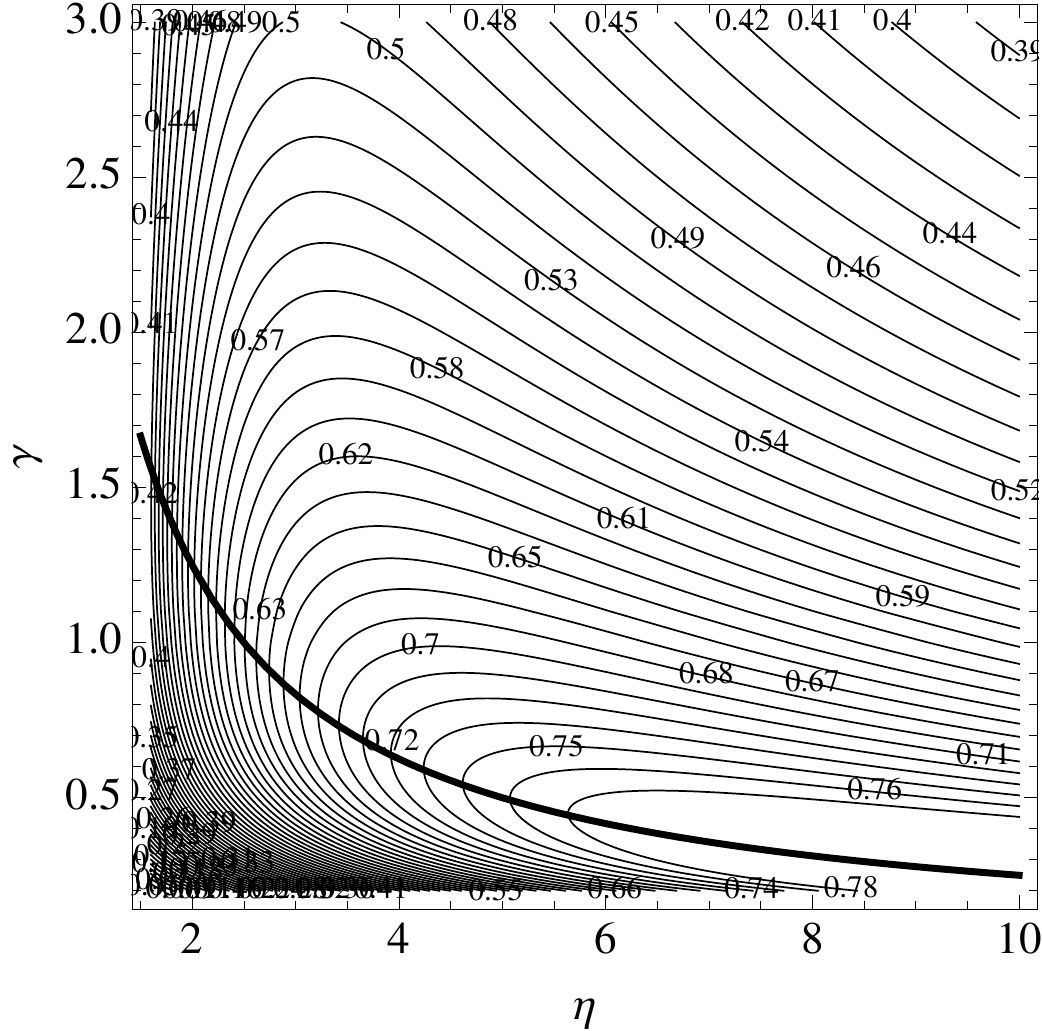}
\caption{Top: contours of modal PDF of GIGa distributions with mean $1$. Thin lines: contours of modal PDF at mode. Thick line: $\ga=2.1/\eta$. Bottom: contours of total half-widths of GIGa distributions with mean $1$. Thick line: $\ga=2.5/\eta$. }
\label{ST:fig:GIGa_contour}
\end{figure}

\begin{figure}[htp]
\centering
\includegraphics[width=0.345\textwidth]{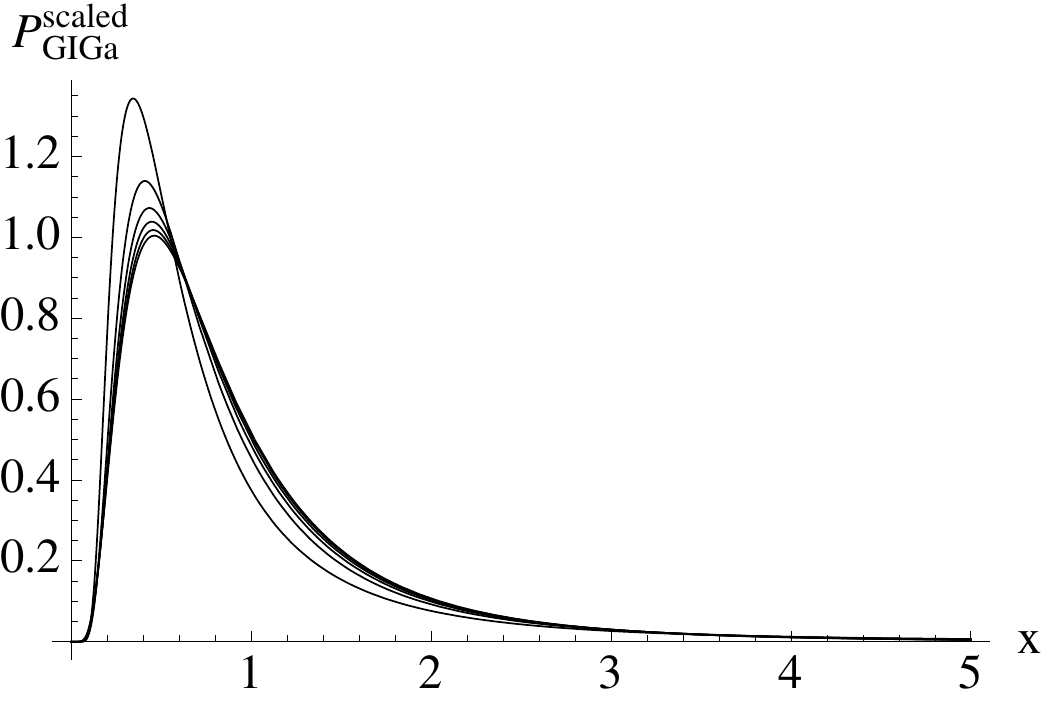}
\caption{Scaled PDF of GIGa distributions with mean $1$. In the plots, $\ga=2.1/\eta$. Six lines correspond to $\eta=2,3,...,7$}
\label{ST:fig:GIGa_PDF_list_overlay}
\end{figure}

\begin{figure}
\centering
\includegraphics[width=0.23\textwidth]{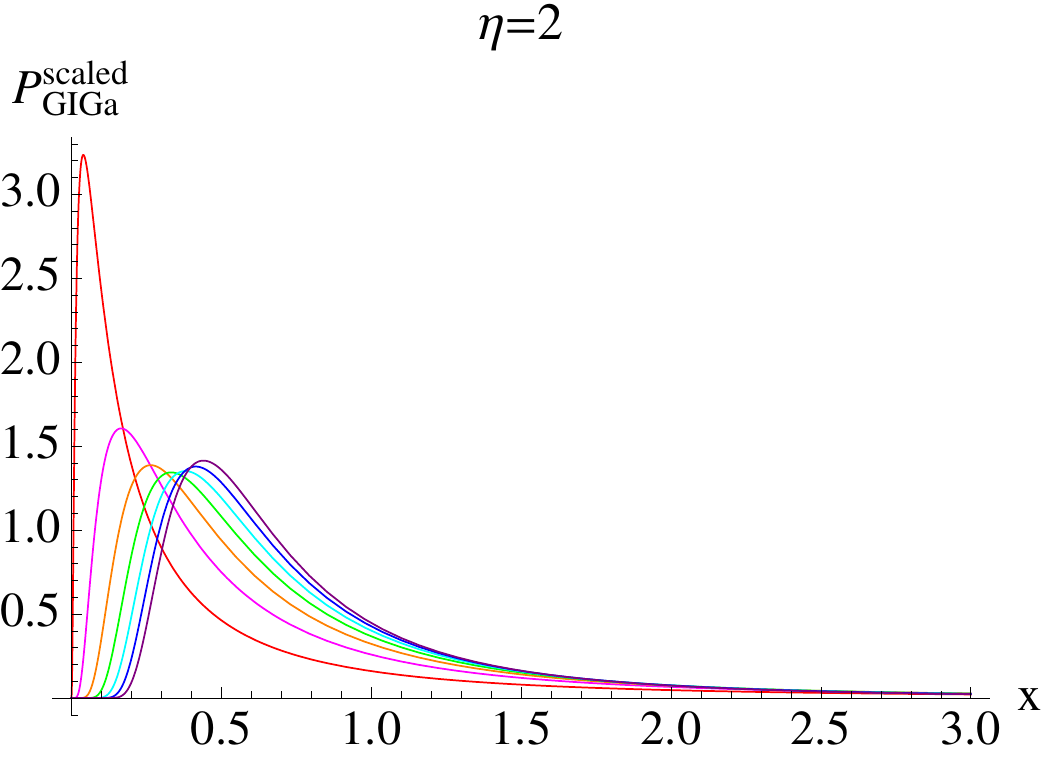}
\includegraphics[width=0.23\textwidth]{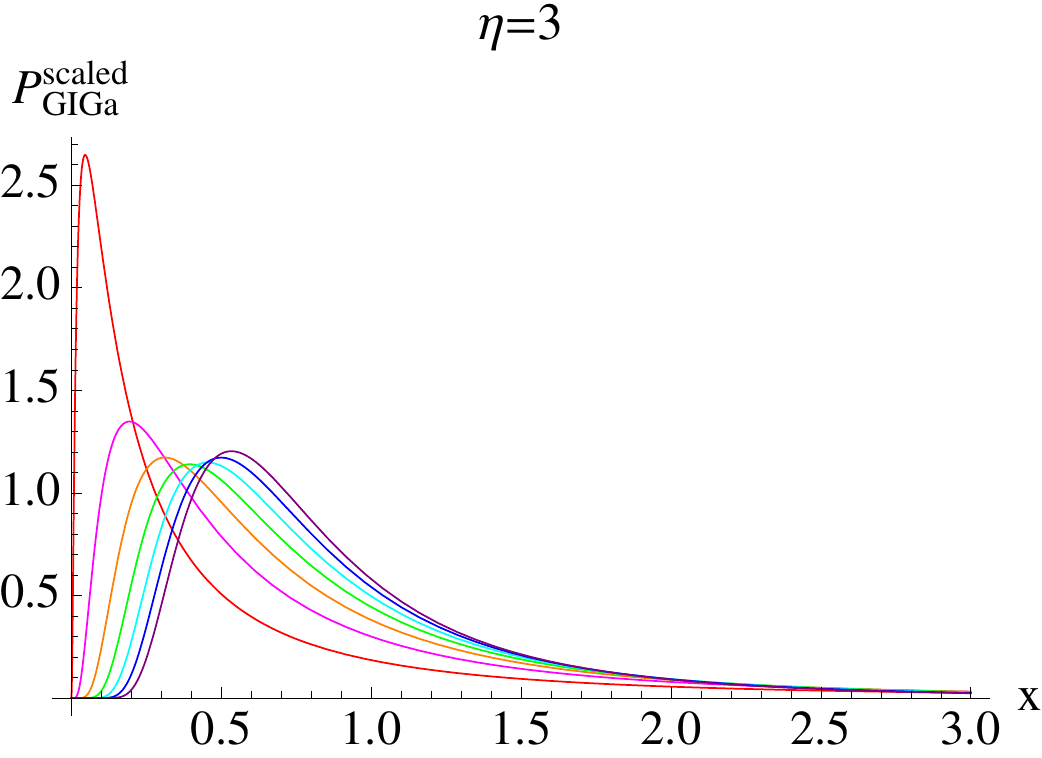}\\
\includegraphics[width=0.23\textwidth]{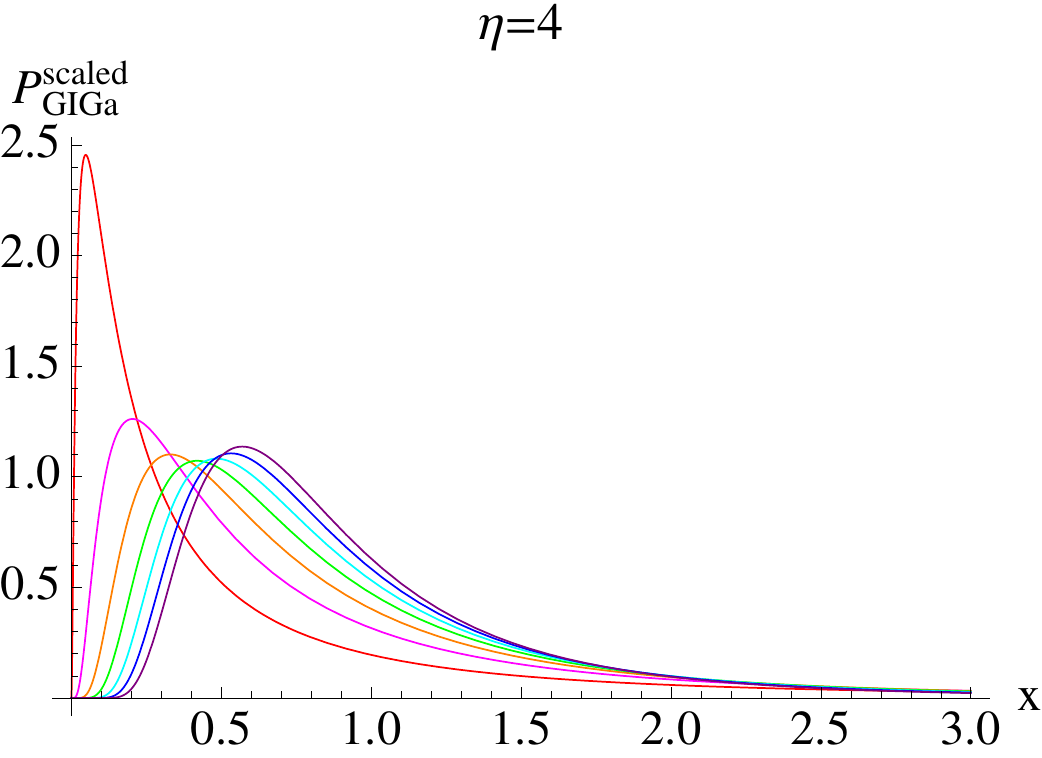}
\includegraphics[width=0.23\textwidth]{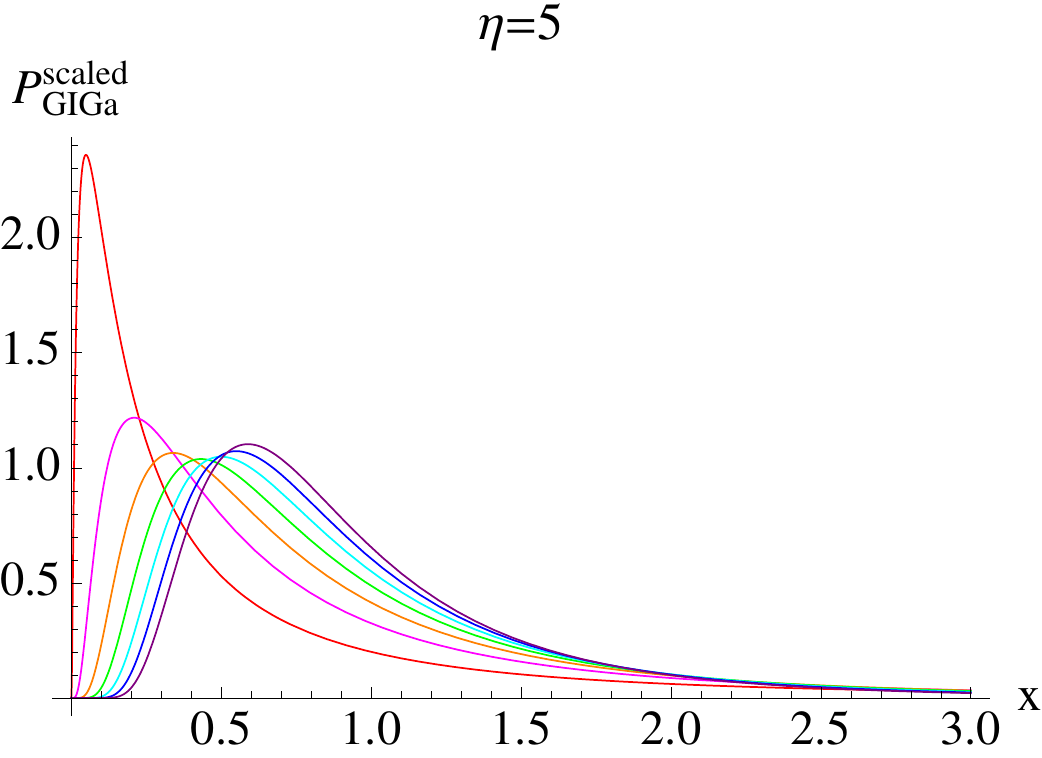}\\
\includegraphics[width=0.23\textwidth]{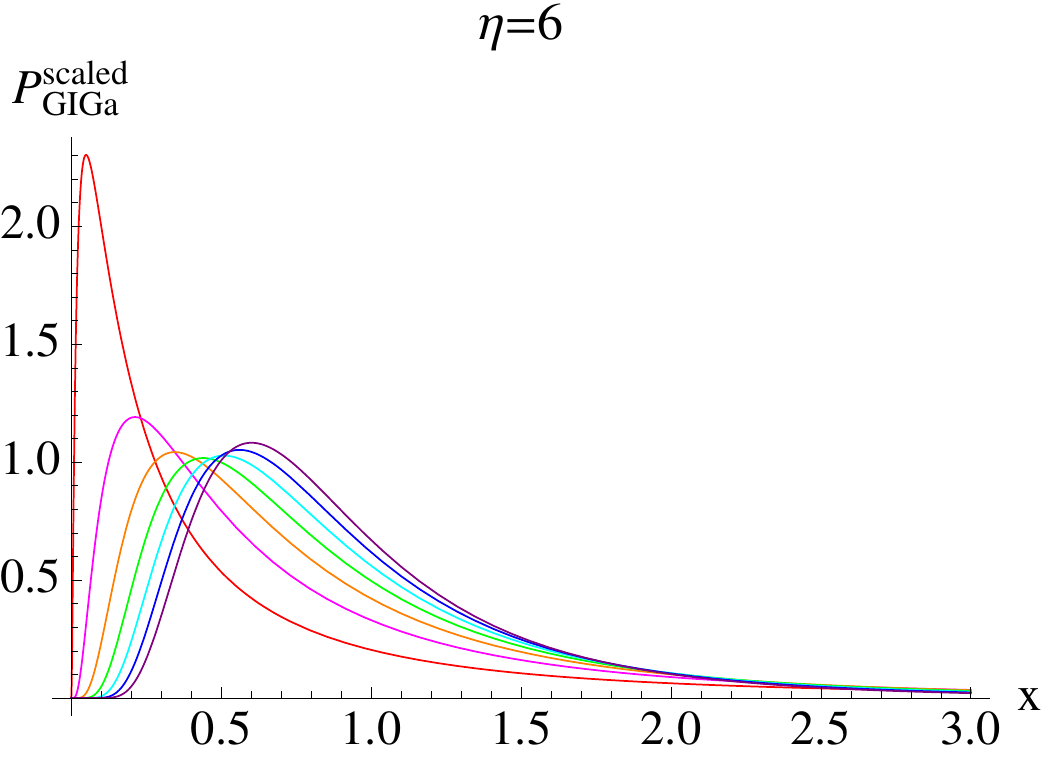}
\includegraphics[width=0.23\textwidth]{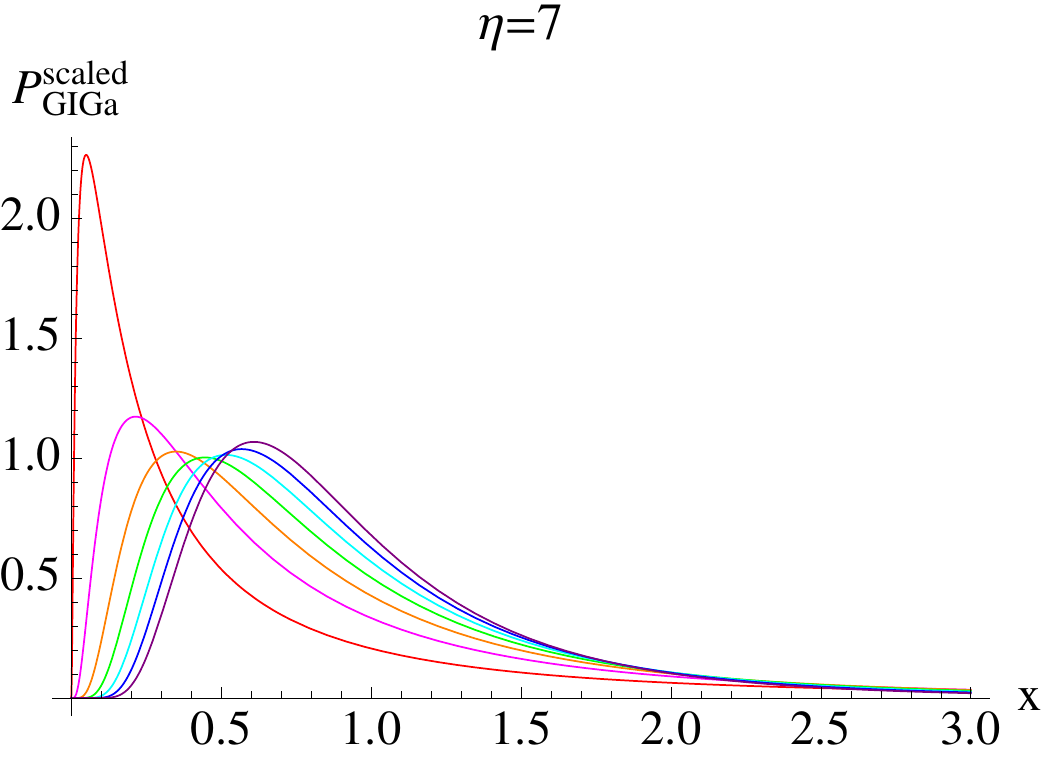}
\caption{Scaled PDF of GIGa distributions with mean $1$. In each subplot with constant $\eta$, from left to right, $\ga=0.5/\eta, 1/\eta, 1.5/\eta, 2/\eta, 2.5/\eta, 3/\eta$, and $3.5/\eta$, corresponding to red, magenta, orange, green, cyan, blue, and purple lines. }
\label{ST:fig:GIGa_PDF_list}
\end{figure}

We now turn to GIGa distribution and the effect of parameter $\ga$. In Fig. \ref{ST:fig:GIGa_contour} we give the contour plots of modal PDF and total half-widths in the $(\eta, \ga)$ plane, where $\eta = \al\ga$ and $-1-\eta$ is the exponent of the power law tail.  We observe an interesting \emph{scaling property} of GIGa: for $\ga\approx 2.1/\eta$, the dependence of the PDF on $\eta$ is very weak, as demonstrated in Fig.\ref {ST:fig:GIGa_PDF_list_overlay}, where it is plotted for integer $\eta$ from 2 to 7. An alternative way to illustrate this is to plot PDF for a fixed $\eta$ and variable $\ga$, as shown in Fig. \ref{ST:fig:GIGa_PDF_list}. Following the thick line we notice that, for $\eta > 3$, mode and half-width change very little with $\eta$. The key implication of the scaling property is that IGa contains all essential features pertinent to GIGa.

\section{Parametrization of the GIGa family of distributions}\label{GIGa_LN}

This Appendix is a self-contained re-derivation of a LN limit of GIGa.  \cite{weibullcom} The three-parameter GIGa distribution is given by
\begin{equation}\label{ST:eq:PDF_GIGa}
\GIGa(x; \al, \be, \ga)
= \fr{\ga}{\be\Ga(\al )} e^{-\lf(\fr{\be }{x}\rg)^\ga} \lf(\fr{\be }{x}\rg)^{1+\al\ga}
\end{equation}
for $x>0$ and 0 otherwise.
We require that $\al, \be, \ga >0$. IGa is the the $\ga=1$ case of GIGa:
\begin{equation}\label{ST:eq:PDF_IGa}
\IGa(x; \al, \be)
= \fr{1}{\be\Ga(\al )} e^{-\fr{\be }{x}} \lf(\fr{\be }{x}\rg)^{1+\al} .
\end{equation}
Note that GIGa and IGa have power-law tails $x^{-1-\al\ga}$ and $x^{-1-\al}$ respectively for $x \gg \be$.

We proceed to rewrite GIGa in the following form:
\begin{equation}
\begin{split}
\GIGa(x; \al, \be, \ga)
&= \fr{\ga}{x \Ga(\al )} \exp\lf[\al \ln\lf(\fr{x}{\be}\rg)^{-\ga} - \lf(\fr{x}{\be}\rg)^{-\ga}\rg] .
\end{split}
\end{equation}
A re-parameterization
\begin{eqnarray}\label{ST:eq:reparameterization}
\mu &=& \ln \be - \fr{1}{\ga} \ln\fr{1}{\lambda^2}  \\
\si &=& \fr{1}{\ga \sqrt{\al}} \\
\lambda &=& \fr{1}{\sqrt{\al}} ,
\end{eqnarray}
with $\si>0$ and $\lambda>0$, allows to express the old parameters in terms of the new:
\begin{eqnarray}
\al &=& \fr{1}{\lambda^2} \\
\be &=& e^\mu \lambda^{-\fr{2\si}{\lambda}} \\
\ga &=& \fr{\lambda}{\si} ,
\end{eqnarray}
leading, in turn, to
\begin{equation}\label{ST:eq:exp_expansion}
\lf(\fr{x}{\be}\rg)^{-\ga} = e^{-\fr{\lambda}{\si} (\ln x - \mu)} \lambda^{-2}
\end{equation}
\begin{equation}
\ln\lf(\fr{x}{\be}\rg)^{-\ga} = -\fr{\lambda}{\si} (\ln x - \mu) + \ln(\lambda^{-2}) \\
\end{equation}
and
\begin{equation}\label{ST:eq:GIGaLN1}
\al \ln\lf(\fr{x}{\be}\rg)^{-\ga} - \lf(\fr{x}{\be}\rg)^{-\ga} \\
\approx \fr{\ln(\lambda^{-2}) - 1}{\lambda^2} - \fr{(\ln x - \mu)^2}{2\si^2} ,\\
\end{equation}
where we have used the Taylor expansion of the $\exp$ term in Eq. (\ref{ST:eq:exp_expansion}), which depends on $\lambda/\si = \ga \rightarrow 0^+.$
We can also prove that
\begin{equation}\label{ST:eq:GIGaLN2}
\fr{\ga}{\Ga(\al)} \exp\lf[ \fr{\ln(\lambda^{-2}) - 1}{\lambda^2} \rg] = \fr{1}{\sqrt{2\pi}\si} ,
\end{equation}
based on the Stirling's approximation when we let $\lambda^{-2} = \al \rightarrow +\infty$.

Upon substitution of Eqs. (\ref{ST:eq:GIGaLN1}) and (\ref{ST:eq:GIGaLN2}) into eq. (\ref{ST:eq:PDF_GIGa}), we obtain the LN distribution
\begin{equation}\label{ST:eq:LN}
\LN(x; \mu, \si) = \fr{1}{\sqrt{2\pi}\si x} \exp \lf[ - \fr{(\ln x - \mu)^2}{2 \si^2} \rg] .
\end{equation}
In conclusion, GIGa has the limit of LN when $\lambda$ tends to $0$ in such a way that $\alpha$ tends to $+\infty$ quadratically and $\gamma$ tends to $0$ linearly.

GIGa (IGa) are also transparently related to GGa (Ga) distribution: $\GGa(x; \al, \be, \ga) \xleftrightarrow{\ga\leftrightarrow -\ga} -\GGa(x; \al, \be, -\ga) = -\GIGa(x; \al, \be, \ga)$ and $\GGa(x; \al, \be, \ga) \leftrightarrow \GIGa(1/{x}; \al, 1/{\be}, \ga)$. Note, finally, that Lawless \cite{Lawless1982} derived the LN limit of GGa in a manner similar to ours, which solidifies the concept of the``family'' that unites these distributions.

\section{Stochastic ``birth-death'' model }\label{Birth_Death}
Many natural and social phenomena fall into a stochastic ``birth-death'' model, described by the equation 
\begin{equation}
dx = c_1 x^{1-\ga} dt - c_2 x dt + \si x dW ,
\end{equation}
where $x$ can alternatively stand for additive quantities such as wealth, \cite{ma2013distribution} body mass of a species, \cite{west2012}  human response time, \cite{ma12RT} etc., and volatility variance in this work.

The second term in the rhs describes an exponentially fast decay, such as the loss of wealth and mass due to the use of one's own resources, or the reduction of volatility in the absence of competing inputs and of response times due to learning. The first rhs term may alternatively describe metabolic consumption, acquisition of wealth in economic exchange, plethora of market signals, and variability of cognitive inputs. 

The third, stochastic term is the one that changes the otherwise deterministic dynamics, characterized by the saturation to a final value of the quantity, with the probabilistic distribution of the values - as it were, GIGa in the steady-state limit. Furthermore, just as the wealth model has microscopic underpinnings in a network model of economic exchange, \cite{ma2013distribution} it is likely that stochastic ontogenetic single body mass growth \cite{west2012} could be described by analogous network model based on capillary exchange. A network analogy may be possible for cognitive response times and volatility as well.

\section{Product distribution of GIGa and GGa with normal distribution}\label{Product_Distribution}

Given two distributions of $x$ and $y$ with PDF $f(x)$ and $g(y)$ respectively, the product distribution is defined as
\begin{equation}
z=xy ,
\end{equation}
whose PDF is given by
\begin{equation}
\int_{-\infty}^\infty \fr{1}{|x|} f(x) g(\fr{z}{x}) dx .
\end{equation}
We are interested in the circumstance when the distribution of $x$ is generated from a stochastic volatility model, such as (\ref{dsigma}), and $y$ is normally distributed with 0 mean, such as assumed in (\ref{dSS}). Since the standard deviation of $y$ can always be absorbed in $x$, it can be set to 1 without loss of generality so that $y$ has the standard normal distribution $\text{N}(0,1)$.

The closed form of $\GGa(\al,\be,\ga) * N(0,1)$ or $\GIGa(\al,\be,\ga) * \text{N}(0,1)$ cannot be obtained in the general case. Below we give expressions for several important limits:
\begin{equation}\label{GGa*N}
\GGa(\al,\be,2)*\text{N}(0,1)
= \fr{\sqrt{\fr{2}{\pi }} \lf(\fr{|z|}{\sqrt{2} \be }\rg)^{\al -\fr{1}{2}} K_{\al
   -\fr{1}{2}}\lf(\fr{\sqrt{2} |z|}{\be }\rg)}{\be  \Ga (\al )} ,
\end{equation}
where $K$ is the modified Bessel function of the second kind. 
\begin{equation}
\GIGa(\al,\be,2)*\text{N}(0,1)
= \fr{\Gamma\lf(\fr{1}{2}+\al\rg)}{\sqrt{2\pi} \be\Gamma(\al)} \lf(\fr{2\be^2}{z^2+2\be^2}\rg)^{\fr{1+2\al}{2}} ,
\end{equation}
which is the generalized Student's \emph{t}-distribution T$(0,\fr{\be}{\sqrt\al}, 2\al)$. 
\begin{equation}\label{GIGa*N}
\IGa(\al,\be)*\text{N}(0,1)
= \fr{2^{-\fr{\al }{2}-1} \al  \lf(\fr{\be }{z}\rg)^{\al } U\lf(\fr{\al
   +1}{2},\fr{1}{2},\fr{\be ^2}{2 z^2}\rg)}{\sqrt{\pi } z} ,
\end{equation}
where $U$ is Tricomi's confluent hypergeometric function. 

Conversely, the mean and variance of variable $z$ can be analytically evaluated in the general case:
\begin{equation}
\begin{split}
&\text{mean} \, z = 0 \\
&\text{var} \, z = \fr{\be ^2 \Ga \lf(\al \pm \fr{2}{\ga}\rg)}{\Ga (\al )} , 
\end{split}
\end{equation}
where the plus and minus correspond to distributions $\GGa(\al,\be,\ga)*N(0,1)
$ and $\GIGa(\al,\be,\ga)*N(0,1)$ respectively. Notice also that the variance of the GGa/GIGa itself is given by 
\begin{equation}
\fr{\be ^2 \lf[\Ga (\al ) \Ga \lf(\al \pm \fr{2}{\ga}\rg)-\Ga \lf(\al
   \pm \fr{1}{\ga}\rg)^2\rg]}{\Ga (\al )^2} . 
\end{equation}
Finally, it can be shown that \cite{MaThesis2013} 
\begin{equation}
\GGa(\al,\be,\ga)*\text{N}(0,1)
\rightarrow |z|^{-1-\al\ga}, \, |z| \rightarrow \infty ,
\end{equation}
that is the tail of the GIGa carries through into its product distribution with the normal.

\section{Stochastic differential equations of volatility}\label{SDE_Volatility}

Per Eq. (\ref{P_integral}), any transformation $f \rightarrow f \si^{2\al}, g\rightarrow f \si^{\al}$ does not change the functional form of the integral. In turn, this means that there exists a family of stochastic differential equations (SDE) for each type of the distribution, such as GIGa, GGa, etc. However, as already mentioned, we are interested in a number of specific SDE rooted in modeling of various phenomena. With this in mind, we discuss several SDE for volatility and the corresponding steady state distributions.

\subsection{GIGa}
The equation for volatility, which was obtained from NDL of GARCH(1,1), was already discussed in Section II and is added here for completeness. The steady-state, normalized solution of  
\begin{equation}\label{ST:eq:GIGa_SDE_apx}
d\si = J(\te \si^{1-\ga}-\si)dt + \Si \si dW ,
\end{equation}
is given by 
\begin{equation}\label{ST:eq:GIGa_from_SDE_apx}
\begin{split}
P(\si) &= \GIGa\lf(\si; \al,\be,\ga\rg) \\
&=\GIGa\lf(\si; \lf(1+\fr{2J}{\Si^2}\rg) \ga^{-1}, \lf(\fr{\te}{\ga}\fr{2J}{\Si^2}\rg)^{1/\ga},\ga\rg), 
\end{split} 
\end{equation}
where
\begin{equation}\label{ST:eq:GIGa_SDE_theta_apx}
\te 
= \fr{\ga\Si^2}{2J} 
\lf[ \fr{\overline{\si} \Ga((1+\fr{2J}{\Si^2}) \ga^{-1}) }{\Ga(\fr{2J}{\Si^2} \ga^{-1})} \rg]^\ga . 
\end{equation}
and $\overline{\si}$ is the mean value of $\si$. Above, $\overline{\si}$, $J$ and $\Si$ are positive constants. Similar assumptions are made throughout this section. Since $\te(1)=\overline{\si}$, for $\ga=1$ the distribution reduces to
\begin{equation}
P(\si) = \IGa\lf(1+\fr{2J}{\Si^2}, \fr{2J\overline{\si}}{\Si^2}\rg) ,
\end{equation}

\subsection{GGa}

A natural generator of GGa is Eq. (\ref{ST:eq:GIGa_SDE}) with reversed signs of $J$ and $\ga$, 
\begin{equation}\label{ST:eq:GGa2_SDE}
d\si = J(\si - \te \si^{1+\ga})dt + \Si \si dW  .
\end{equation}
Its steady-state solutions is given by
\begin{equation}\label{ST:eq:GGa_from_another_SDE}
P(\si)=\GGa\lf(\si; (-1+\fr{2J}{\Si^2})\ga^{-1}, \lf( \fr{\ga}{\te} \fr{\Si^2}{2J}  \rg)^{1/\ga}, \ga\rg) .
\end{equation}
Taking into account the scaling property discussed at the top of the Appendix, another two possible SDEs are
\begin{equation}\label{ST:eq:GGa_SDE}
d\si = J(1-\te \si^\ga)dt + \Si \sqrt\si dW ,
\end{equation}
and 
\begin{equation}\label{ST:eq:GGa_yet_another_SDE}
d\si = J(\si^{-1}-\te \si^{\ga-1})dt + \Si dW .
\end{equation}
The latter equation is particularly significant since it is a direct consequence of the Heston model for $\ga=2$. 

Indeed, the Heston model \cite{heston1993,gatheral2006} for volatility variance $V$ reads 
\begin{equation}\label{ST:eq:Ga_SDE_heston}
dV = J\lf(\overline{V} - V\rg) dt + \phi \sqrt V dW .
\end{equation}
Absorbing ${\overline{V}}$ in $J$, we rewrite the equation as 
\begin{equation}\label{ST:eq:Ga_SDE_variation}
dV = J\lf(1 - \fr{V}{\overline{V}}\rg) dt + \phi \sqrt V dW .
\end{equation}
The latter yields a steady-state distribution given by
\begin{equation}
P(V) = \Gadist\lf(V; \fr{2J}{\phi^2},\fr{\phi^2 \overline{V}}{2J}\rg) ,
\end{equation}
whose mean is $\overline{V}$. Changing variable to volatility $\si=\sqrt{V}$, Ito calculus yields 
\begin{equation}
d\si = \lf[\fr{1}{2} J \lf(\si^{-1} - \fr{\si}{\overline{V}}\rg) - \fr{\phi^2}{8}\si^{-1}\rg] dt 
+ \fr{1}{2}\phi dW, 
\end{equation}
whose steady-state distribution is given by
\begin{equation}
P(\si) = \GGa\lf(\si; \fr{2J}{\phi^2}, \phi\sqrt{\fr{\overline{V}}{2J}}, 2\rg) .
\end{equation}

\subsection{LN}

An Ornstein-Uhlenbeck process 
\begin{equation}
dx = \te (\mu-x)dt + \si dW  
\end{equation}
yields a normal steady state distribution
\begin{equation}
P(x) = \sqrt{\fr{\te}{\pi\si^2}} \exp\lf[ -\te \fr{(x-\mu)^2}{\si^2} \rg] ,
\end{equation}
whose mean is $\mu$. 
A change of variable $x=\log X$ leads, with Ito calculus, to \cite{scott1987,wiggins1987}
\begin{equation}\label{ST:eq:LN_SDE}
dX = \te X \lf(\mu - \log X \rg) dt + \fr{1}{2}\si^2 X dt + X \si dW 
\end{equation}
and the steady state distribution 
\begin{equation}
P(X) = \sqrt{\fr{\te}{\pi}} \fr{1}{\si X} \exp\lf[ -\te \fr{(\log X-\mu)^2}{\si^2} \rg]  ,
\end{equation}
whose mean is $\exp\lf(\mu+{\si^2}/{2\te}\rg)$.

Just as before, when we showed that LN distribution can be obtained as a limit of GIGa distribution, we can show that LN SDE can be obtained as a limit of GIGa SDE. Changing notations in (\ref{ST:eq:GIGa_SDE}), we rewrite it as 
\begin{equation}\label{ST:eq:LN_SDE2}
dY 
= J(\Theta Y^{1-\ga}-Y)dt + Y \Si .
\end{equation}
 It can be shown that (\ref{ST:eq:LN_SDE}) is a limiting case of (\ref{ST:eq:LN_SDE2}) if we set 
 \begin{equation}
\begin{split}
&J\ga = \te \\
&\overline{Y} = \exp\lf( \mu+\fr{2\si^2}{2\te} \rg) \\
&\Si = \si
\end{split}
\end{equation}
and let $\ga\rightarrow 0^+$ linearly and $({2J}/{\Si^2}) \ga^{-1}$ ($\al$ in GIGa$(\al,\be,\ga)$) tend to $+\infty$ quadratically. Details of the derivation can be found in \cite{MaThesis2013}.

\section{Relaxation time}\label{Relax_Time}

Consider an IGa process defined as 
\begin{equation}\label{ST:eq:IGa_relaxation_time}
dX = J(1-X) dt + \Si X dW ,
\end{equation}
where $J$ and $\Si$ are constants and $dW$ is Wiener process. This is the process described by Eq. (\ref{ST:eq:GIGa_SDE}) (Eq. (\ref{ST:eq:GIGa_SDE_apx})) for GIGa with $\ga=1$ and $\overline{X}=1$. As previously pointed out in Appendix \ref{GIGa_Scale}, a GIGa process can be understood from that of IGa.] The stationary distribution of $X$ is an IGa distribution,
\begin{equation}\label{stationary}
P_{s}(X) =\IGa (X ; 1+\fr{2J}{\Si^2}  ,\fr{2J}{{\Si^2}}) .
\end{equation}
The purpose of this Appendix is to estimate the mean and the standard deviation of the relaxation time on approach to the steady-state distribution and to test these results numerically.

The existence of the stationary distribution is possible due to the first term in Eq. (\ref{ST:eq:IGa_relaxation_time}). For $J=0$, on the other hand, it reduces to a lognormal process described by the time dependent distribution (obtained with Ito calculus) given by
\begin{equation}\label{t-dependent}
P_{t}(X,t)=\fr{1}{X \sqrt{2 \pi \Si t}} \exp \left[ - \fr{(\ln X + g^{2} t /2)^{2}} {2gt} \right] .
\end{equation}
Clearly, (\ref{t-dependent}) describes a normalized distribution which tends to zero for every $X$ as time tends to infinity.

The mean relaxation time can be defined as the time scale $t$ such that $\overline{\ln X}_{s} \approx  \overline{\ln X}_{t}$, where the mean are evaluated with distributions (\ref{stationary}) and (\ref{t-dependent}) respectively. Simple calculation yields
\begin{equation}\label{ST:eq:relaxation_time_fixed_sigma_mean}
\text{mean} = -\fr{2 c_{1}}{\si^2} \left[  - \psi^{(0)}\lf(\fr{2J}{\si^2}\rg) + \ln\lf(\fr{2J}{\si^2}\rg) \right] ,
\end{equation}
where $\psi^{(0)}$ is the digamma function and $c_{1}$ is a constant to account for the approximate nature of the estimate.\footnote{The same result can be obtained by equating the modes of the two distributions} Note that when $2J/\si^2\gg 1$, (\ref{ST:eq:relaxation_time_fixed_sigma_mean}) becomes 
\begin{equation}\label{ST:eq:relaxation_time_approximation}
\text{mean} \approx \fr{c_{1}}{2J}. 
\end{equation} 

Similarly, the rms of relaxation time can be estimated from $\overline{(\ln X_{s}-\overline{\ln X}_{s})^{2}} \approx  \overline{(\ln X_{t}-\overline{\ln X}_{t})^{2}}$
\begin{equation}\label{ST:eq:relaxation_time_fixed_sigma_STDEV}
\text{STDEV}=\fr{c_{2}}{\si^2} \psi^{(1)}\lf( 1+\fr{2J}{\si^2} \rg) ,
\end{equation}
where $\psi^{(1)}$ is the polygamma function of order 1 and $c_2$ is a constant. 

\begin{figure}[htp]
\centering
\includegraphics[width=0.23\textwidth]{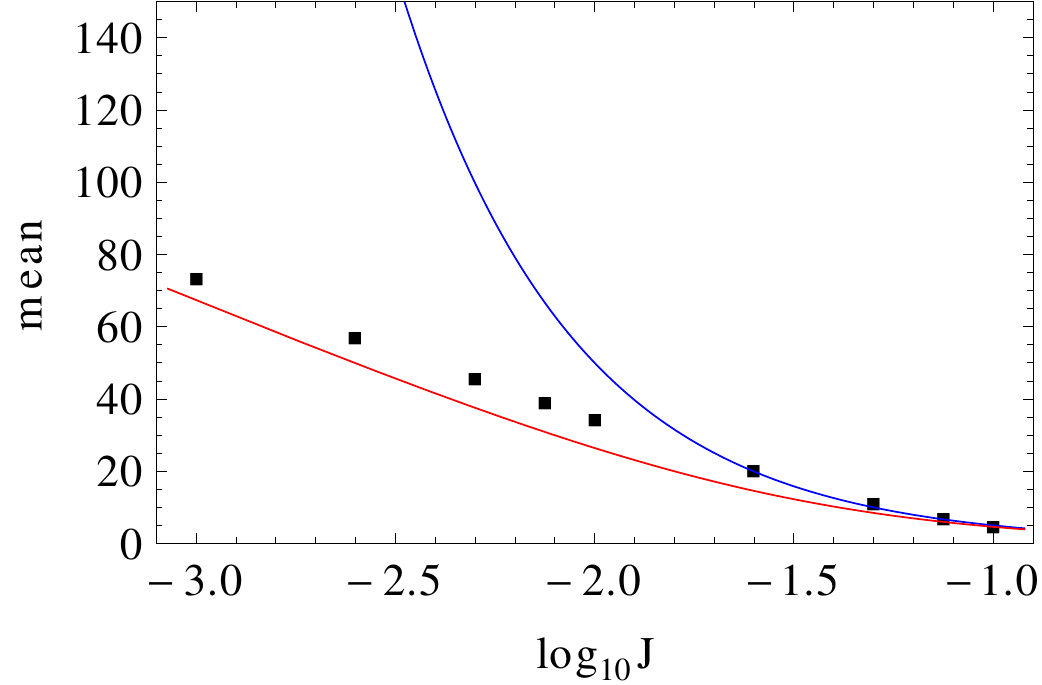}
\includegraphics[width=0.23\textwidth]{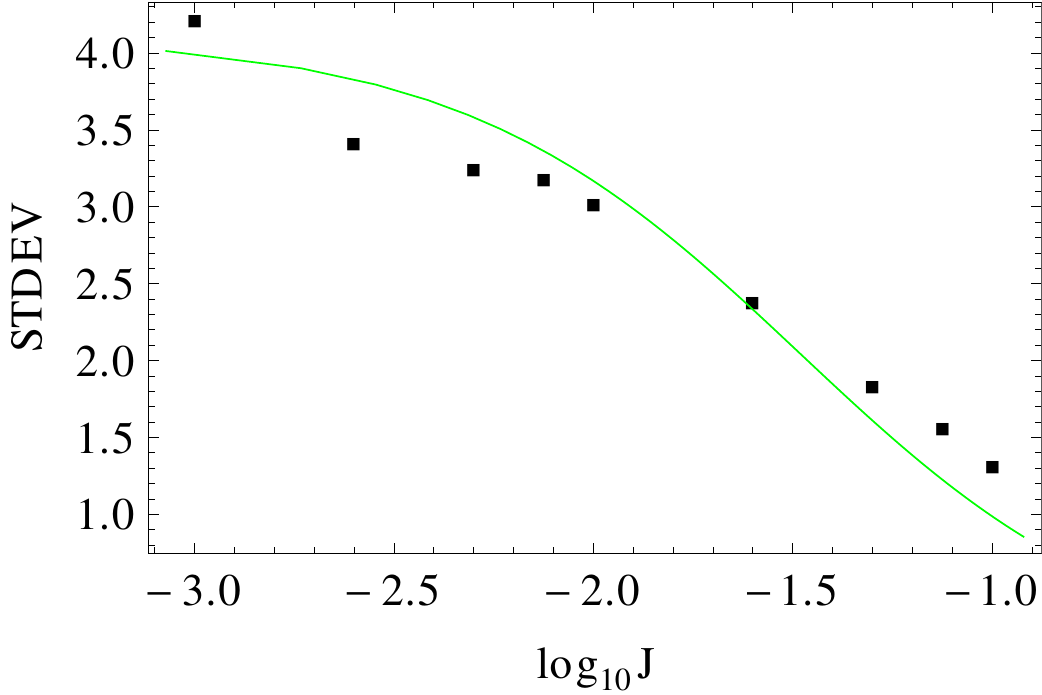}\\
\includegraphics[width=0.23\textwidth]{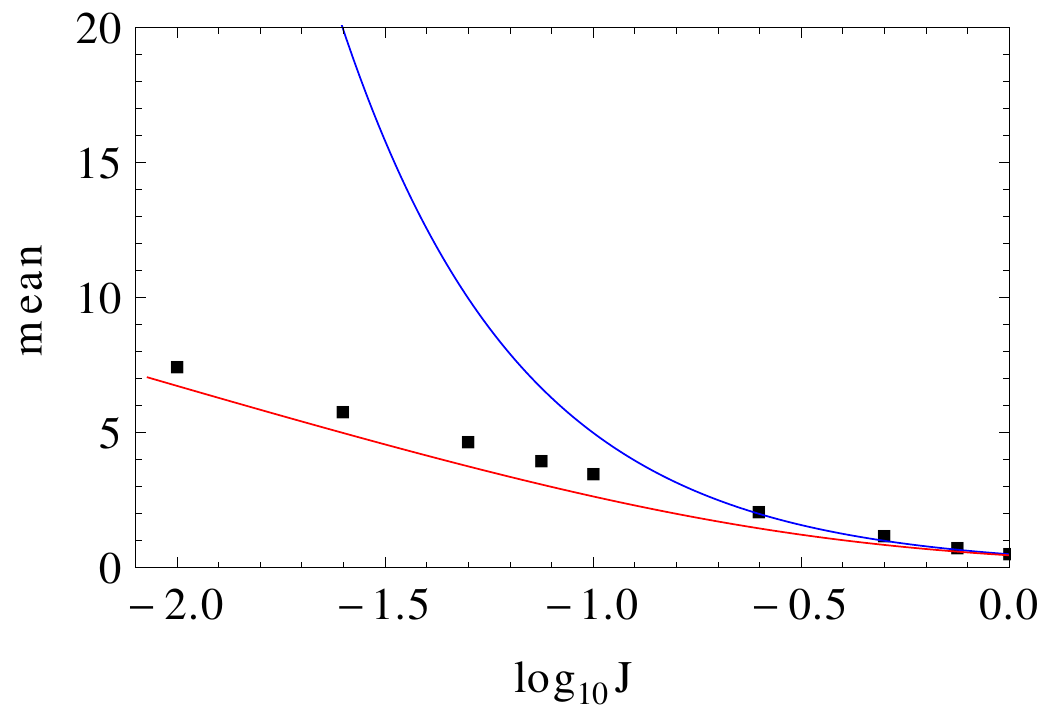}
\includegraphics[width=0.23\textwidth]{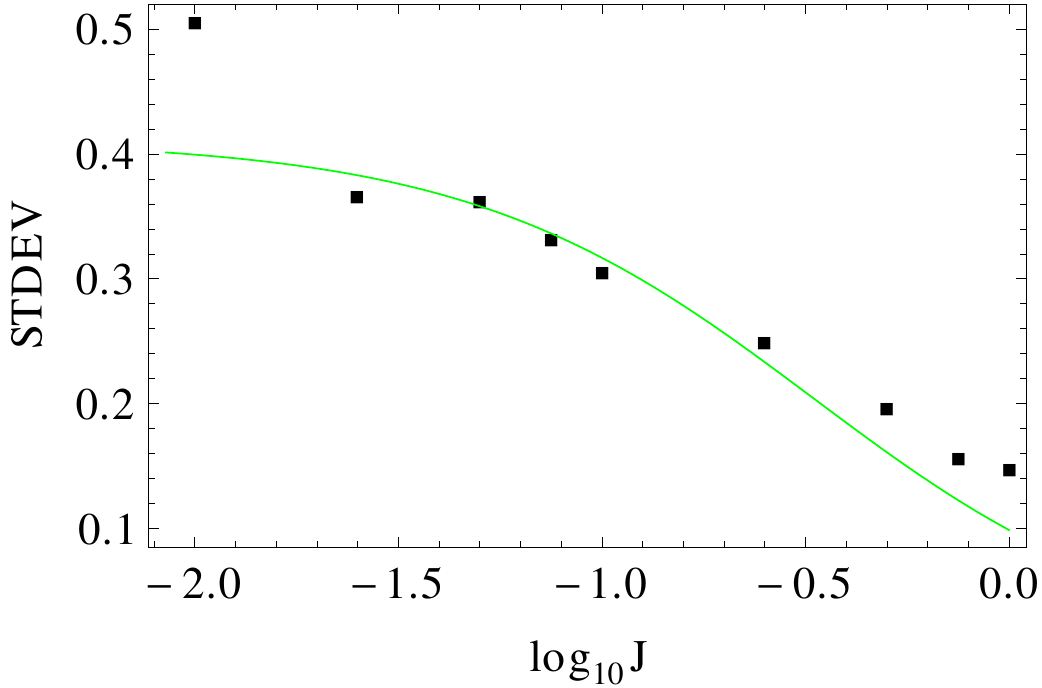}
\caption{Relaxation times of an inverse gamma process. Top: $\si = \sqrt{0.1}$ and bottom: $\si = 1$. Mean: black squares: numerical results; red line: theoretical estimate (\ref{ST:eq:relaxation_time_fixed_sigma_mean}) with $c_{1} = 1$; blue line: theoretical estimate (\ref{ST:eq:relaxation_time_approximation}). STDEV: black squares: numerical results; green line: theoretical estimate (\ref{ST:eq:relaxation_time_fixed_sigma_STDEV}) with $c_{2} = 1/4$}
\label{ST:fig:relaxation_time_mean_STDEV}
\end{figure}

Numerically, we consider an ensemble of paths described by (\ref{ST:eq:IGa_relaxation_time}). The relaxation time is then defined as such when the p-value of the ensemble of $X$ conforming to the IGa distribution is larger than 0.1. In our computation, 5000 paths are considered for each relaxation time. Our results are shown in Fig. \ref{ST:fig:relaxation_time_mean_STDEV}. Clearly, our estimates (\ref{ST:eq:relaxation_time_fixed_sigma_mean}) - (\ref{ST:eq:relaxation_time_fixed_sigma_STDEV}) with $c_{1} = 1$ and $c_{2} = 1/4$ fit the data quite well.

\section{Maximum likelihood estimation of GGa and GIGa}\label{Max_Likelihood}

For PDF $f(x|\te)$, with parameter(s) $\te$ and a dataset $\{x_i\}$ of size $n$, the likelihood function is
\begin{equation}
\prod_{i=1}^n f(x_i|\te)
\end{equation}
and the log-likelihood function is 
\begin{equation}
\fr{1}{n} \sum_{i=1}^n \log f(x_i|\te) . 
\end{equation}
The maximum likelihood estimation of $\te$ should maximize the log-likelihood function. Here we consider the case of GGa and GIGa. Generalized Student's \emph{t}-distribution is discussed in \cite{MaThesis2013}.

Since, as already mentioned, the PDFs of GGa and GIGa formally follow $\GGa(x; \al, \be, \ga) \xleftrightarrow{\ga\leftrightarrow -\ga} -\GGa(x; \al, \be, -\ga) = -\GIGa(x; \al, \be, \ga)$, it is sufficient to consider GGa. Setting to zero partial derivatives over $\al$, $\be$, and $\ga$ of the log-likelihood function for $\GGa(x;\al,\be,\ga)$ gives
\begin{eqnarray}
(\fr{1}{n} \sum_{i=1}^n \log x_i^\ga) - \log\be^\ga - \psi(\al) = 0 \\
\fr{\fr{1}{n} \sum_{i=1}^n x_i^\ga}{\be^\ga} - \al = 0 \\
1 - \fr{1}{n} \sum_{i=1}^n \lf( \fr{x_i^\ga}{\be^\ga} -\al \rg) \log\fr{x_i^\ga}{\be^\ga} = 0. 
\end{eqnarray}
With the definition
\begin{equation}
\fr{1}{n} \sum_{i=1}^n f(x_i) = \overline{f(x)}, 
\end{equation}
we obtain
\begin{eqnarray}
\overline{\log x^\ga} - \log\be^\ga - \psi(\al) = 0 \label{ST:eq:GGa_MLE1}\\
\fr{\overline{x^\ga}}{\be^\ga} - \al = 0 \label{ST:eq:GGa_MLE2}\\
1 - \overline{\lf( \fr{x^\ga}{\be^\ga} -\al \rg) \log\fr{x^\ga}{\be^\ga}} = 0. \label{ST:eq:GGa_MLE3}
\end{eqnarray}
Substitution of Eq. (\ref{ST:eq:GGa_MLE2}) into (\ref{ST:eq:GGa_MLE1}) and Eq. (\ref{ST:eq:GGa_MLE3}) yileds
\begin{equation}\label{ST:eq:GGa_MLE_program1}
\overline{\log x^\ga} - \log\overline{x^\ga} + \log\al - \psi(\al) = 0  
\end{equation}
and
\begin{equation}\label{ST:eq:GGa_MLE_program2}
\al = \left[{\overline{\lf( \fr{x^\ga \log x^\ga}{\overline{x^\ga}} - \log x^\ga \rg)}}\right]^{-1} .
\end{equation}
Eqs. (\ref{ST:eq:GGa_MLE_program1}) and (\ref{ST:eq:GGa_MLE_program2}) form the basis of a maximum likelihood estimation program. Given a $\ga$, from (\ref{ST:eq:GGa_MLE_program2}), we calculate $\al$ and then insert $\al$ into the lhs of (\ref{ST:eq:GGa_MLE_program1}). A bisection method can be realized over $\ga$. For more details, see \cite{MaThesis2013}. We note that Eqs. (\ref{ST:eq:GGa_MLE_program1}) and (\ref{ST:eq:GGa_MLE_program2}) result in either GGa ($\ga>0$), or GIGa ($\ga<0$). \cite{MaThesis2013}.

\section{Log-log plot of distribution tails}\label{Tail_Power}

\begin{figure}[htp]
\centering
\includegraphics[width=0.345\textwidth]{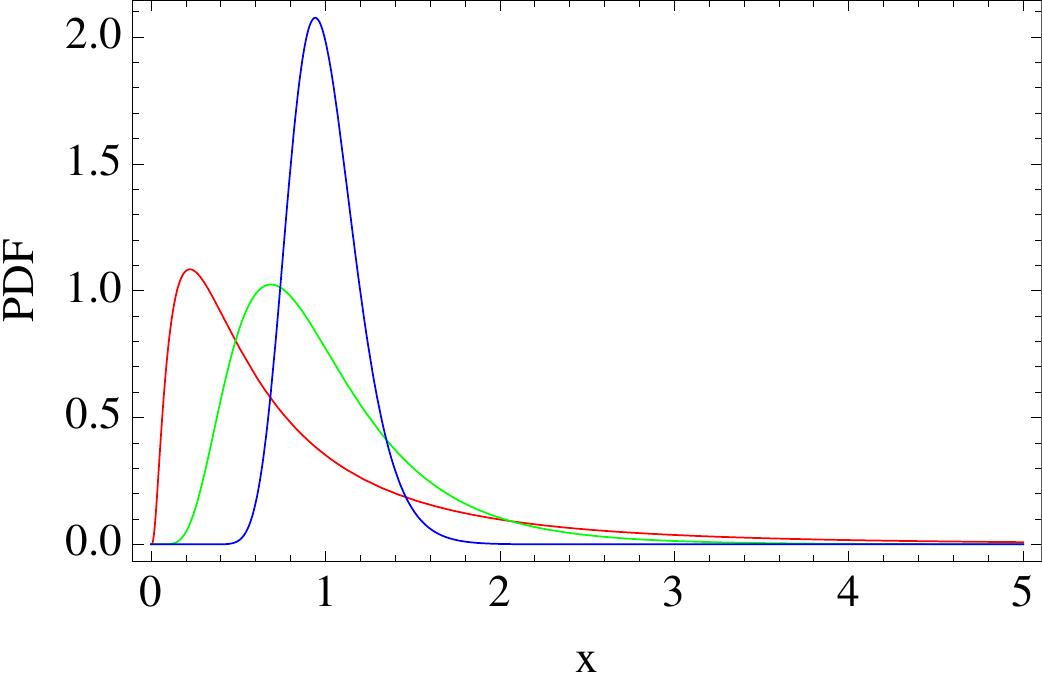}
\includegraphics[width=0.345\textwidth]{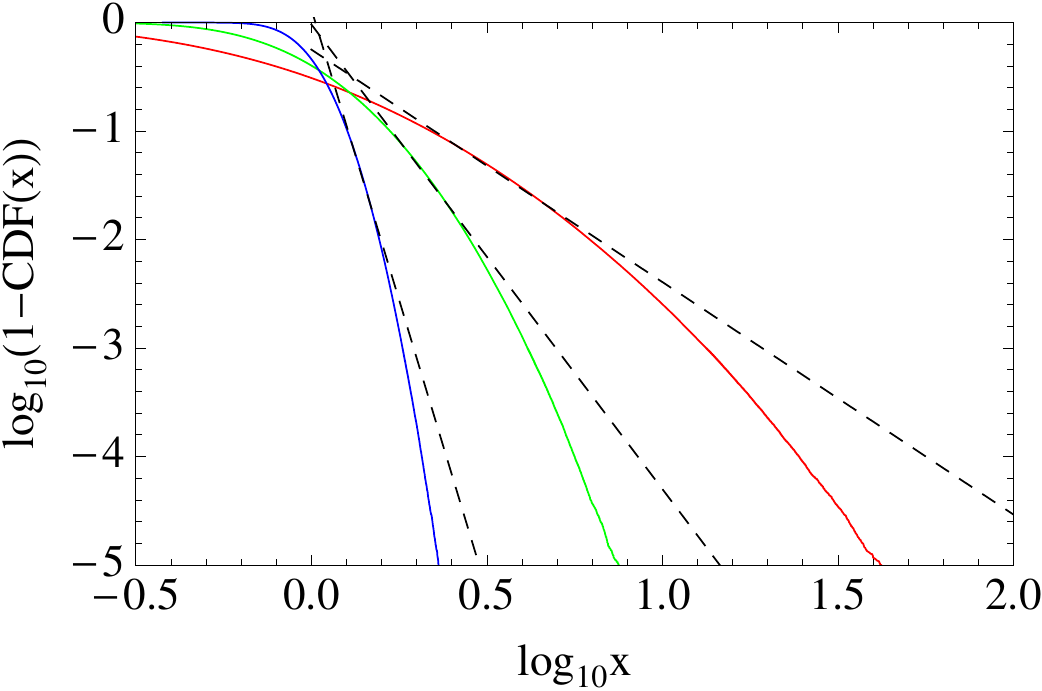}
\caption{Top: plots of PDF of $\LN(x; \mu,\si)$ with mean 1. The left red, middle green, and right blue curves correspond to parameter $\si = 1, 0.5$, and $0.2$ respectively. Bottom: log-log plots of simulated data sampled from the LN distributions. Below $-1$ of the y-axis, the left blue, middle green, and right red curves correspond to $\si = 0.2, 0.5$, and $1$ respectively. The dashed lines are fitting of $\log_{10}(1-\text{CDF}(x))$ vs. $\log_{10} x$ in a range of CDF from 0.9 to 0.99.}
\label{ST:fig:LN_simulation_loglogplot}
\includegraphics[width=0.345\textwidth]{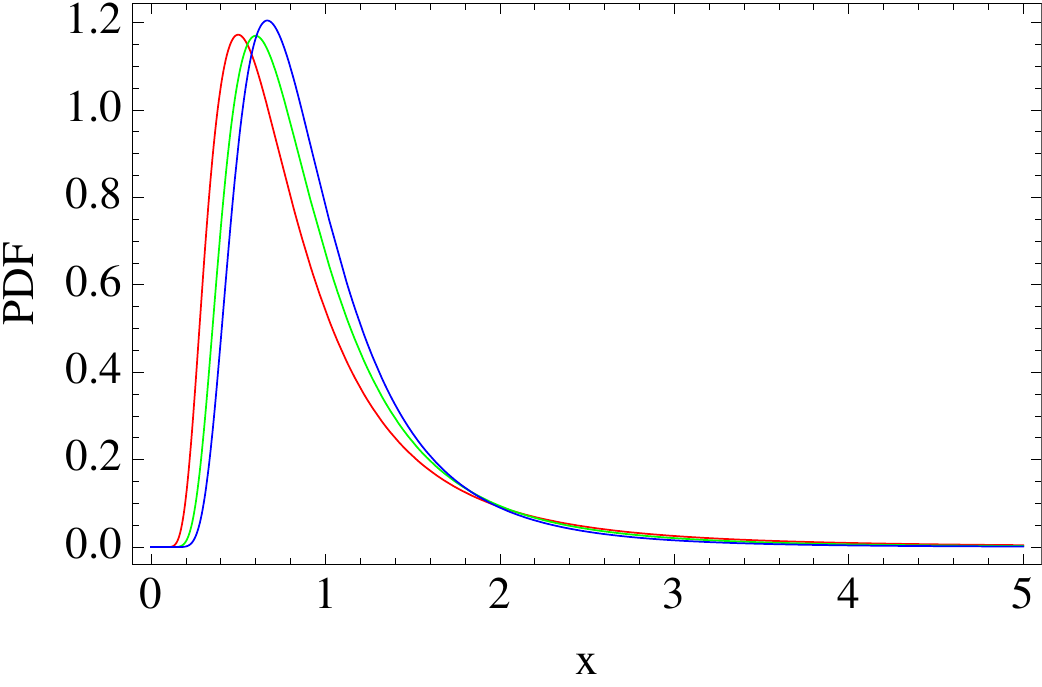}
\includegraphics[width=0.345\textwidth]{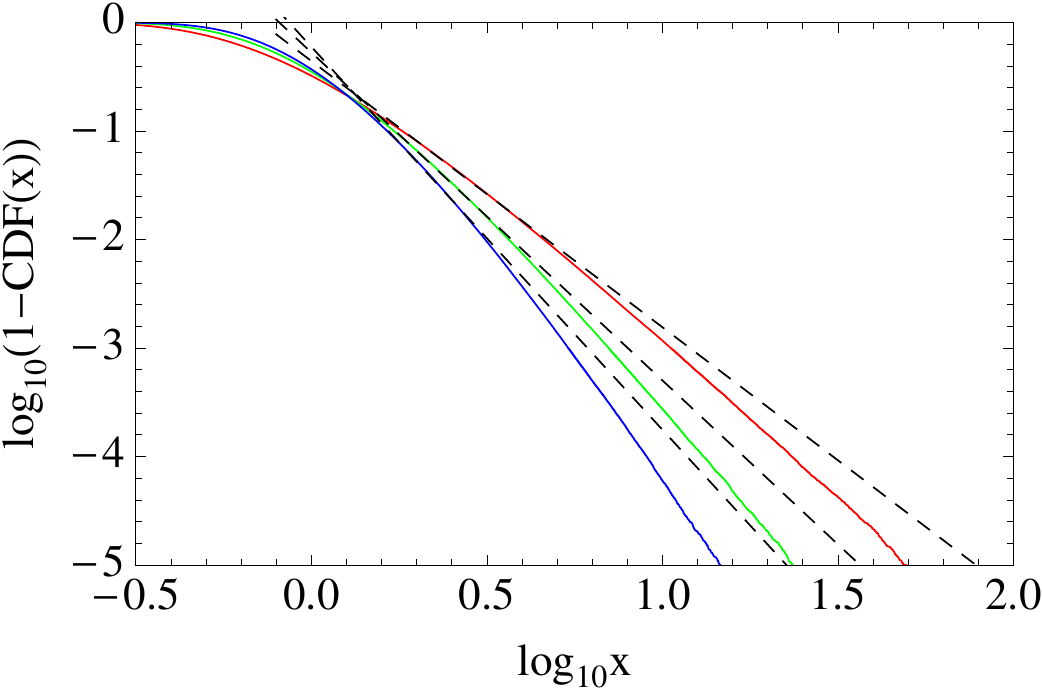}
\caption{Top: plots of PDF of $\IGa(x; \al,\be)$ with mean 1. The left red, middle green, and right blue curves correspond to $\al = 3, 4$, and $5$ respectively. Bottom: log-log plots of simulated data sampled from the IGa distributions. Below $-1$ of the y-axis, the left blue, middle green, and right red curves correspond to $\al = 5, 4$, and $3$ respectively. The dashed lines with slopes $-3.5, -3.0$, and $-2.5$ respectively are fitting of $\log_{10}(1-\text{CDF}(x))$ vs. $\log_{10} x$ in a range of CDF from 0.9 to 0.99. }
\label{ST:fig:IGa_simulation_loglogplot}
\end{figure}

\begin{figure}[htp]
\centering
\includegraphics[width=0.345\textwidth]{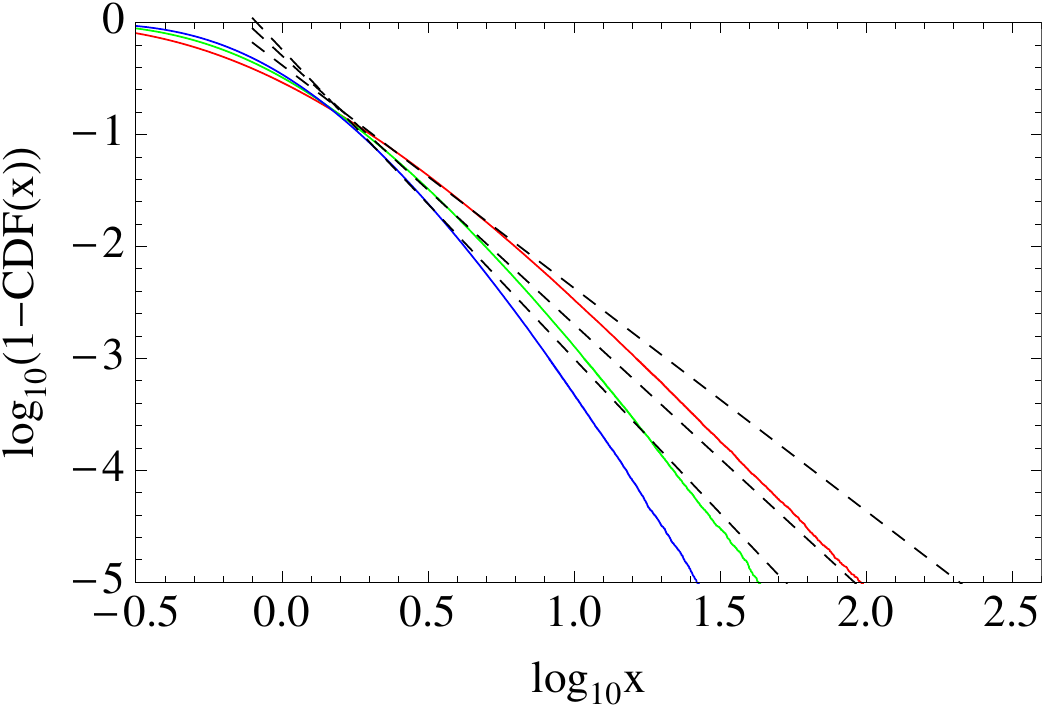}
\includegraphics[width=0.345\textwidth]{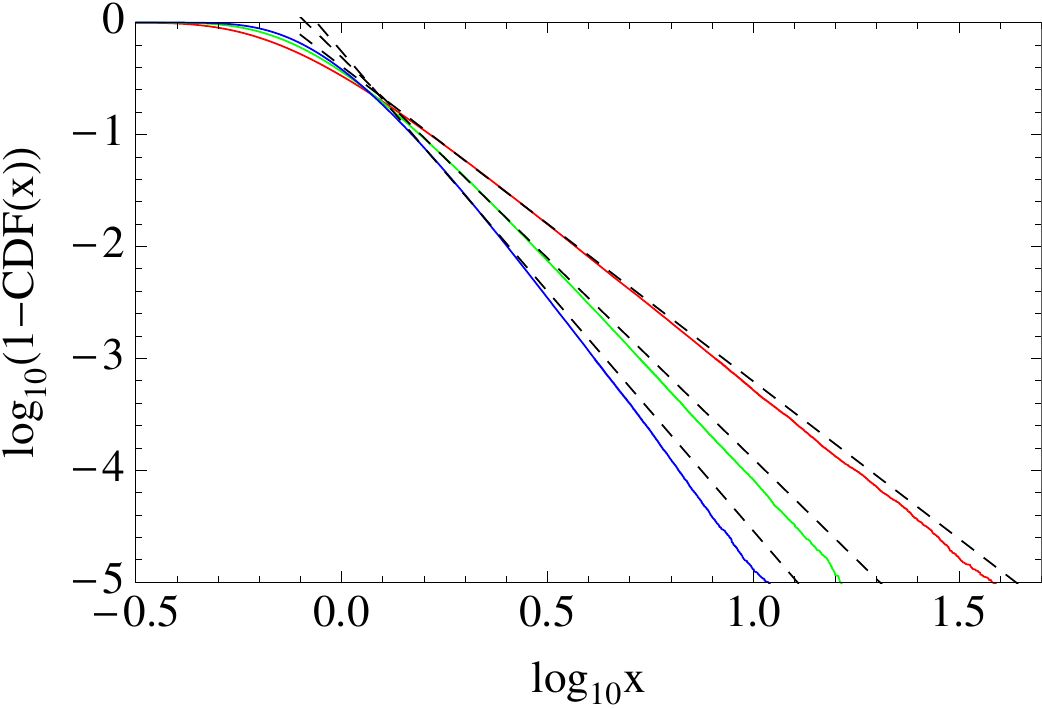}
\caption{Log-log plots of simulated data sampled from GIGa distributions $\GIGa(x; \al,\be, 0.5) (top)$ and $\GIGa(x; \al,\be,2)$ (bottom) with mean 1. Below $-1$ of the y-axis, the left blue, middle green, and right red curves correspond to $\al=2.5, 2$, and $1.5$ respectively. The dashed lines with slopes $-2.8, -2.4$, and $-2.0$ respectively (top) and $-4.3, -3.6$, and $-2.8$ (bottom) are fitting of $\log_{10}(1-\text{CDF}(x))$ vs. $\log_{10} x$ in a range of CDF from 0.9 to 0.99. }
\label{ST:fig:GIGa_ga_half_simulation_loglogplot}
\end{figure}

The exponent of a power law tail can be easily calculated once we notice that 
\begin{equation}
1-\text{CDF}(x) = \int_x^{+\infty} \text{PDF}(x) dx . 
\end{equation}
If $\text{PDF}(x) \propto x^{-1-k}$ with $x\gg 1$, then \footnote{When calculating the emperical CDF of a sorted sequence $\{x_1, x_2, ..., x_n\}$, we set the empirical CDF as $\lf\{ \fr{1}{n+1}, \fr{2}{n+1},\cdots, \fr{n}{n+1} \rg\}$ for two reasons. First empirical CDF $\lf\{ \fr{1}{n}, \fr{2}{n},\cdots, \fr{n}{n} \rg\}$ renders $\log(1-\text{CDF})$ meaningless for $x_n$. Second, it is symmetrical as needed for symmetrical distributions such as (generalized) Student's \emph{t}-distribution.}

\begin{equation}
\log( 1-\text{CDF}(x) ) \propto \text{const} - k \log x . 
\end{equation}
In Figs. \ref{ST:fig:LN_simulation_loglogplot} and \ref{ST:fig:IGa_simulation_loglogplot}, we show the log-log plot of the tail of LN and IGa distributions respectively. Clearly, a straight line fit is considerably better for the latter, even though the fitted slope does not agree with the theoretical value. Towards this end, in Fig. \ref{ST:fig:GIGa_ga_half_simulation_loglogplot}, we show log-log plots of the tail of GIGa distributions for $\ga=0.5$ and $\ga=2$. The empirical trend emerging form the IGa and GIGa plots is that the straight line fits of log-log plots become progressively better as $\ga$ gets larger. 

\begin{figure}[htp]
\centering
\includegraphics[width=0.45\textwidth]{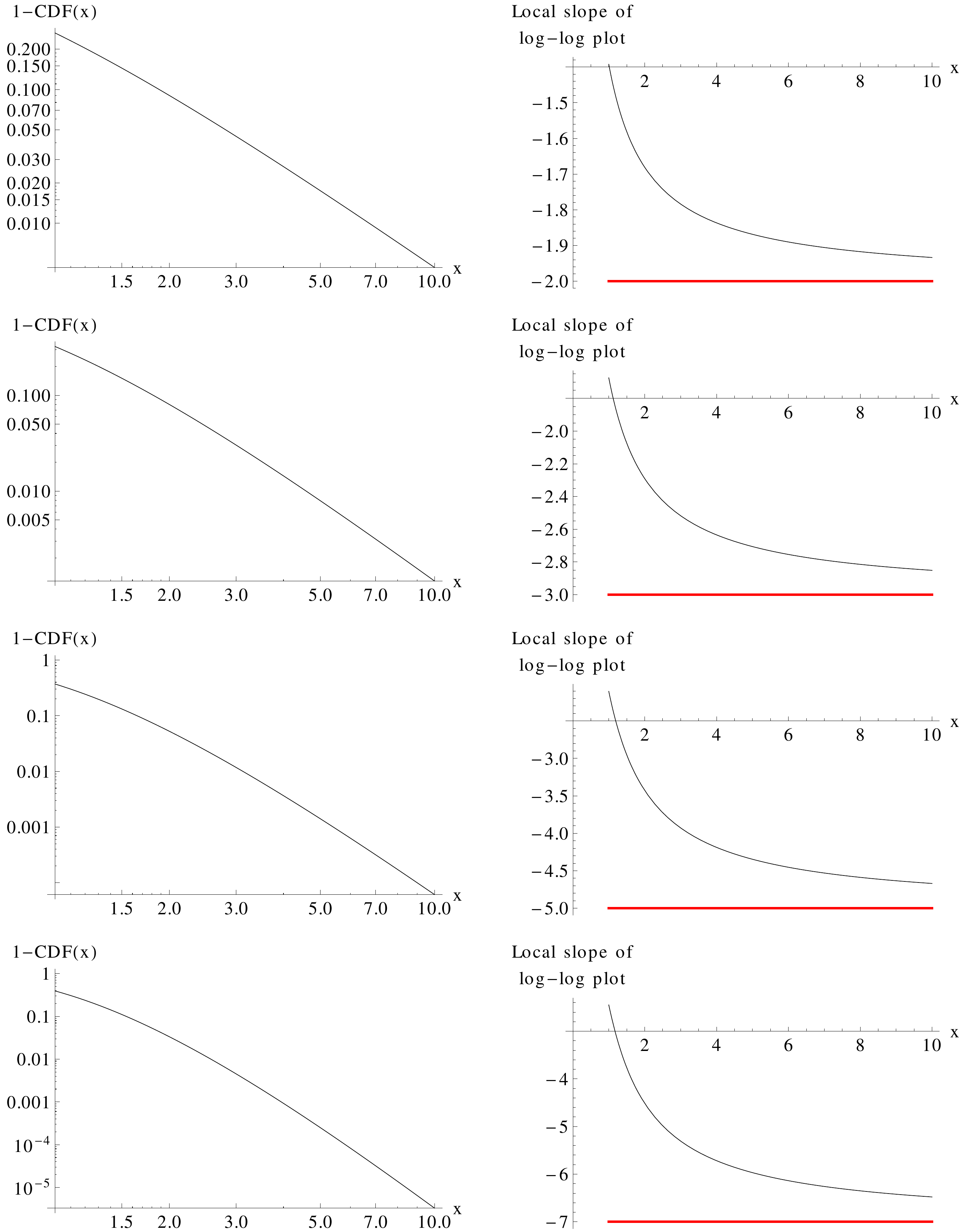}
\caption{Local slope of log-log plot of IGa distribution IGa$(x; \al, \be)$ with mean 1 ($\be=\Ga(\al)/\Ga(\al-1)$). The left column is the log-log plot and the right one is the local slope of the log-log plot from Eq. (\ref{RT:eq:local_slope_GIGa}). $\al$ is 2, 3, 5, and 7 for the first, second, third, and fourth rows respectively. The red lines are $-\al$: the limit of the local slope when $x\rightarrow\infty$.}
\label{RT:fig:local_slope_IGa}
\end{figure}

\begin{figure}[htp]
\centering
\includegraphics[width=0.45\textwidth]{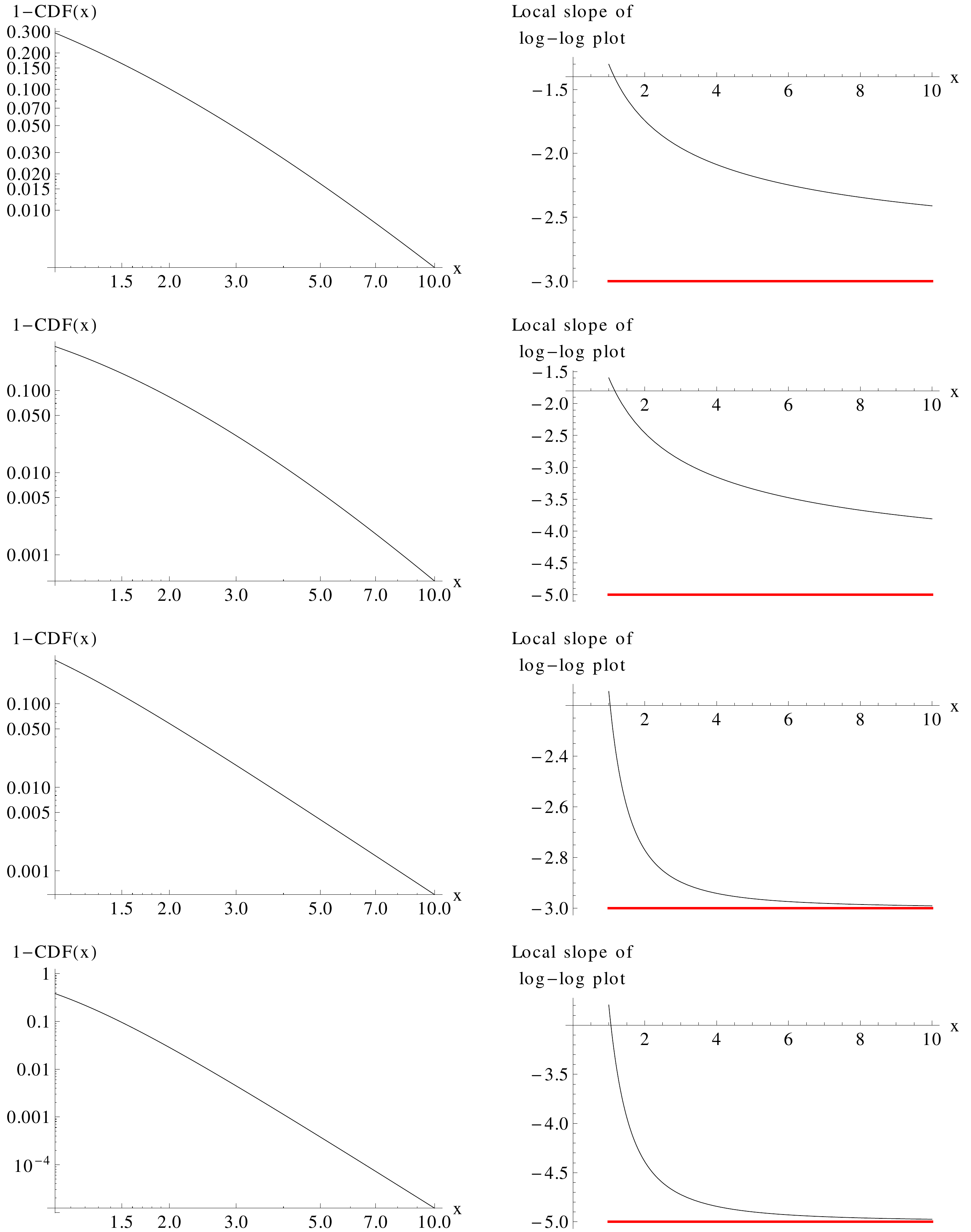}
\caption{Local slope of log-log plot of GIGa$(x; \al, \be, \ga)$ with mean 1 ($\be=\Ga(\al)/\Ga(\al-1/\ga)$). The left column is the log-log plot and the right one is the local slope of the log-log plot from Eq. (\ref{RT:eq:local_slope_GIGa}). $\{\al,\ga\}$ is $\{6,0.5\}$, $\{10,0.5\}$, $\{1.5,2\}$, and $\{2.5,2\}$ for the first, second, third, and fourth rows respectively. The red lines are $-\al\ga$: the limit of the local slope when $x\rightarrow\infty$.}
\label{RT:fig:local_slope_GIGa}
\end{figure}

To understand this $\ga$-dependence the difference between the theoretical and fitted slope, we consider the local slope of the log-log plot.
\begin{equation}\label{RT:eq:local_slope_def}
\fr{d\log(1-\text{CDF}(x))}{d\log x} .
\end{equation}
For GIGa (and IGa, $\ga=1$), the local slope is given by
\begin{equation}\label{RT:eq:local_slope_GIGa}
\fr{\ga  e^{-\lf(\fr{\be}{x}\rg)^{\ga}} \lf(\fr{\be}{x}\rg)^{\al\ga}}{\Gamma
   (\al) \lf(Q\lf(\al ,\lf(\fr{\be}{x}\rg)^{\ga}\rg)-1\rg)}
\end{equation}
with the regularized gamma function $Q(s,x)={\Ga(s,x)}/{\Ga(s)}$, where $\Ga(s,x)\equiv \int_x^\infty t^{s-1}e^{-t}dt$ is the incomplete gamma function. The local slopes are shown, as function of $x$ in Figs. \ref{RT:fig:local_slope_IGa} and \ref{RT:fig:local_slope_GIGa} respectively. It is clear that the local slope can differ substantially from its limiting (saturation) value. As $\ga$ becomes larger, the local slope tends closer to its limiting value.  

\begin{figure}[htp]
\centering
\includegraphics[width=0.45\textwidth]{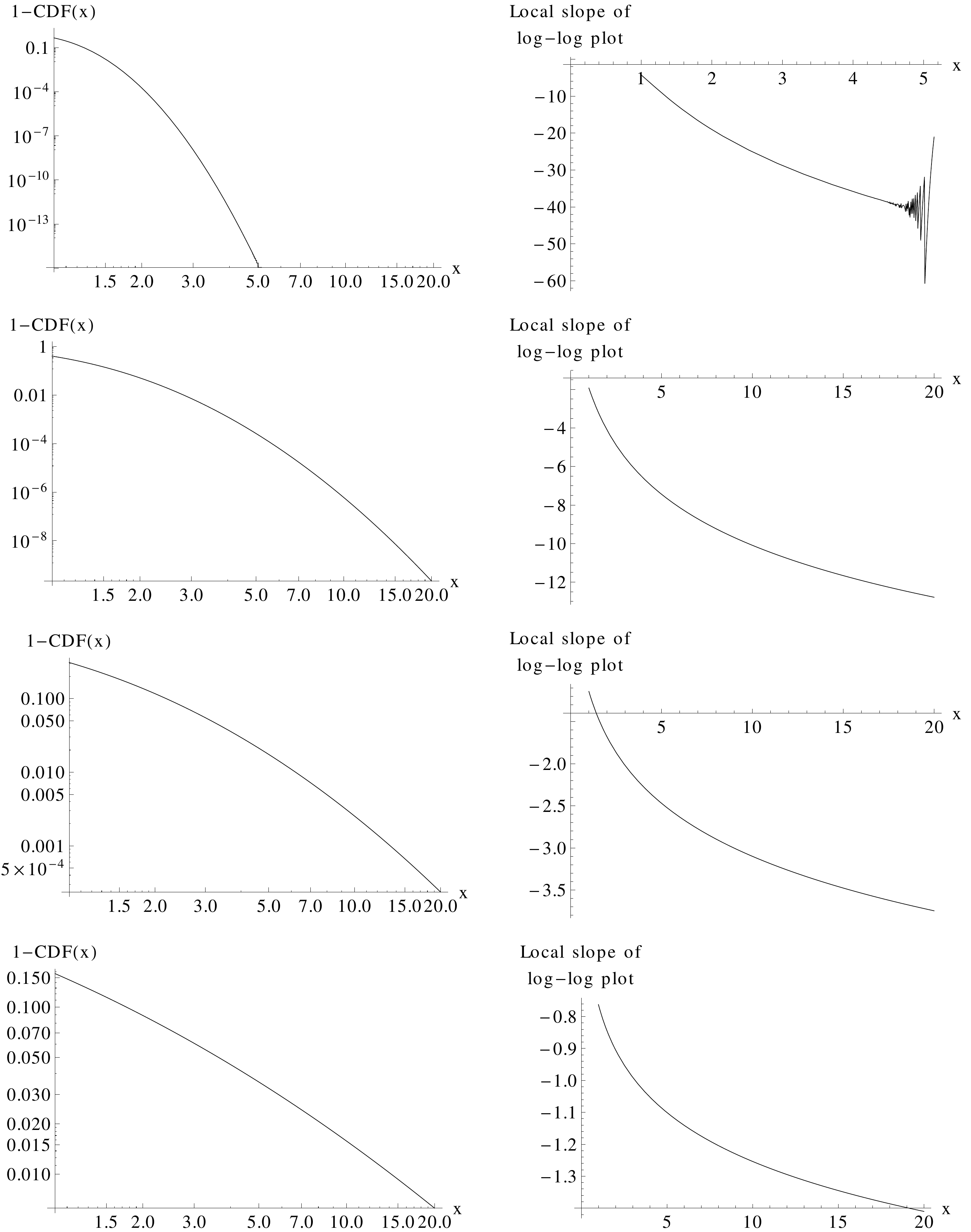}
\caption{Local slope of log-log plot of lognormal distribution. The mean of the distribution is set as 1 through $\mu = -\si^2/2$. The left column is the log-log plot and the right one is the local slope of the log-log plot in Eq. (\ref{RT:eq:local_scope_LN}). $\si$ is 0.2, 0.5, 1, and 2 for the first, second, third, and fourth row respectively. The jagged part of the top right plot is due to computational precision.}
\label{RT:fig:local_slope_LN}
\end{figure}

For the LN distribution, the local slope is given by
\begin{equation}\label{RT:eq:local_scope_LN}
\fr{\sqrt{\fr{2}{\pi}} e^{-\fr{(\log x  - \mu)^2}{2 \si ^2}}}{\si \lf(1 + \text{erf}\lf(-\fr{\log x - \mu}{\sqrt{2}\si}\rg)\rg)} ,
\end{equation}
which slowly decreases with $x$. But as is clear from (\ref{RT:eq:local_scope_LN}) and Fig. \ref{RT:fig:local_slope_LN}, the local slope does not saturate when $x\rightarrow \infty$. 

\end{document}